\documentclass[12pt,preprint]{aastex}
\usepackage{psfig,natbib,float}
\shortauthors{Bower et al.}
\shorttitle{PiGSS II}
\begin{document}

\newcommand\degd{\ifmmode^{\circ}\!\!\!.\,\else$^{\circ}\!\!\!.\,$\fi}
\newcommand{\etal}{{\it et al.\ }}
\newcommand{\uv}{(u,v)}
\newcommand{\rdm}{{\rm\ rad\ m^{-2}}}
\newcommand{\msuny}{{\rm\ M_{\sun}\ y^{-1}}}
\newcommand{\mylesssim}{\stackrel{\scriptstyle <}{\scriptstyle \sim}}
\newcommand{\lsim}{\stackrel{\scriptstyle <}{\scriptstyle \sim}}
\newcommand{\gsim}{\stackrel{\scriptstyle >}{\scriptstyle \sim}}
\newcommand{\sci}{Science}
\newcommand{\boo}{Bo\"{o}tes\ }


\title{The Allen Telescope Array Pi GHz Sky Survey II.  
Daily and Monthly Monitoring for Transients and Variability in the \boo Field}

\author{
 Geoffrey C.\ Bower,
 David Whysong,
 Samantha Blair,
 Steve Croft, 
 Garrett Keating,
 Casey Law,
 Peter K.G. Williams,
 Melvyn C.H. Wright\altaffilmark{1}
}
\altaffiltext{1}{University of California, Berkeley, 601 Campbell Hall \#3411, Berkeley, CA 94720, USA; gbower@astro.berkeley.edu }

\begin{abstract}
We present results from daily radio continuum
observations of the \boo field as part of the Pi GHz Sky
Survey (PiGSS).  These results are part of a systematic and unbiased
campaign to characterize 
variable and transient sources in the radio sky.
The observations include 78 individual epochs distributed over 5 months
at a radio frequency of 3.1 GHz with a median
RMS image noise in each epoch of 2.8 mJy.
We produce 5 monthly images with a median RMS of 0.6 mJy.
No transient radio sources are detected in the daily or monthly images.
At 15 mJy, we set an
upper limit ($2\sigma$) to the surface density of 1-day radio transients 
at 0.025 deg$^{-2}$.  At 5 mJy, we set an upper limit ($2\sigma$) to the surface density of
1-month radio transients at 0.18 deg$^{-2}$.  We also produce
light curves for 425 sources and explore the variability properties 
of these sources.  Approximately 20\% of the sources exhibit some variability
on daily and monthly time scales.  The maximum RMS fractional modulations
on the 1 day and 1 month time scales for sources brighter than 10 mJy are 2 and 0.5, 
respectively.  The probability of a daily fluctuation for all sources and all epochs
by a factor of 10 is less than $10^{-4}$.
We compare the radio to mid-infrared 
variability for sources in the field and find no correlation.  Finally, we apply the
statistics of transient and variable populations to constrain 
models for a variety of source classes.
\end{abstract}


\section{Introduction}

The temporal properties of the radio wavelength sky are poorly understood.  
Much of what is known results from follow-up of events discovered at 
optical, X-ray, or $\gamma$-ray wavelengths such as supernovae, gamma-ray
burst, and X-ray binaries.  An important exception is the sub-second domain,
which has been systematically, if incompletely, explored, leading to the
discovery of pulsars \citep{1969Natur.224..472H}, rotating radio transients
\citep{2006Nature...mcl}, and other unexplained
phenomena \citep{2007Sci...318..777L,2009GeoRL..3613202R,2010MNRAS.402..855B,2011arXiv1104.2727K}.  Longer-time 
scale searches, which tend to be sensitive to
synchrotron sources, are rarer and the sky coverage is less complete.

Systematic searches on long timescales for radio transients (RTs) are undergoing a renaissance
that is defining the available parameter space.  These searches are motivated
by the specific promise of orphan gamma-ray burst afterglows
\citep{2008MNRAS.390..675R}, radio
supernovae \citep{2009A&A...499L..17B}, tidally-disrupted stars around massive black holes
\citep{2011arXiv1102.1429G,2011arXiv1103.4328B},
intrinsic AGN activity such as found in III Zw 2 \citep{1999ApJ...514L..17F},
extreme scattering events \citep{1987Natur.326..675F},
stellar events
\citep{2007ApJ...663L..25H}, and counterparts to gravity-wave events
\citep{2011arXiv1102.1020N}.  Additionally, these searches are motivated
by the desire to fully characterize the radio sky and discover the unexpected
and unpredicted.  Archival searches have proven fruitful in 
this search, uncovering a number of transients that are not easily explained
with known source classes
\citep{2002ApJ...576..923L,2006ApJ...639..331G,2007ApJ...666..346B,2010AJ....140..157B,2011ApJ...728L..14B,2011MNRAS.412..634B,2011arXiv1103.0511B}.
These archival surveys have been powerful for the volume probed but the absence of
the opportunity for real-time follow-up has significantly limited their value in
characterizing the source population.
Synoptic transit surveys have also been carried out 
\citep{2008NewA...13..519K,2009AJ....138..787M}.
In particular, a number of new variable and transient sources without deep radio
or optical counterparts have been found.  \citet{2010ApJ...711..517O} argue that
these may be bursts from galactic neutron stars; other evidence suggests that they could
be strong bursts from low-mass flare stars.  

The success of archival surveys has lead to a growing number of unbiased surveys 
including those carried out by the 
Allen Telescope Array \citep[ATA;][]{2009IEEEP..97.1438W}.  The ATATS 
survey places limits on rare, high flux density transients and variables
\citep{2010ApJ...719...45C,2011ApJ...731...34C}.
A recent VLA/EVLA survey \citep{2011arXiv1103.3010O} has now provided strong limits on the radio
transient and variable population at 5 GHz.  These authors also summarize existing
RT surveys at frequencies $>0.8$ GHz.  Long wavelength transients were searched
for with the Long Wavelength Array \citep{2010AJ....140.1995L}.
These archival and unbiased surveys 
join targeted surveys of specific source classes in characterizing the radio transient sky.
For example, 
the MASIV survey \citep{2003AJ....126.1699L,2008ApJ...689..108L} and 
the University of Michigan monitoring program \citep{1992ApJ...396..469H} provide 
characterization of bright, flat-spectrum radio sources, which are known to have
the strongest variability amplitudes.

The Pi GHz Sky Survey (PiGSS) is a key project of the ATA.
In \citet{2010ApJ...725.1792B}, hereafter Paper I, 
we presented the survey description, methods in observation, data reduction,
and source identification, and initial results on transient statistics.
Briefly, PiGSS is a 3.1 GHz survey of a large area of high galactic latitude sky
conducted synoptically with 100\arcsec\ resolution and milli-Jansky
sensitivity.  The PiGSS emphasis is on two surveys jointly conducted:
1) $10^4$ deg$^2$ of the sky with $b > 30^\circ$; and 2) individual
$\sim 11$ deg$^2$ fields that are observed repeatedly.
Paper I presented results from integrated results for a 11-deg$^2$ field in the
\boo constellation associated with the NOAO Deep Wide Field Survey 
\citep[NDWFS;][]{2004AAS...205.8106J,2009ApJ...701..428A}.  
Numerous sensitive, large area surveys have been
carried out for the \boo field, providing an important baseline for
comparison with PiGSS results.  We found no new radio sources that we could
convincingly identify as RTs.  This sets an upper limit on 
the RT surface density at the milliJansky flux limit.

In this paper,
we present daily results from the \boo field.  These consist
of 78 individual epochs over five months.
These observations provide a very sensitive probe of transient statistics.  They
also provide one of the first unbiased searches for variability in a sample of mJy
brightness sources.  In \S \ref{sec:obs}, we briefly review 
the observations and data analysis and present the full catalog of
light curves.  In \S \ref{sec:transients}, we 
present results on transient sources.  In \S \ref{sec:var}, we 
characterize the variability of individual sources.  In \S \ref{sec:conc},
we summarize our results.

\section{Observations, Analysis, and Results \label{sec:obs}}

PiGSS observations of the \boo field were conducted
between 2009 May 20 and 2009 September 29.  Typical observations 
consisted of three 3-minute snapshots distributed over an observing epoch,
which was between 4 and 12 hours in duration.  The \boo
observations were interleaved with those of a larger field.
Multiple hour angles were necessary to achieve good $(u,v)$-coverage
for these complex fields.
The \boo field consists of 7 individual pointings arranged in
a close-packed hexagonal configuration with the central pointing
located at $\alpha=14^h32^m$ and $\delta=34^\circ16^\prime$ (J2000).
Pointings are separated by $\Delta\theta=0.78^\circ$, the voltage 
full width half maximum at 3.14 GHz, which provides nearly uniform
sensitivity in the central region of the field.  Observations were
obtained in two separate frequency bands of 100-MHz width centered
at 3.04 and 3.14 GHz.  The digital correlator provided 1024 channels
for each frequency band with all correlations between parallel- and
cross-hand terms.  The bright calibrator 3C 286 was observed hourly.

Data reduction was performed with standard techniques including
flagging of bad data due to radio frequency interference, amplitude
and phase calibration using 3C 286, and imaging using the 2D FFT.
Most analysis was performed with the MIRIAD package 
\citep{1995adass...4..433S}.
The CLEAN images are convolved with a synthesized beam of $100\arcsec \times 100\arcsec$.
The majority of the reduction was performed with pipeline software
\citep{2010SPIE.7740E..39K} although some of it required human intervention.
Individual pointings were stitched together into images using
linear-mosaicking techniques.  We refer the reader to Paper I for
further details of the reduction technique.

\begin{figure}[H]
\psfig{figure=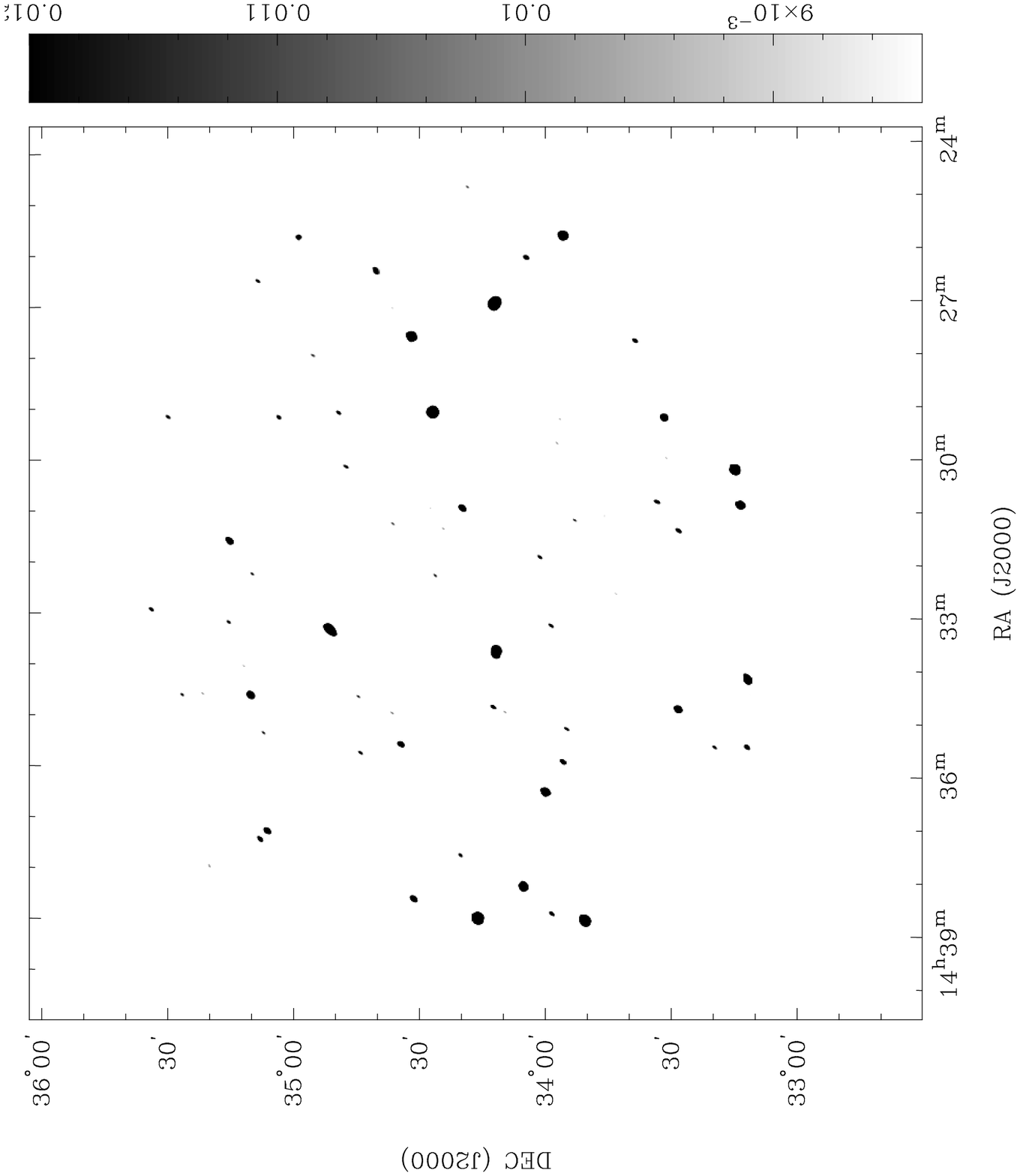,width=\textwidth}
\caption[]{Single-epoch image of the field from 30 August 2009.  The gray scale ranges from 8.4 to 12 mJy.
The image RMS is 1.4 mJy for this epoch.
\label{fig:image}}
\end{figure}

\begin{figure}[H]
\mbox{\psfig{figure=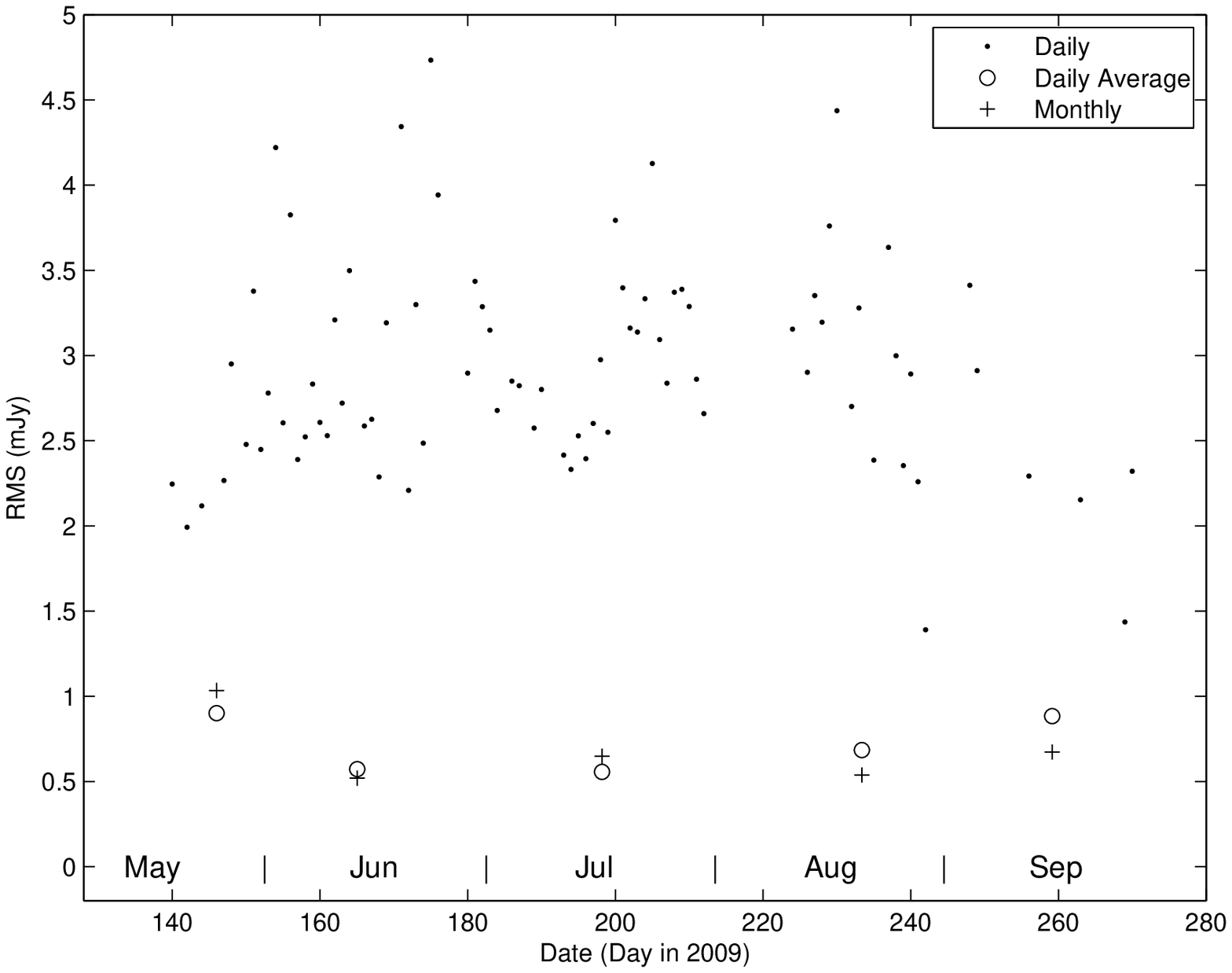,width=0.45\textwidth}\hspace*{0.1\textwidth}\psfig{figure=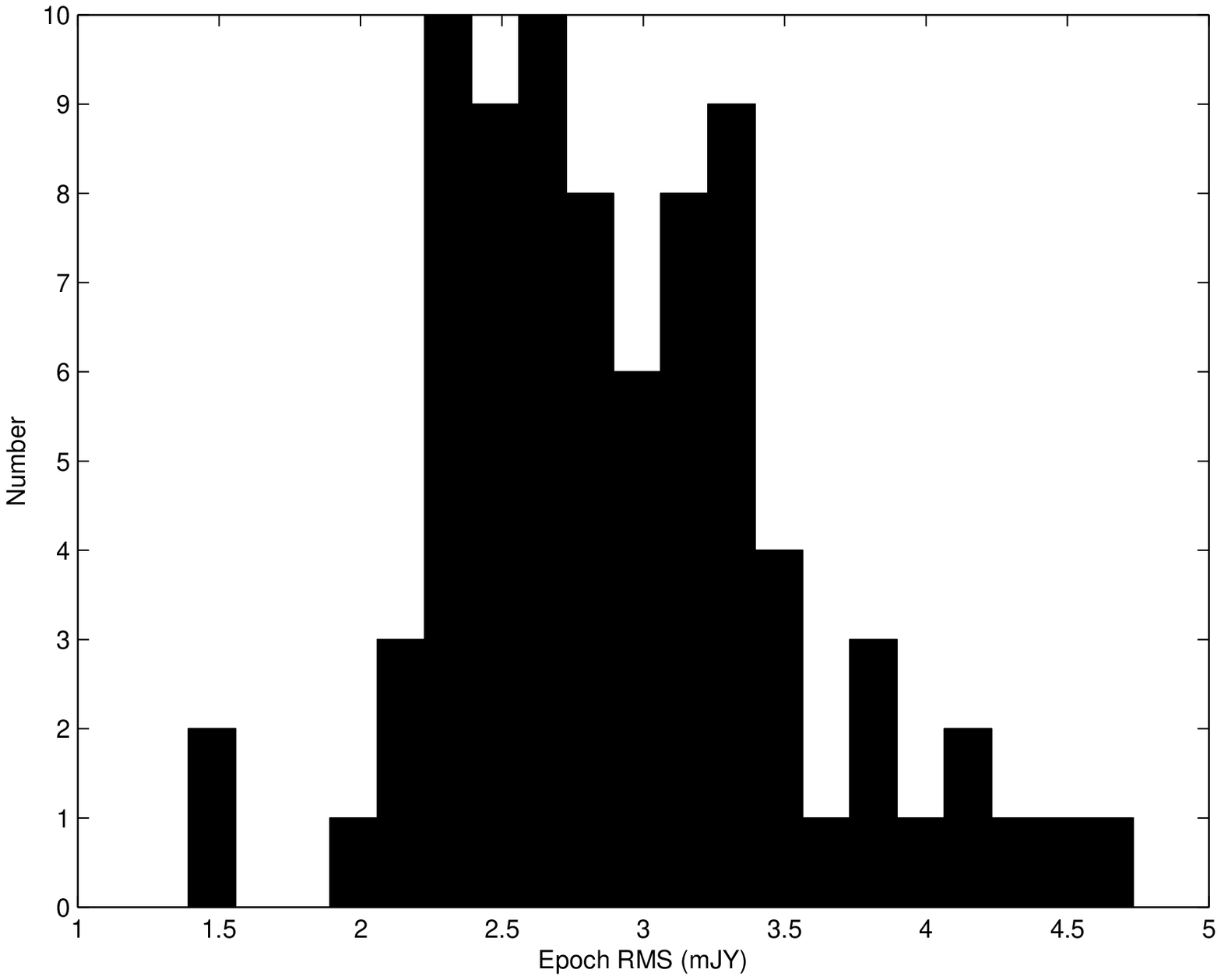,width=0.45\textwidth}}
\caption[]{({\em left}) RMS as a function of date for daily epoch images (dots), monthly average
of the daily epoch results (circles), and monthly averages (crosses).
The date is given in days in year 2009.  We also identify the months; vertical bars indicate the 
first of each month.  Note that the symbols representing the monthly data are aligned with the mean observing
epoch during each month.
({\em right}) Histogram of the daily RMS.
\label{fig:rms}}
\end{figure}

We show a single-epoch image in Fig.~\ref{fig:image}.
The RMS flux density in 
images from individual epochs has a minimum of 1.4 mJy, median of 2.8 mJy, and 
maximum of 4.7 mJy.  90\% of all images
have an RMS below 3.7 mJy (Fig.~\ref{fig:rms}).  These RMS statistics differ 
from those presented in Paper I due to a change in how we performed the CLEAN 
on the images.  We used a larger number of iterations (3000) for daily images
in Paper I, which reduced the image RMS below theoretical expectations and
lead to 
a systematic reduction of source flux densities (by as much as a factor of 3 at flux densities
around 10 mJy).  In this paper, we stopped
CLEAN when the first negative component is reached.  This typically occurred at fewer than
100 iterations.  As we show below, the agreement between daily and deep flux densities
is good.  There is a bias of $\sim 10\%$ in the daily flux densities relative
to the deep flux densities, which we think is an effect of $(u,v)$-coverage.  We do
not correct for that effect in these data because we are primarily interested in
transient detections and light curve variability, which are insensitive to the absolute
flux density.  The median number of sources detected at
a $6\sigma$ threshold per individual epoch is 30 and ranges from 17 to 69.
Systematic variations in the RMS with time are real and are due to changing performance of
the array, small variations in the number of observations per epoch, and the variable interference environment.  

We also create monthly images of the field that are integrated over all epochs in
the five-month span of these observations (2009 May through September).  These
images have a significant improvement in imaging quality due to the complete
$(u,v)$-coverage.  The RMS for the monthly images from June to September ranges from 0.52 to
0.67 mJy; the variation is representative of the range of daily epochs
per month.  The RMS for the May epochs is 1.0 mJy due to the smaller number of epochs in this month (7).
The noise statistics from the monthly images are consistent with averaging of
the daily data.
The number of $6\sigma$ sources detected per month ranged from
83 to 171.

\begin{figure}[H]
\psfig{figure=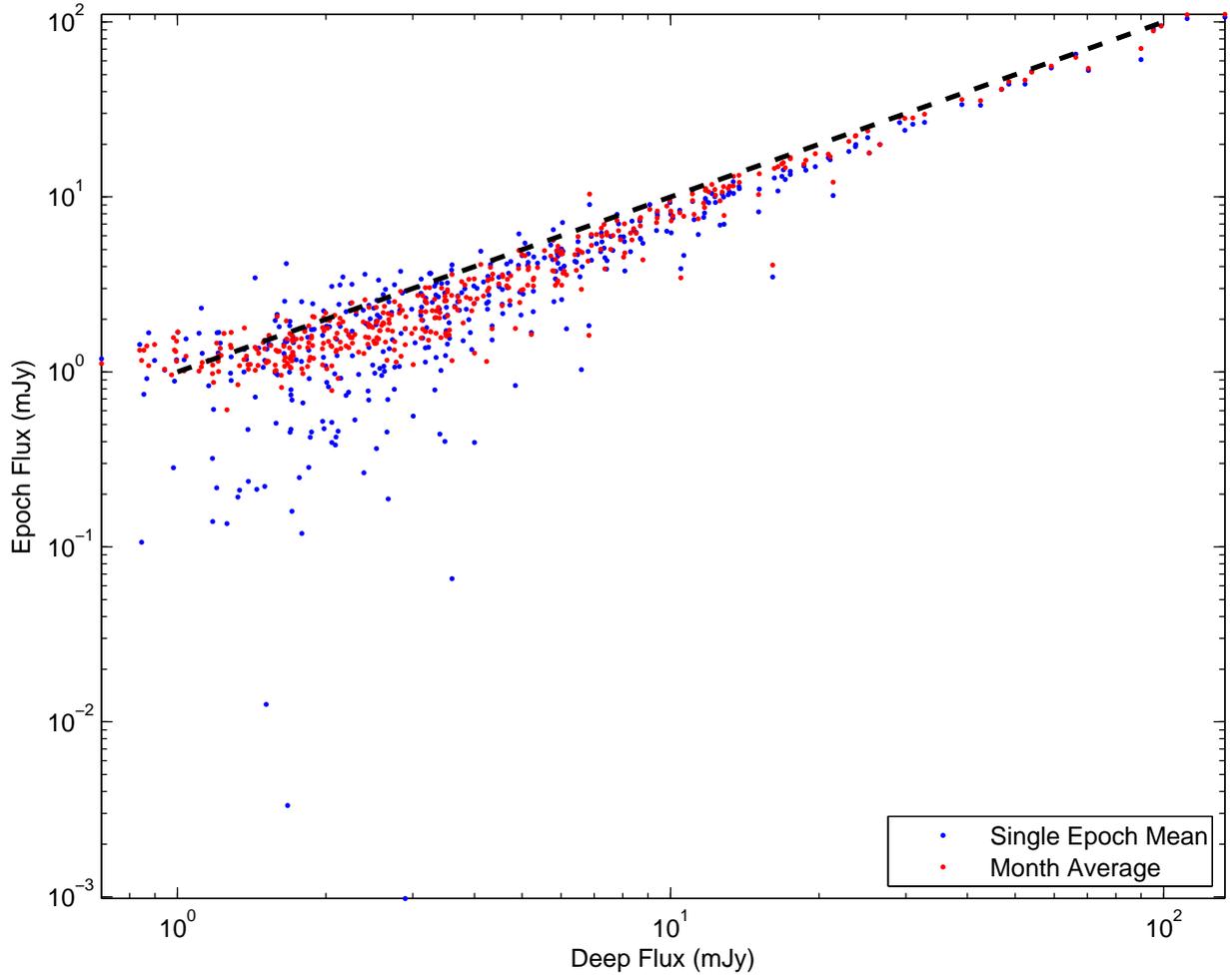,width=\textwidth}
\caption[]{Mean flux for each source from single epoch data and monthly data plotted
against the deep image flux for the source.  The dashed line is one to one.
\label{fig:fluxflux}}
\end{figure}

A table
of 425 sources made from the deep image was presented in Paper I.
Source finding in each epoch image
was carried out using the MIRIAD task {\tt sfind}
\citep{2002AJ....123.1086H}.  We search for point sources
down to a threshold of $4\sigma$; a higher threshold is employed for
identifying transient sources as described below.  
Additionally, light curves for sources from the deep image were 
generated by fitting 2D Gaussians at the fixed positions of the sources
from the deep images.  Errors are the RMS background in the residual following
removal of the best-fit 2D Gaussian model.  
We plot mean flux densities for all sources from daily 
and monthly data against the deep image flux density (Fig.~\ref{fig:fluxflux}).
The plot demonstrates agreement between these three different measures of
source flux density at the 10\% level for most sources.  In some sources,
larger differences may be due to confusion in regions of 
complex sources.

We extensively explored completeness of the deep PiGSS catalog in paper I.  Deviations
from previous catalogs such as NVSS and GB6 were explored for indication of 
transient sources; the absence of these sources is a measure of the completeness
at the threshold we explored.  In this paper, we primarily compare individual
epochs to the deep PiGSS field.  As discussed below, the absence of transient candidates at our
detection threshold indicates that the daily images are complete at the $6\sigma$ threshold.

\section{Candidate Transient Sources and Transient Statistics \label{sec:transients}}

We use a detection threshold of $6\sigma$ for transient sources
in this data set.  With this threshold and 78 epochs of observation, we 
find that the number of false sources expected is $\sim 10^{-3}$.  A 
$5\sigma$ threshold has an expectation of 0.4 false sources.  It is 
important to use an accurate estimate of the background noise for this 
calculation.  A 20\% underestimate of $\sigma$ will convert a $5\sigma$
source to $6\sigma$.  We use the RMS background signal after 
model source subtraction from the sub-image associated with the source to 
provide our best estimate.  For all sources in a typical epoch, the
RMS deviation in the RMS background estimates is $\lsim 7\%$ of the mean RMS
background.  Since this RMS deviation is much less than the 20\% required to
convert $5\sigma$ fluctuations into $6\sigma$ ``detections,'' statistical uncertainties in 
the RMS background or variations in the
RMS background across the field are unlikely to affect our rate of false
positives.  Additionally, a systematic bias across the field towards underestimation of the RMS noise
could introduce a large number of false positives.  The mean of
the RMS backgrounds is comparable to the RMS estimated from a region
without sources in each image; the RMS difference is $\sim 5\%$ of the mean RMS.
Thus, we do not see a significant systematic bias in the RMS estimate.

Our search through all epochs identified four $6\sigma$ sources that are well-fit
as point sources and that are not
associated with steady sources using a match radius of 60\arcsec.  For each
of these sources, we then require sources that are present in both frequency
bands at the same location with consistent flux densities.  This stage
leads to rejection of all sources.  If we reduce the threshold to $5\sigma$, we find
29 point-like sources that are not matched to steady sources.  Of these, 15 
sources survive the dual-frequency threshold.  The much larger number of sources
at this threshold indicates that we are underestimating the noise by $\sim 20\%$
or that errors in calibration or flagging are leading to image defects that we
are detecting as sources.  Given the consistency of RMS measurements discussed above,
we attribute the higher number of $5\sigma$ sources to imaging defects.
We take our threshold of $6\sigma$ with no transient
detections as the definite threshold that we can accept.

\subsection{Search of Monthly Images}

We conducted a search of the monthly images for transients at the $6\sigma$ threshold
by comparing the monthly epoch catalogs with the deep catalog.  
Several candidates are identified, however, inspection of the images reveals that these sources are 
identified in the individual epochs due to a different decomposition of complex sources.
Therefore, we report no RTs with a timescale of 1 month in this data.

\subsection{Candidate RTs:  J143621+334120 and J142554+343552}

\begin{figure}[H]
\psfig{figure=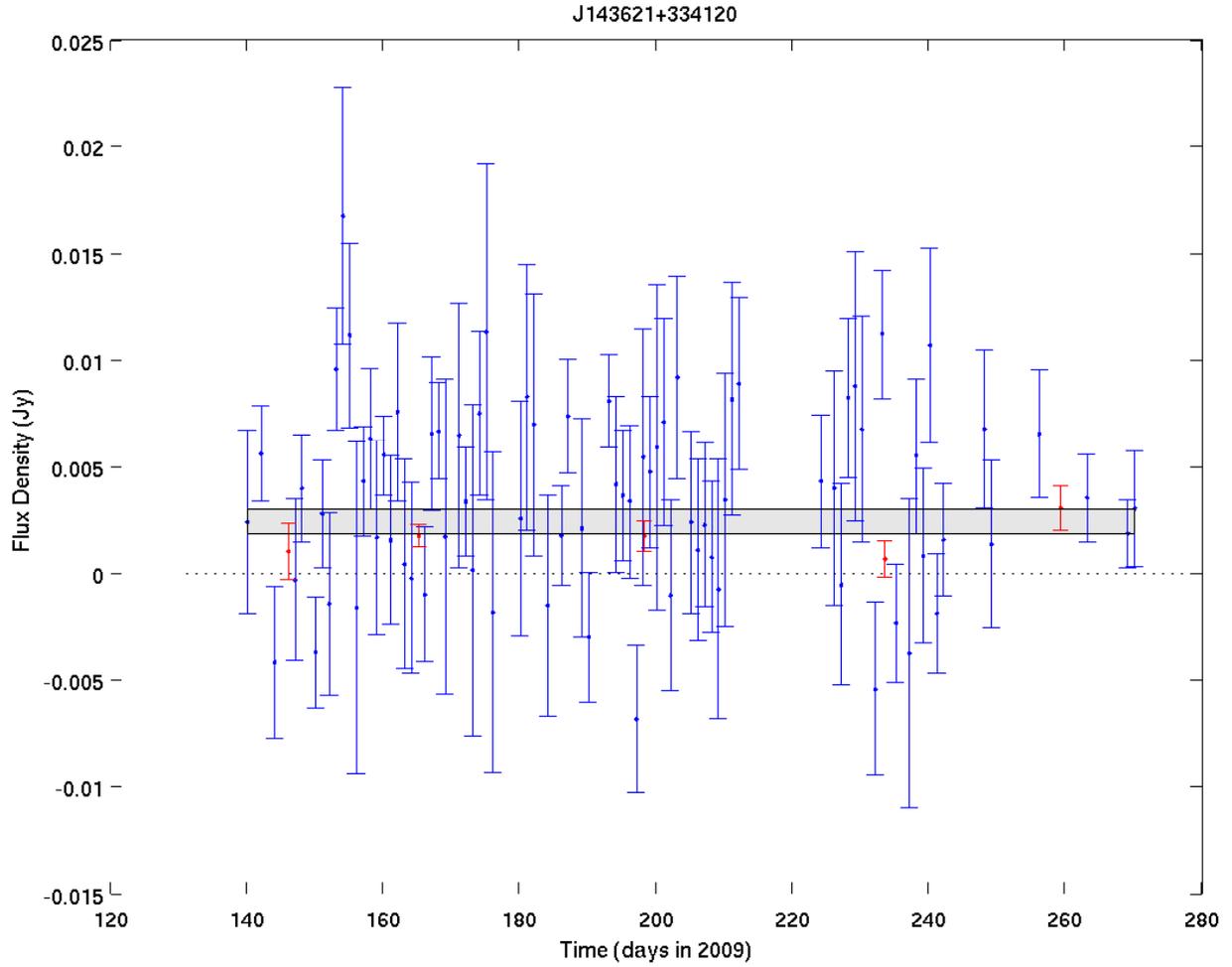,width=\textwidth}
\caption[]{Light curve of candidate RT J143621+334120.
Blue symbols indicate daily flux measurements, red symbols indicate monthly averages,
and the gray bar indicates the deep field flux measurement.
\label{fig:deeprtlc}}
\end{figure}

We also investigate the reality of the candidate RT J143621+334120 identified
in Paper I.  This source was identified in the deep image with flux density of $1.80 \pm 0.42$ mJy (4.3$\sigma$) and had no
counterpart in NVSS or WSRT-1400 catalogs.  We plot the light curve in the daily images
at the position of this source in Figure~\ref{fig:deeprtlc}.  There is no detection $>3\sigma$
of a source in any given epoch or in any month.  The highest flux density at the source position
is in the September monthly image but examination of the image reveals it to be non-point-like
and, therefore, difficult to unambiguously state as a real detection.  It is clear from the light
curve that the detection in the deep image is not the result of a transient event with a duration
of less than 4 months.  If it is a real source in the deep image, then it represents a longer
timescale transient.  The source is within 15\arcsec\ of two stars identified
in SDSS but we cannot conclusively make an association.

In Paper I we reported that the source 
GB6 J142554+343552 had no counterparts in the deep PiGSS image and other
radio surveys.  We note here that the source appears 
in the original GB6 catalog with a warning flag, and is, therefore, likely to be spurious.

\subsection{RT Surface Density}

\begin{figure}[H]
\psfig{figure=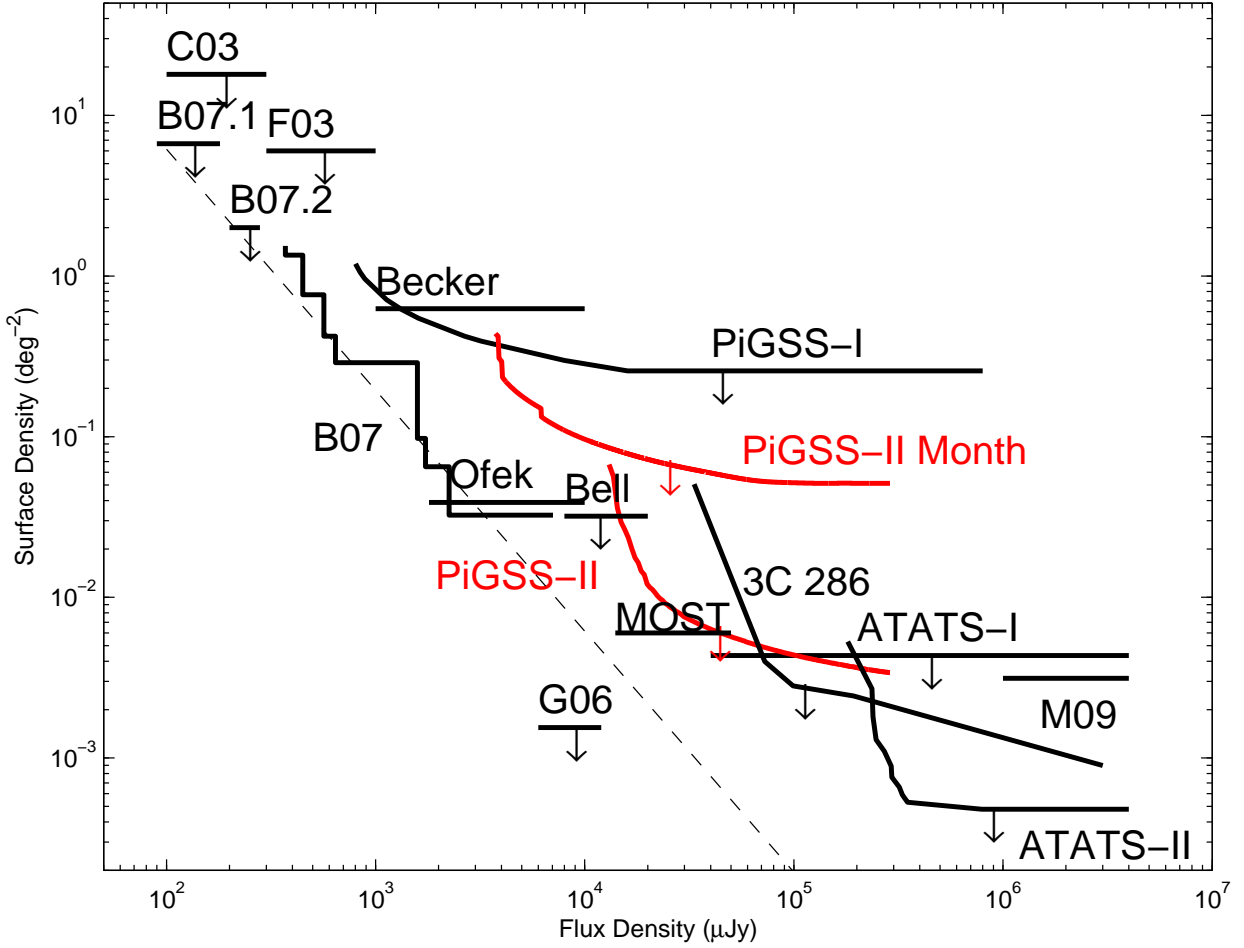,width=\textwidth}
\caption[]{RT surface density from PiGSS and other surveys as a function
of flux density.  The curved solid red lines labeled PiGSS-II
and PiGSS-II month
are the daily and monthly limits from this paper.  PiGSS-I refers to the limits
from \citet{2010ApJ...725.1792B}.
The solid black line with step functions shows the rate from 
\citet{2007ApJ...666..346B}.  The arrows show
$2\sigma$ upper limits for transients from 
\citet{2007ApJ...666..346B} with a
1-year timescale (B07.1), two-month timescale (B07.2), and for transients
from the comparison of the 1.4~GHz NVSS and FIRST surveys
\citep[G06;][]{2006ApJ...639..331G}, from the \citet{2003ApJ...590..192C}
survey (labeled C2003), from the \citet{2003AJ....125.2299F}
survey (labeled F2003), from ATATS \citep[I and II;][]{2010ApJ...719...45C,2011ApJ...731...34C},
from VLA archival survey of the 3C 286 field \citep{2011ApJ...728L..14B},
from MOST archival survey \citep{2011MNRAS.412..634B},
from the VLA calibrator archival survey \citep{2011arXiv1103.0511B},
from the VLA galactic plane survey \citep{2010AJ....140..157B},
from the VLA/EVLA survey \citep{2011arXiv1103.3010O},
and from the \citet{2009AJ....138..787M} survey (labeled M09).  
\label{fig:surfacedensity}
}
\end{figure}

We construct upper limits on the RT surface density using the same formalism developed in
Paper I and other papers \citep[e.g.,][]{2007ApJ...666..346B,2011ApJ...728L..14B}.  
Note that we are using the term surface density for 
clarity rather than two-epoch snapshot-rate, although these terms are interchangeable.
The limit from this search is based on 78 epochs of observation, the RMS as a function
of epoch, the total area observed, and the limit of no RT detections in these data.
The surface density is applicable to the characteristic timescale of $\sim 1$ day
from these observations.  We find a limit for flux densities greater than 15 mJy of 
$0.025$ deg$^{-2}$ and for flux densities greater than 100 mJy of $0.0043$ deg$^{-2}$.
These are $2\sigma$ limits on the transient surface density.  These limits represent
a nearly two-order of magnitude improvement in the surface density limit on the results
reported in Paper I.  From the monthly data, we estimate $2\sigma$ upper limits of
0.18 deg$^{-2}$ at 5 mJy and $0.05$ deg$^{-2}$ at 100 mJy.
We plot the surface density limits against other surveys in 
Figure~\ref{fig:surfacedensity}.

The upper limits on surface density from this search are consistent with upper limits
from previous searches with the ATA 
\citep{2010ApJ...725.1792B,2010ApJ...719...45C,2011ApJ...731...34C}
and with other instruments and archival data
\citep{2006ApJ...639..331G,2007ApJ...666..346B,2011MNRAS.412..634B,2011ApJ...728L..14B,2011arXiv1103.0511B,2011arXiv1103.3010O}.
The upper limits confirm the previous result of rejection of the detection
of 1-Jy RTs reported in numerous papers
\citep{2007ApJ...657L..37N,2007PASP..119..122K,2007AJ....133.1441M,2009AJ....138..787M}.  
\citet{2011arXiv1103.3010O} 
argue that the surface density reported in these papers is 
mis-stated; they estimate a rate that is two orders of magnitude lower than reported.  
As we have noted in the previous papers, the ATATS and 3C 286 surveys clearly 
rule out the quoted rate.  The new PiGSS results also rule out the stated rate.
These results also explore similar parameter space to
work based on the MOST archive  \cite{2011MNRAS.412..634B}.
The PiGSS upper limit is comparable to the detection
rate from MOST.  In fact, the MOST results suggest that PiGSS should have detected
$\sim 3$ RT at 50 mJy.  The absence of these sources could be the result of two factors:
the difference in observing frequency between the two surveys (0.84 GHz versus 3.1 GHz),
and, two, the absence of a deep image in the MOST archival data that prevents 
separation of transient sources from strongly variable sources.  
The PiGSS non-detection are consistent with an extrapolation of
the B07 detection rate to higher flux density thresholds;
that is, the B07 rate predicts no transients in the data reported in this paper.

The limits on surface density presented in Fig.~\ref{fig:surfacedensity} 
ignore a number of significant differences between surveys.  B07, 
\citet{2011arXiv1103.3010O}, and the PiGSS results are all at frequencies
greater than 1.4 GHz, where greater variability is expected.  The surveys all have
different cadences and different sensitivities to certain time scales.
As \citet{2011arXiv1103.4328B} shows for the case of tidal disruption events, 
translation of the surface density into limits on physical
source populations requires consideration of both the observing frequency and
the timescale of observations.

\section{Light Curve Analysis\label{sec:var}}

We examined the light curves for evidence of systematic variations that are common to
all sources as a function of epoch using two methods.  Neither method improved on the
overall quality of results and for some sources appeared to introduce distortions
in the light curves that worsened the results.  In the first method, we used a sum
of the flux densities of the brightest sources to normalize the flux densities of
all sources.  We varied the number of sources from the brightest five to the brightest
fifty.  There were reductions at the 20\% in the scatter for the brightest sources
but at the cost of potentially introducing distortions in the light curves,
and required assuming
constant flux density for these sources.  In the second method, we performed a 
linear fit between the deep flux density and the epoch flux density for all sources
brighter than 10 mJy in each epoch.  The RMS of the corrections to the flux densities
was 4\%, which would only be significant for the brightest sources and introduced
similar issues to the normalization scheme described above.  Therefore, we
decided not to apply these corrections.  This analysis does confirm a bias in the daily flux densities
of $\sim 10\%$, as discussed in \S~\ref{sec:obs}.

We summarize the statistics that we compute below for all sources in Table~\ref{tab:stats}.
Columns are (1) the source name, (2) primary-beam corrected
flux density from the deep image in mJy, $S$,
(3) primary beam gain, (4) daily reduced $\chi^2$, (5) monthly reduced $\chi^2$,
(6) daily RMS fractional modulation, (7) monthly RMS fractional modulation,
(8-9) minimum and maximum of daily fractional modulation, and (10-11)
minimum and maximum of monthly fractional modulation.  See below for definition of
these terms.

\subsection{$\chi^2$ Analysis}

\begin{figure}[H]
\mbox{\psfig{figure=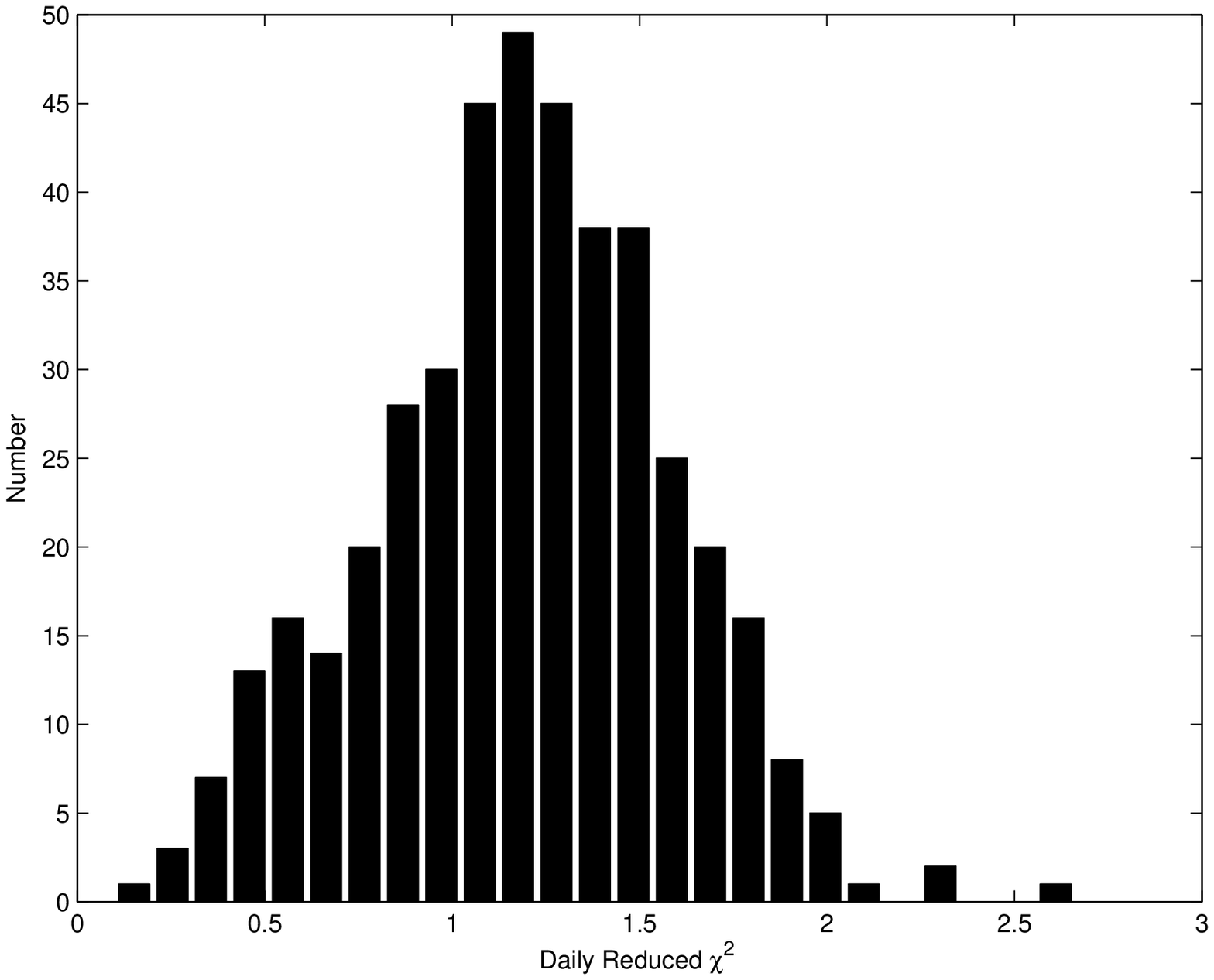,width=0.45\textwidth}\psfig{figure=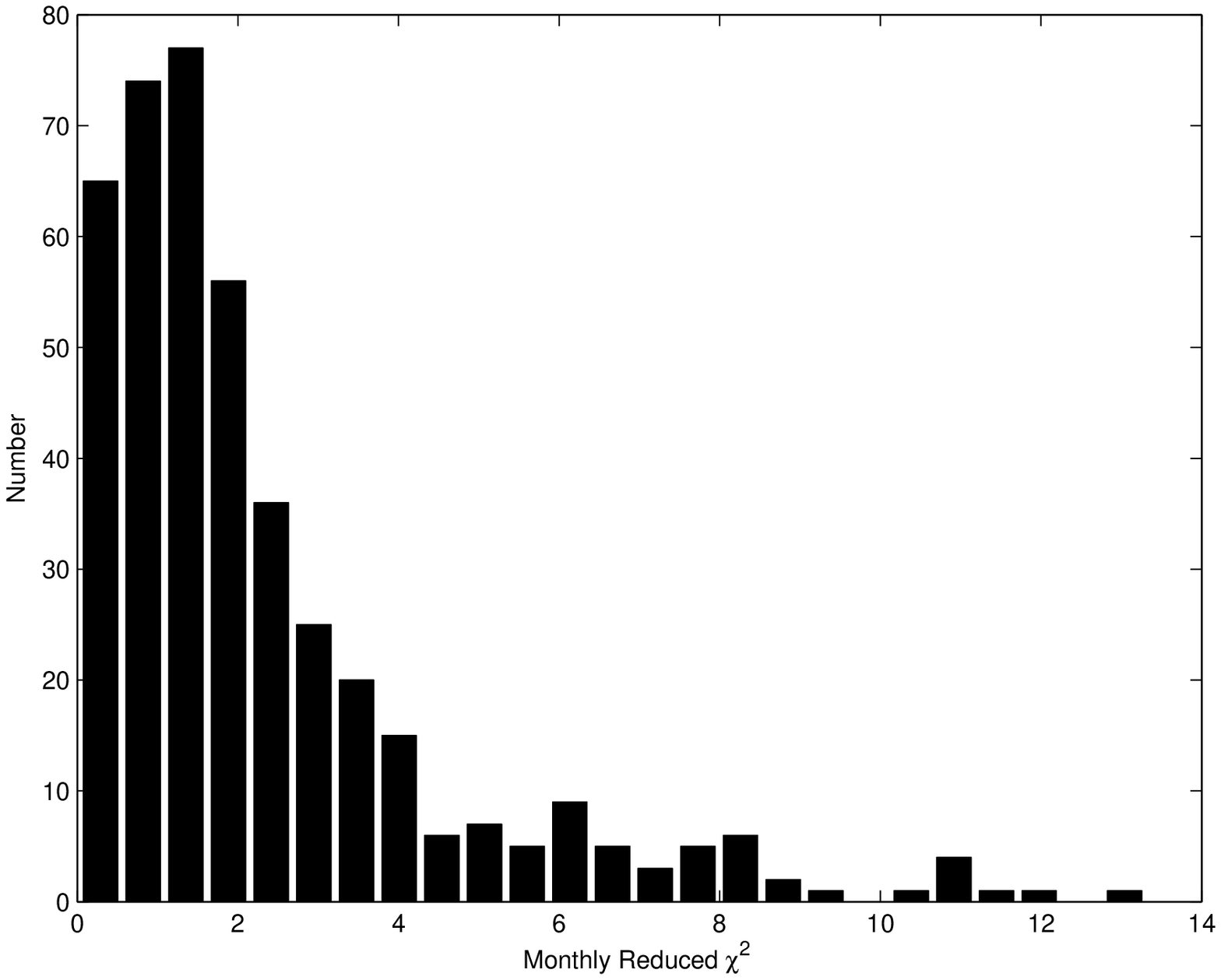,width=0.45\textwidth}}
\caption[]{
Reduced $\chi^2$ for the daily ({\em left}) and monthly ({\em right})
light curves under the hypothesis of constant flux density.  Values of
1.5 in the daily $\chi^2$ and 4.0 in the monthly $\chi^2$ correspond to the 99.7\%
confidence level.
\label{fig:chi2}}
\end{figure}

The reduced $\chi^2$ for the hypothesis of constant flux density is an established measure of 
variability.  We plot histograms of the reduced $\chi^2$ for the daily and monthly 
epochs in Fig.~\ref{fig:chi2}.  The histogram for the daily data is dominated by
the large number of sources that are undetected or marginally detected.  The peak
of the distribution is $>1$ and there is a tail to $>2.5$.  For the large number of
degrees of freedom in this data, even small deviations from 1 are significant.  A
value of $\chi^2>1.5$ corresponds to greater than 99.7\% certainty in rejection of
the hypothesis.  Thus, 88/425 sources qualify as significantly variable on a daily
timescale.  The monthly data have lower errors and, therefore, more clearly identify
variable sources.  63/425 sources have $\chi^2>4$, corresponding again to 99.7\%
certainty of variability.  Thus, we conclude that $\sim 20\%$ of sources are variable.
This is consistent with the conclusions of 
\citet{2011arXiv1103.3010O} that $\gsim 30\%$ of sources brighter than 1.8 mJy are variable.

\subsection{RMS Fractional Modulation}

\begin{figure}[H]
\mbox{\psfig{figure=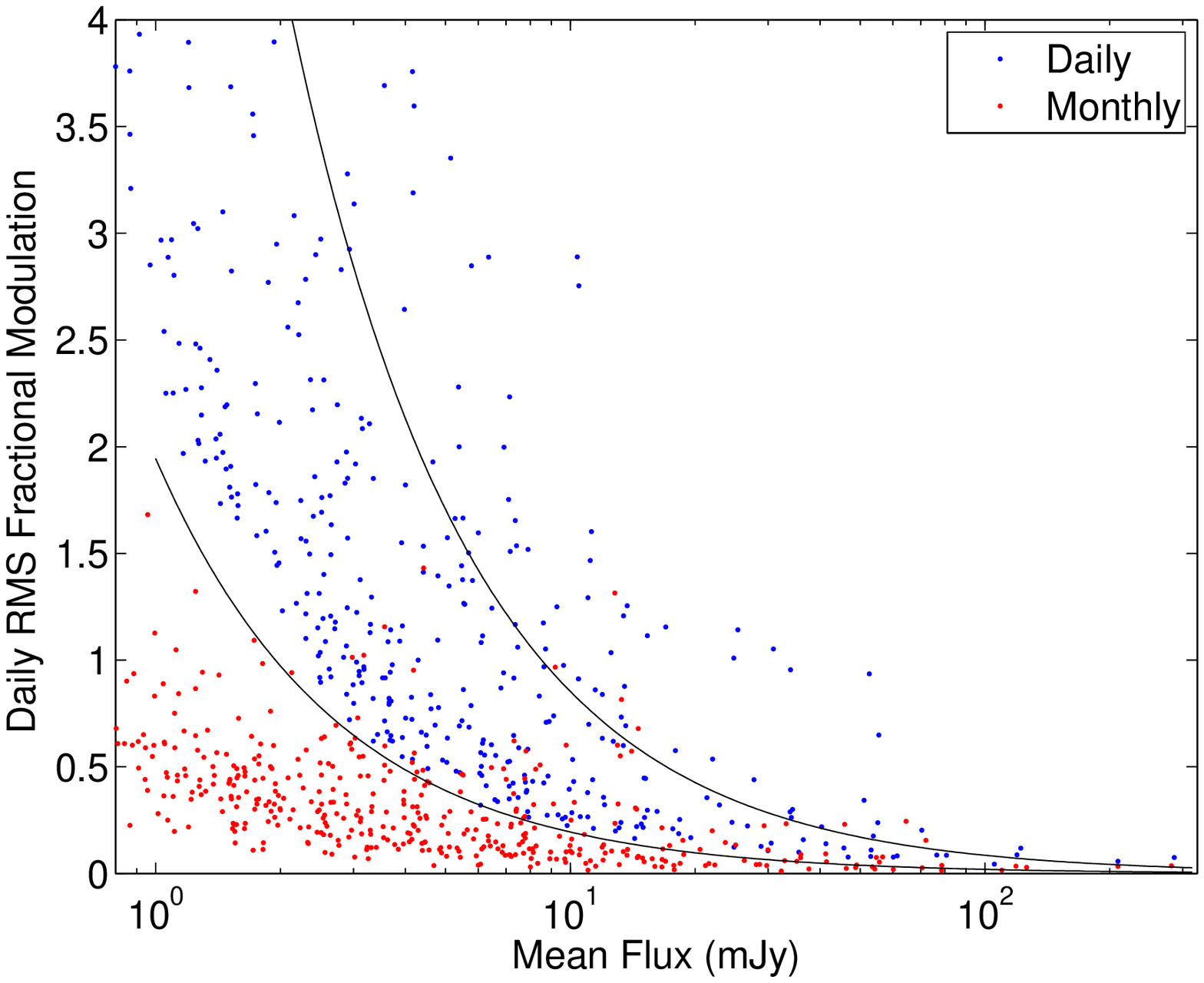,width=0.45\textwidth}\psfig{figure=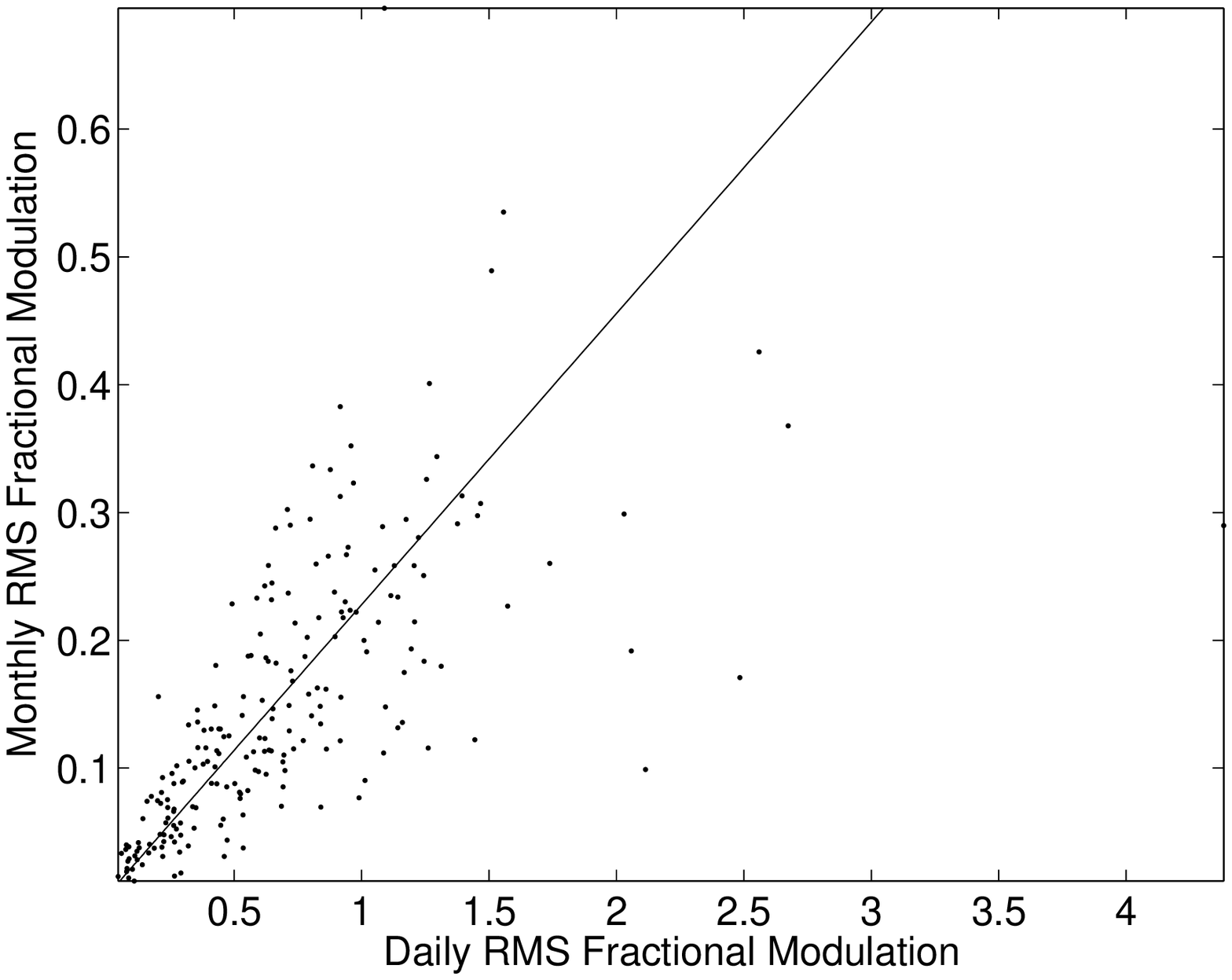,width=0.45\textwidth}}
\caption[]{
({\em left})
RMS fractional modulation as a function of flux density on daily ($\sigma_{m,d}$) 
and monthly ($\sigma_{m,m}$) timescales.  The solid lines
indicate expected modulation fraction for 3 times the median RMS noise in the daily (top) and monthly
(bottom) epochs.
({\em right})
Daily RMS fractional modulation versus monthly RMS fractional modulation.  The solid line is based on
the median RMS noise in the daily and monthly epochs.
\label{fig:modulation}}
\end{figure}

The RMS fractional modulation is the ratio of the RMS flux density to the mean flux density:
\begin{equation}
\sigma_m = { {\Sigma ( S_i - \bar{S} ) w_i } \over {\bar{S} \Sigma w_i} },
\end{equation}
where $S_i$ is the flux density in individual epochs,
$\bar{S}$ is the mean flux density and
$w_i = \sigma_i^{-2}$ is a weighting factor equal to the inverse RMS noise squared.  
We calculate this value for the monthly ($\sigma_{m,m}$) and daily ($\sigma_{m,d}$) data (Fig.~\ref{fig:modulation}).  The daily statistic is the same as the ratio of the
standard deviation to flux density reported by \citet{2011arXiv1103.3010O}.
For the majority of sources, $\sigma_{m,d} > \sigma_{m,m}$.
The dominant effect producing this is the lower signal to noise ratio and the more limited
$(u,v)$ coverage of the daily data.  The solid curves in the plot correspond to RMS 
fractional
modulations that are three times the median RMS noise for the daily and monthly data.  
Strongly variable sources are those that are significantly above these curves.

We also see that the RMS fractional modulation
on the two time scales is correlated, with $\sigma_{m,m} \sim \sigma_{m,d}/4$
(Fig.~\ref{fig:modulation}).  This ratio is expected based on the ratios of the median
RMS between monthly and daily epochs (which is also consistent with the noise decreasing
as the number of epochs squared).  Here we plot only the sources for which the mean monthly
flux exceeds 2 mJy.  This excludes sources that are only marginally detected in the monthly images
but permits inclusion of sources strongly variable on daily timescales.  Sources that fall far
off this line are of interest as strongly variable on one timescale but not the other.

\begin{figure}[H]
\mbox{\psfig{figure=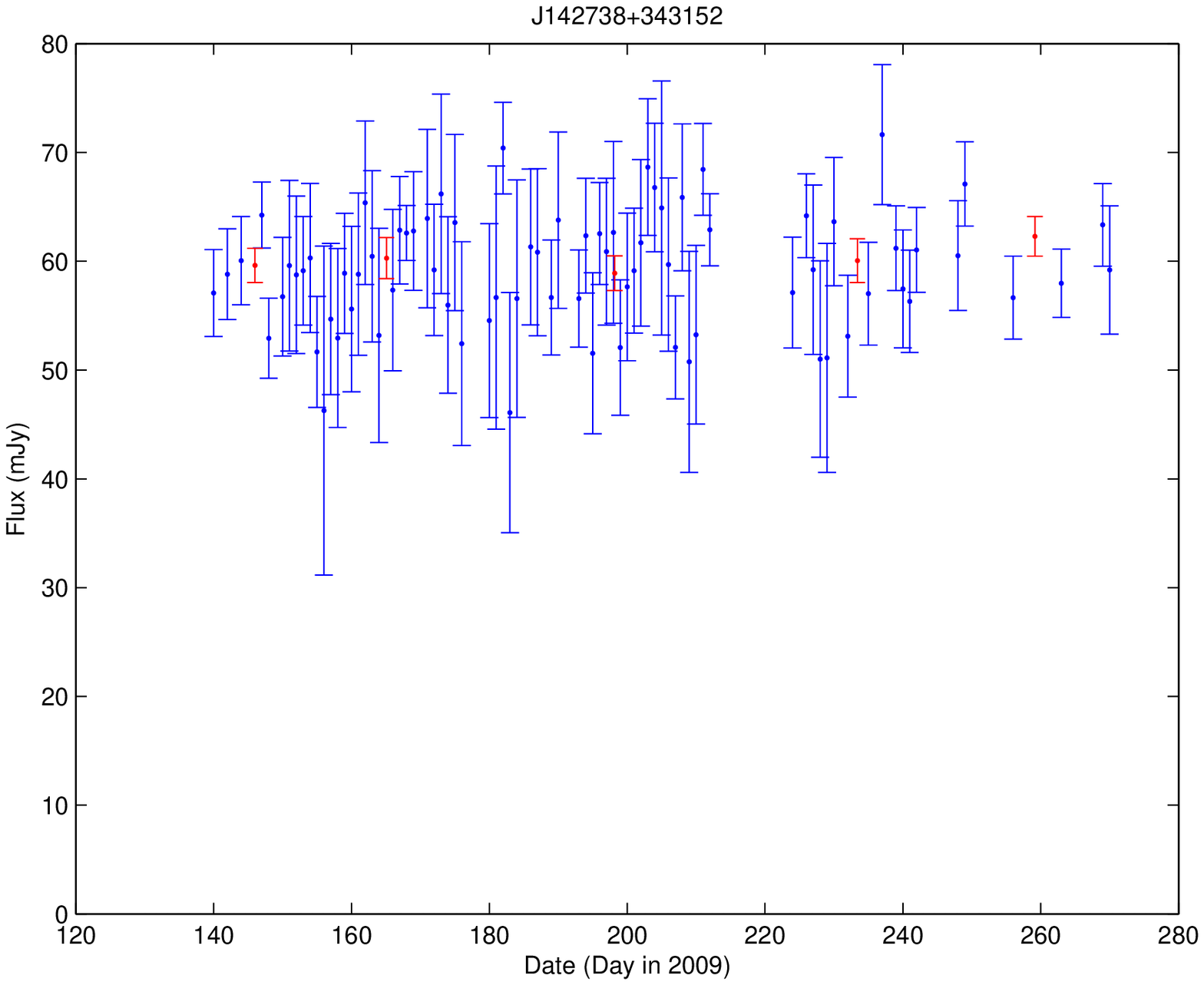,width=0.35\textwidth}\psfig{figure=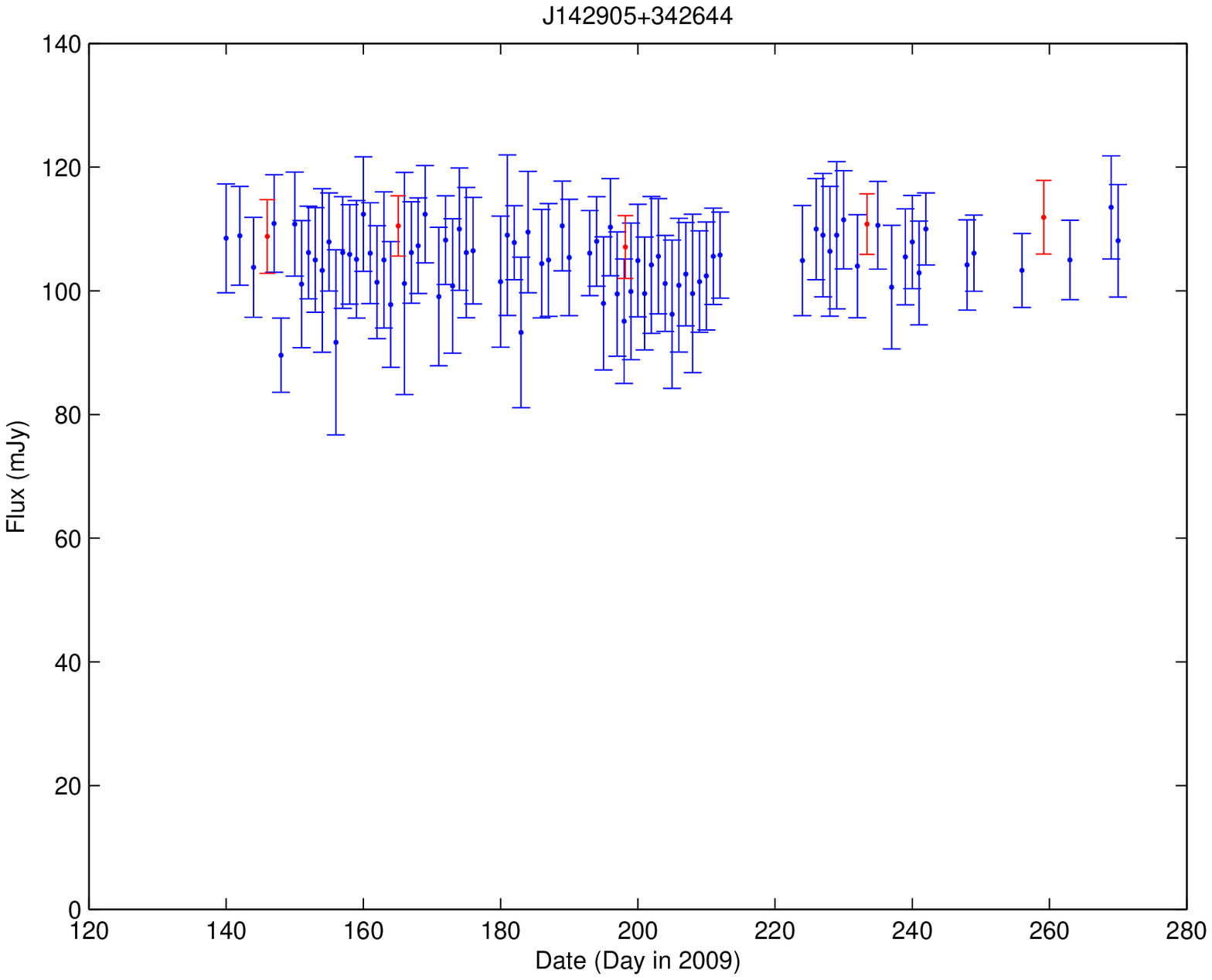,width=0.35\textwidth}}
\mbox{\psfig{figure=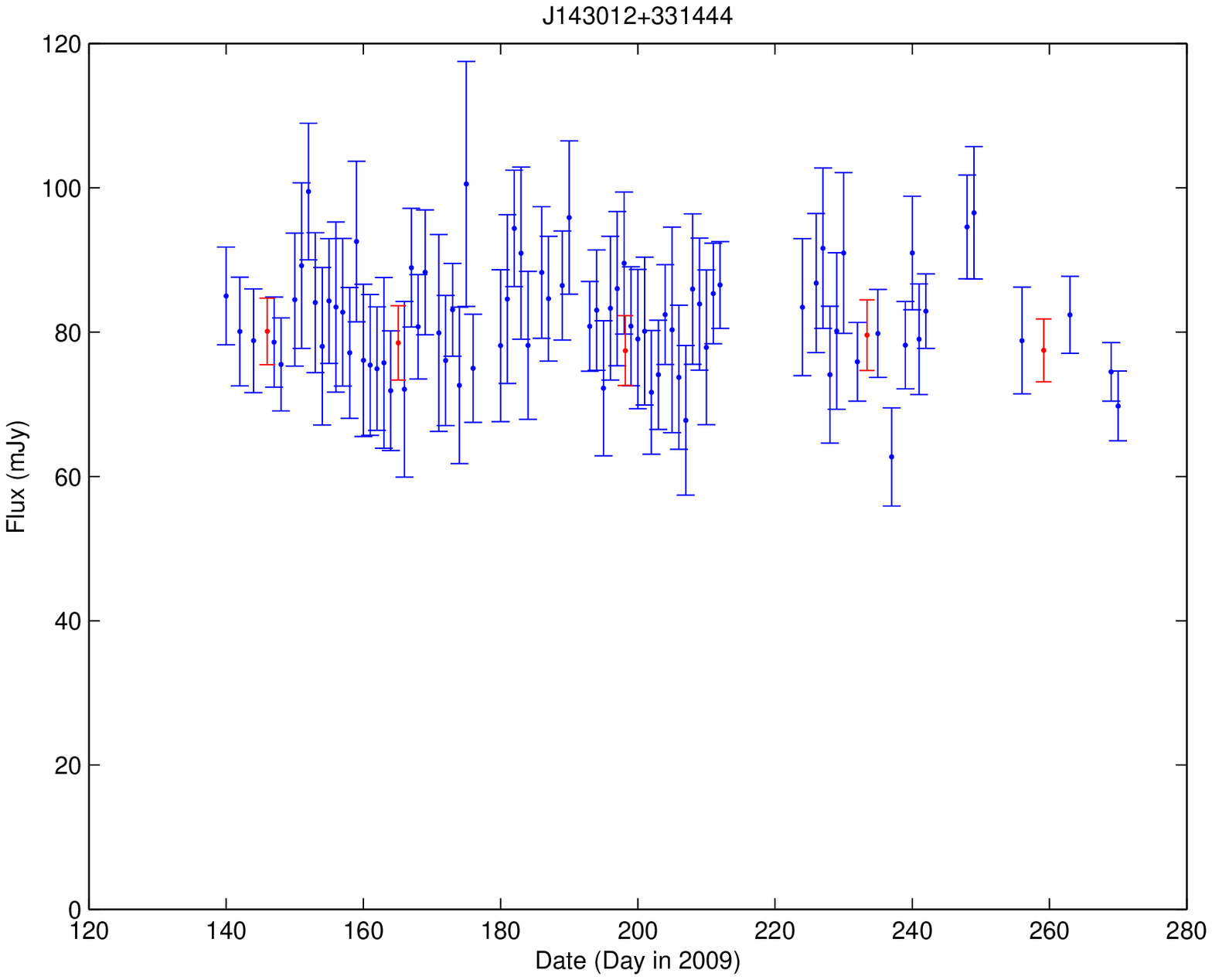,width=0.35\textwidth}\psfig{figure=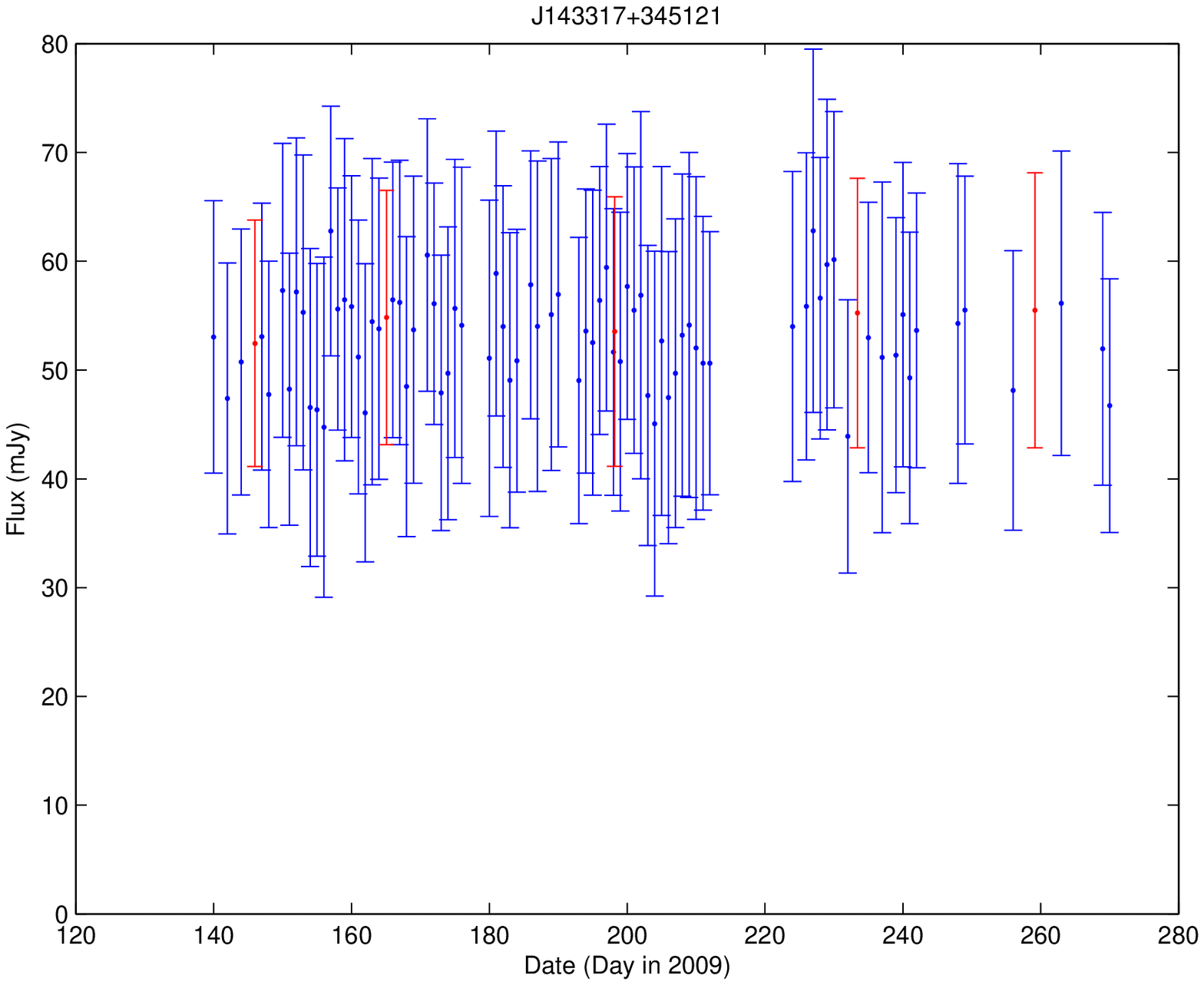,width=0.35\textwidth}}
\mbox{\psfig{figure=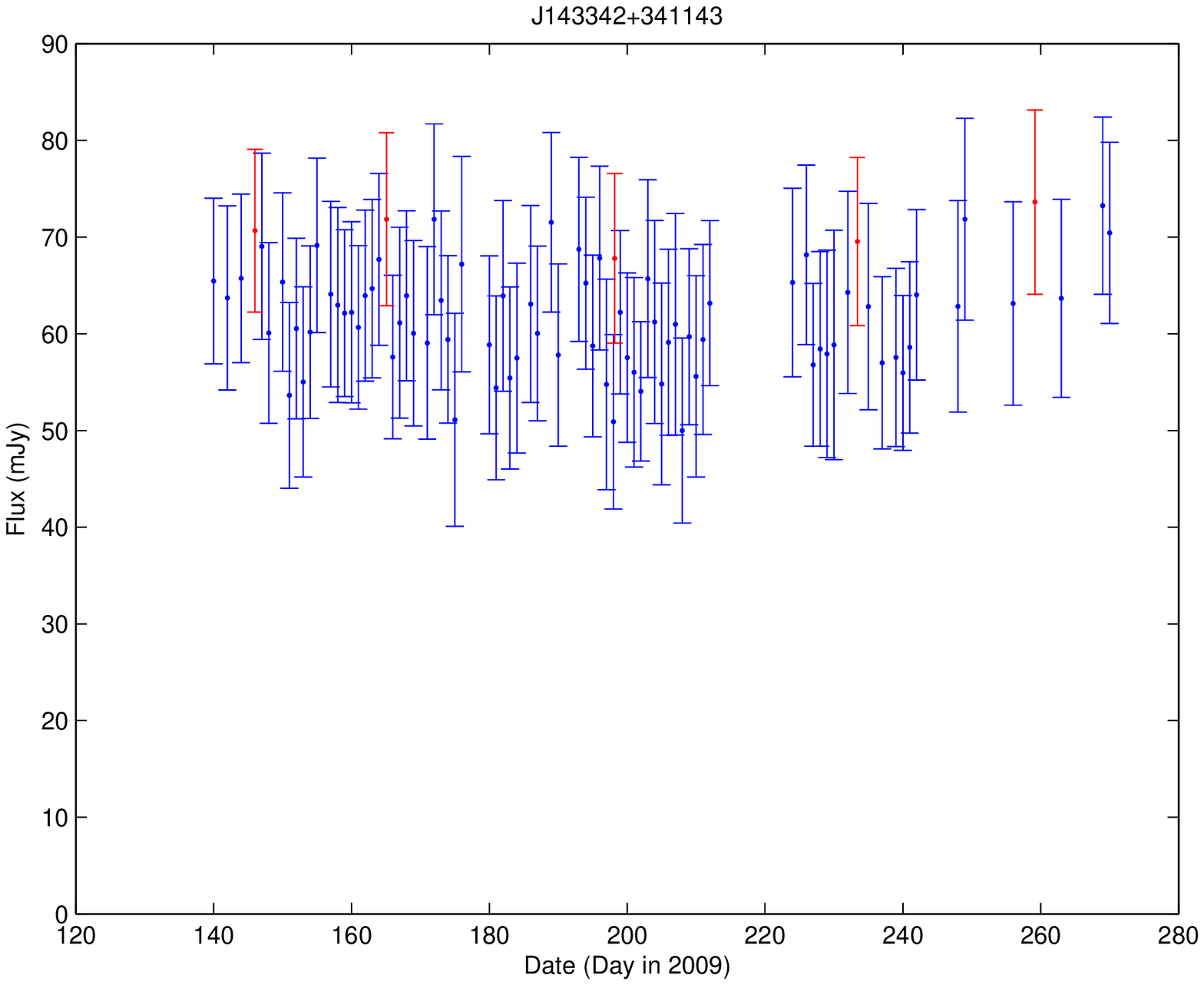,width=0.35\textwidth}\psfig{figure=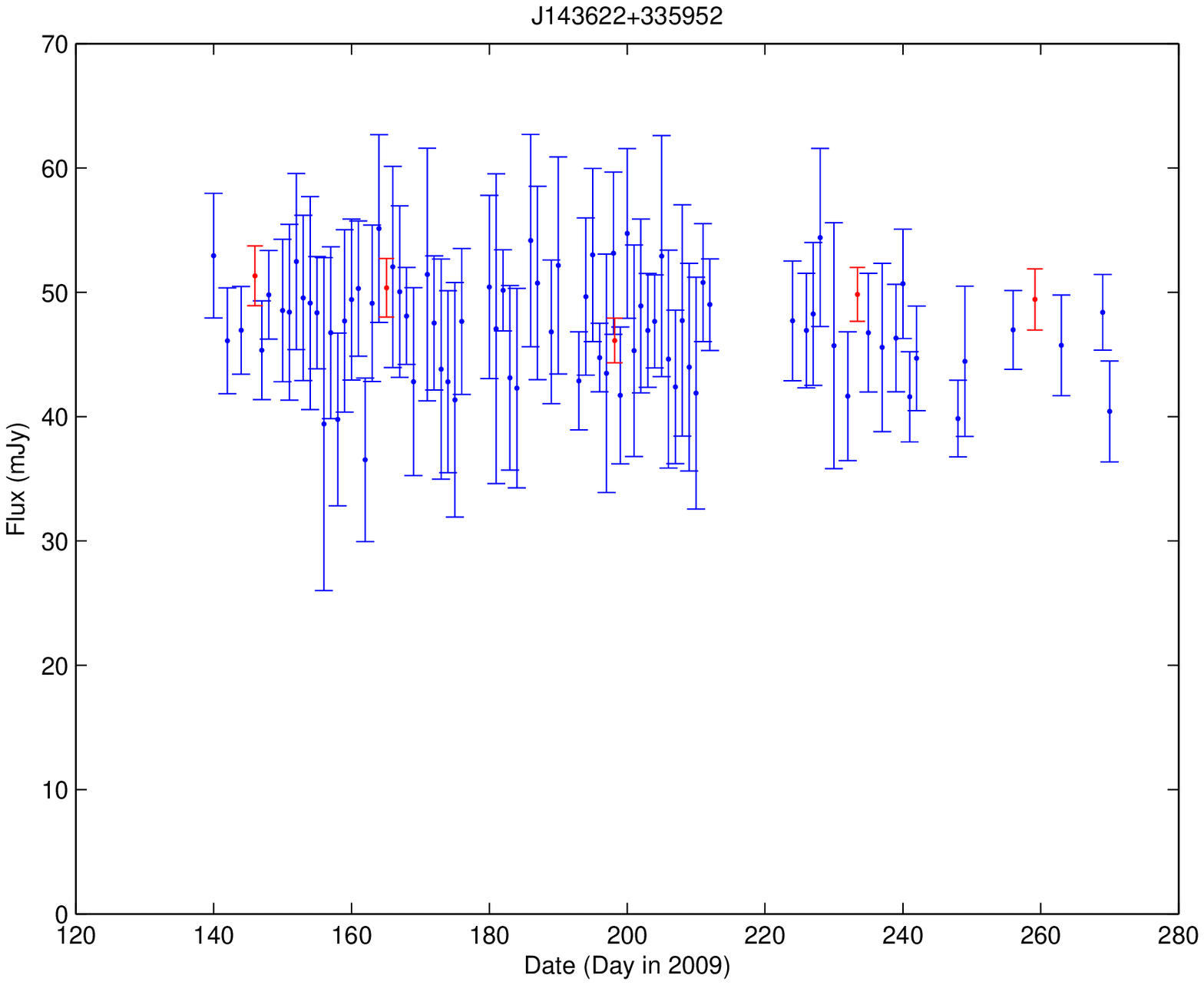,width=0.35\textwidth}}
\mbox{\psfig{figure=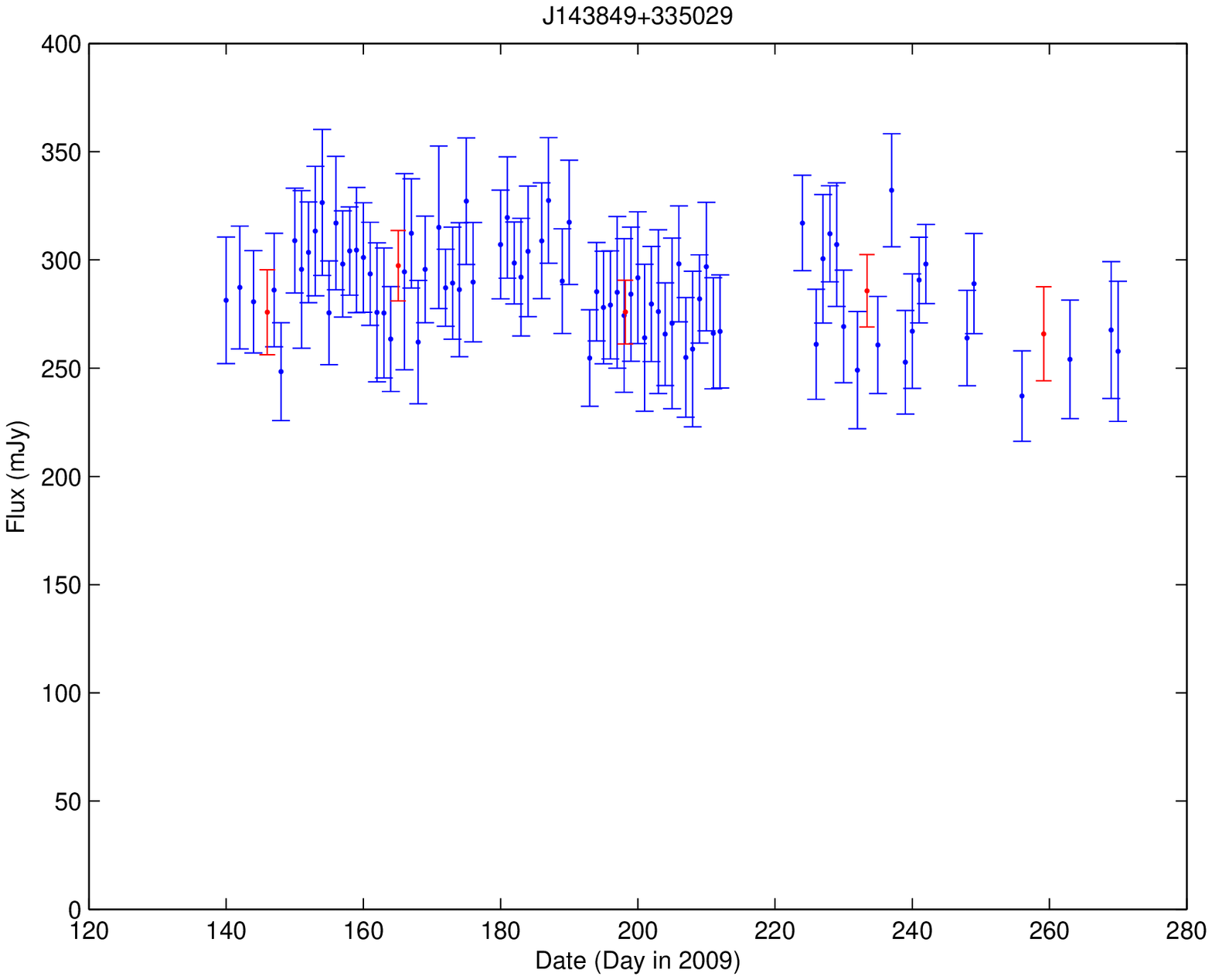,width=0.35\textwidth}\psfig{figure=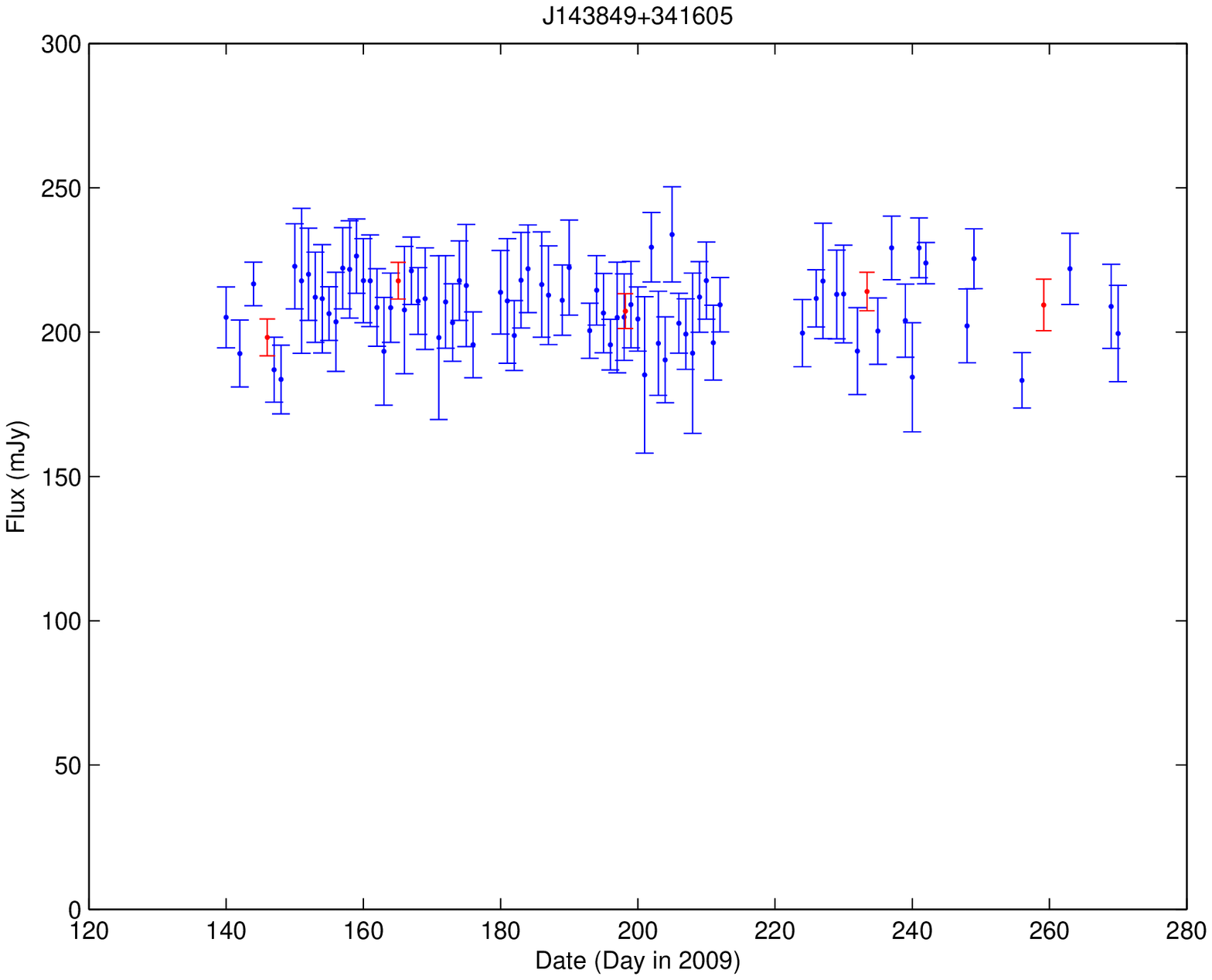,width=0.35\textwidth}}
\caption{Light curves for eight sources with the lowest daily RMS modulation fractions.  Blue
dots indicate daily flux densities; red dots indicate monthly flux densities.
\label{fig:steady}}
\end{figure}

In Figure~\ref{fig:steady} we show example light curves for sources
that are not strongly variable in our data.
These four sources were selected as those showing the smallest $\sigma_{m,d}$.  This
selection biases in favor of the brightest sources in the sample. These sources
are also among the least active on monthly time scales as well, although the relationship between
daily and monthly RMS fractional modulation is not one-to-one.  
For instance, J143012+331444 has the minimum $\sigma_{m,m}$ of all sources 
(0.01) but has $\sigma_{m,d} = 0.08$.  Error bars are large in the case of 
J143317+345121 due to confusion with an adjacent bright source.

\begin{figure}[H]
\mbox{\psfig{figure=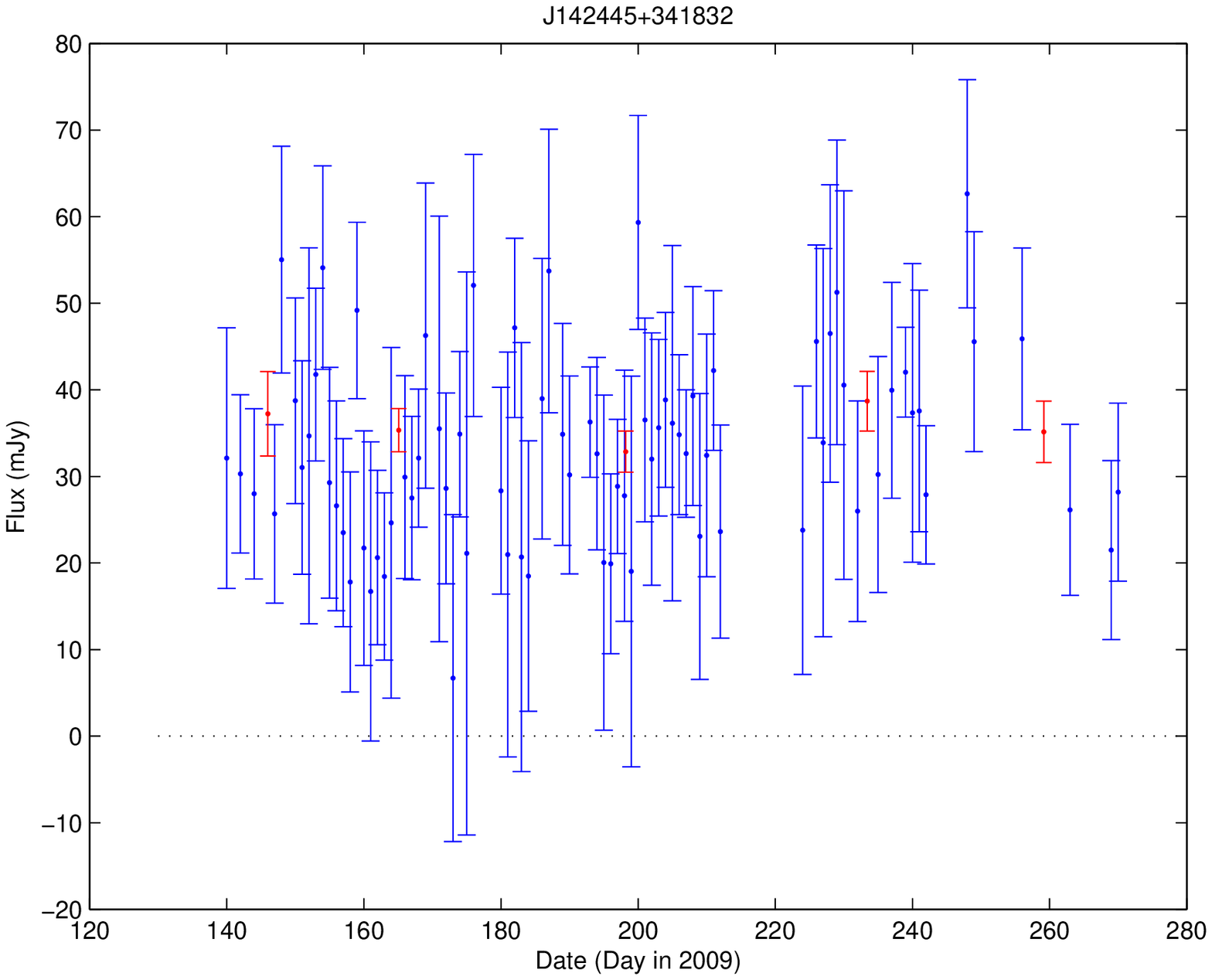,width=0.35\textwidth}\psfig{figure=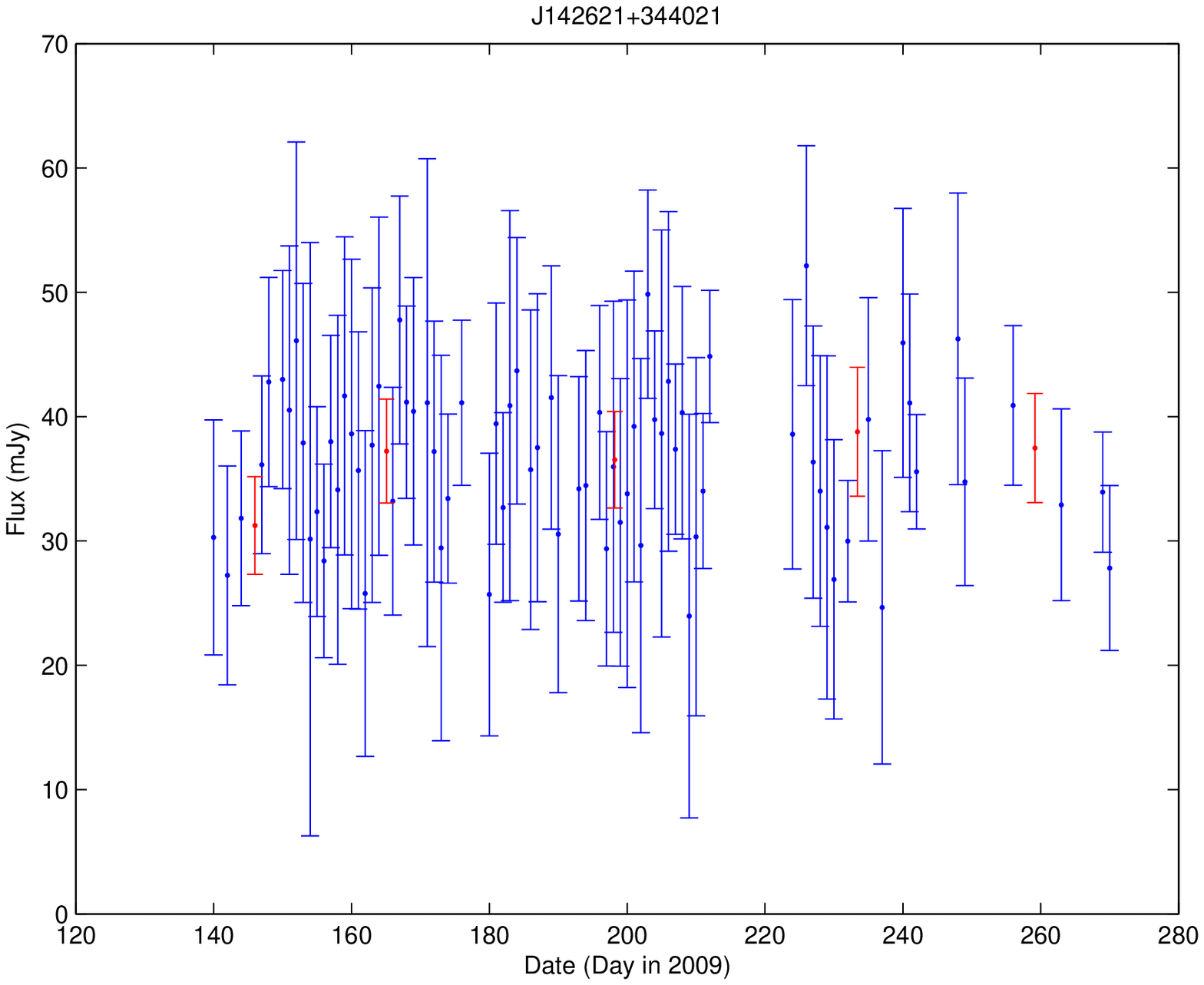,width=0.35\textwidth}}
\mbox{\psfig{figure=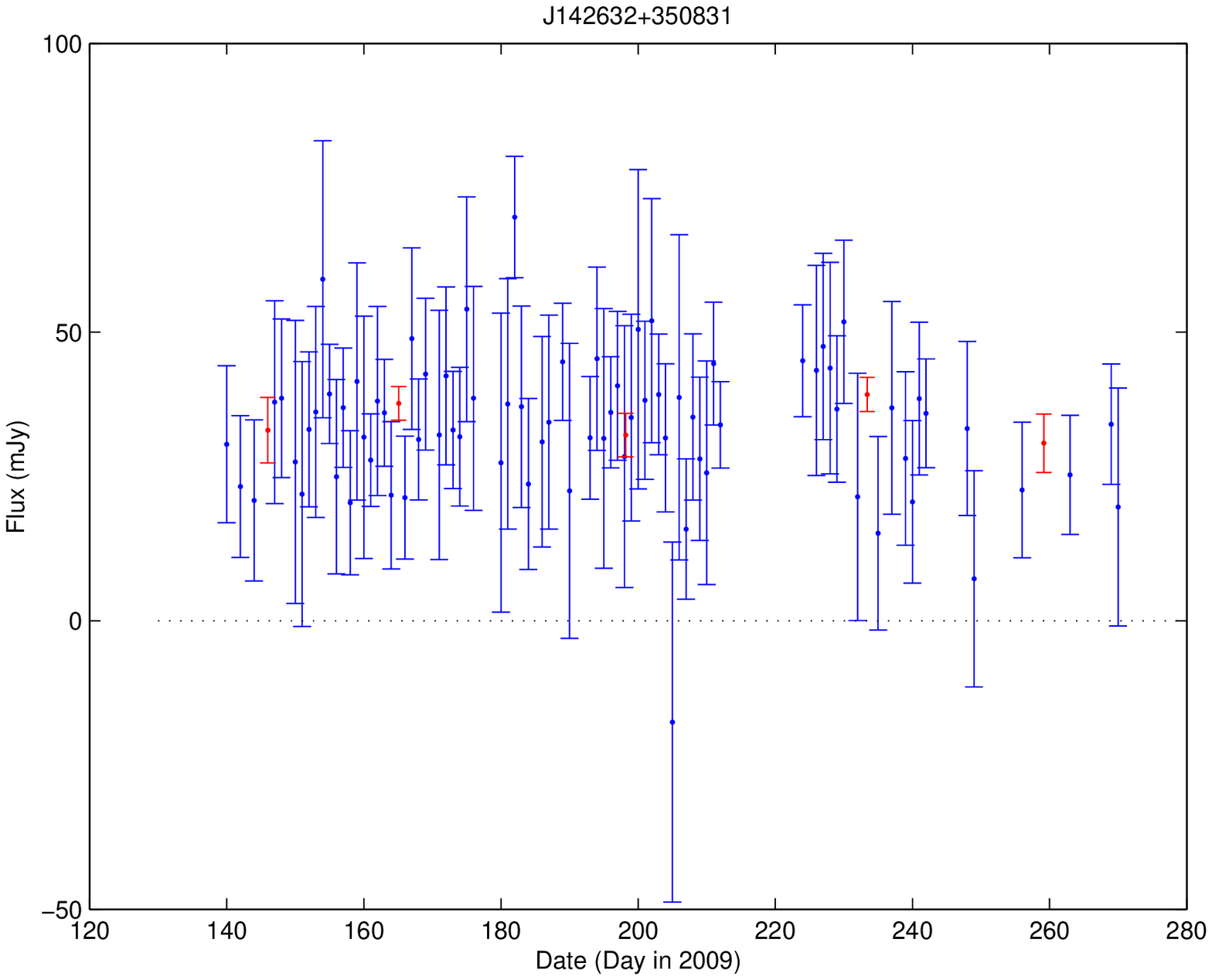,width=0.35\textwidth}\psfig{figure=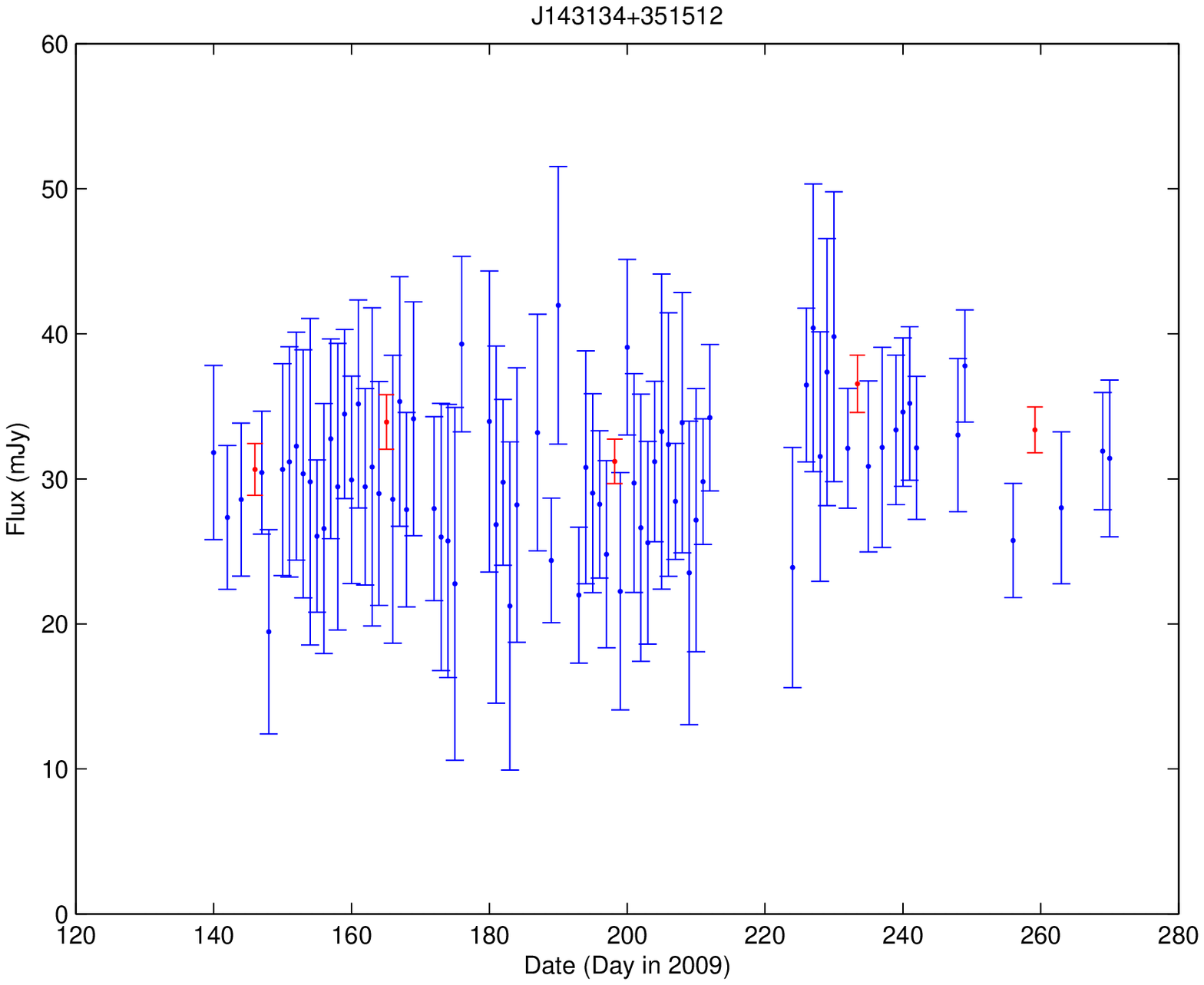,width=0.35\textwidth}}
\mbox{\psfig{figure=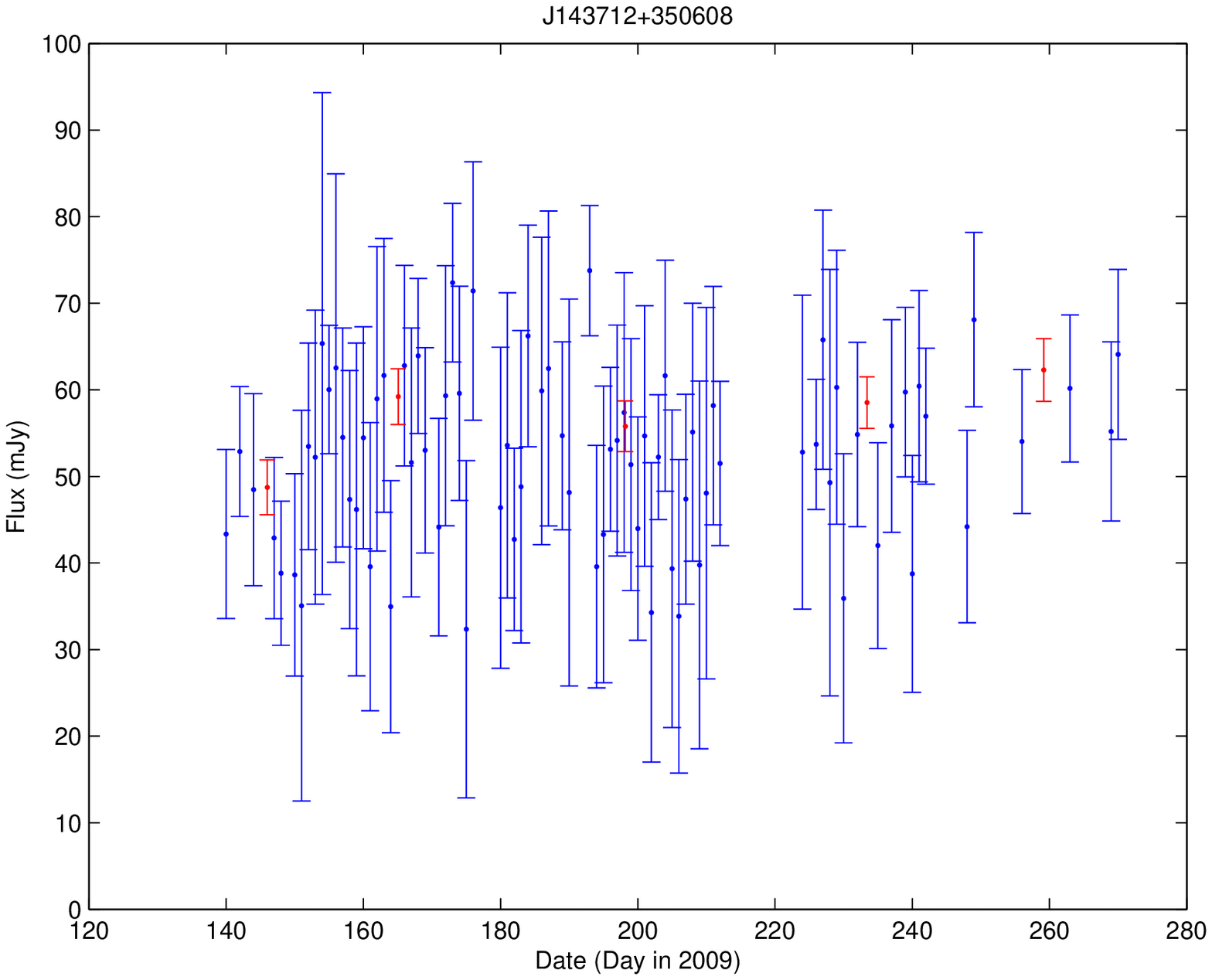,width=0.35\textwidth}\psfig{figure=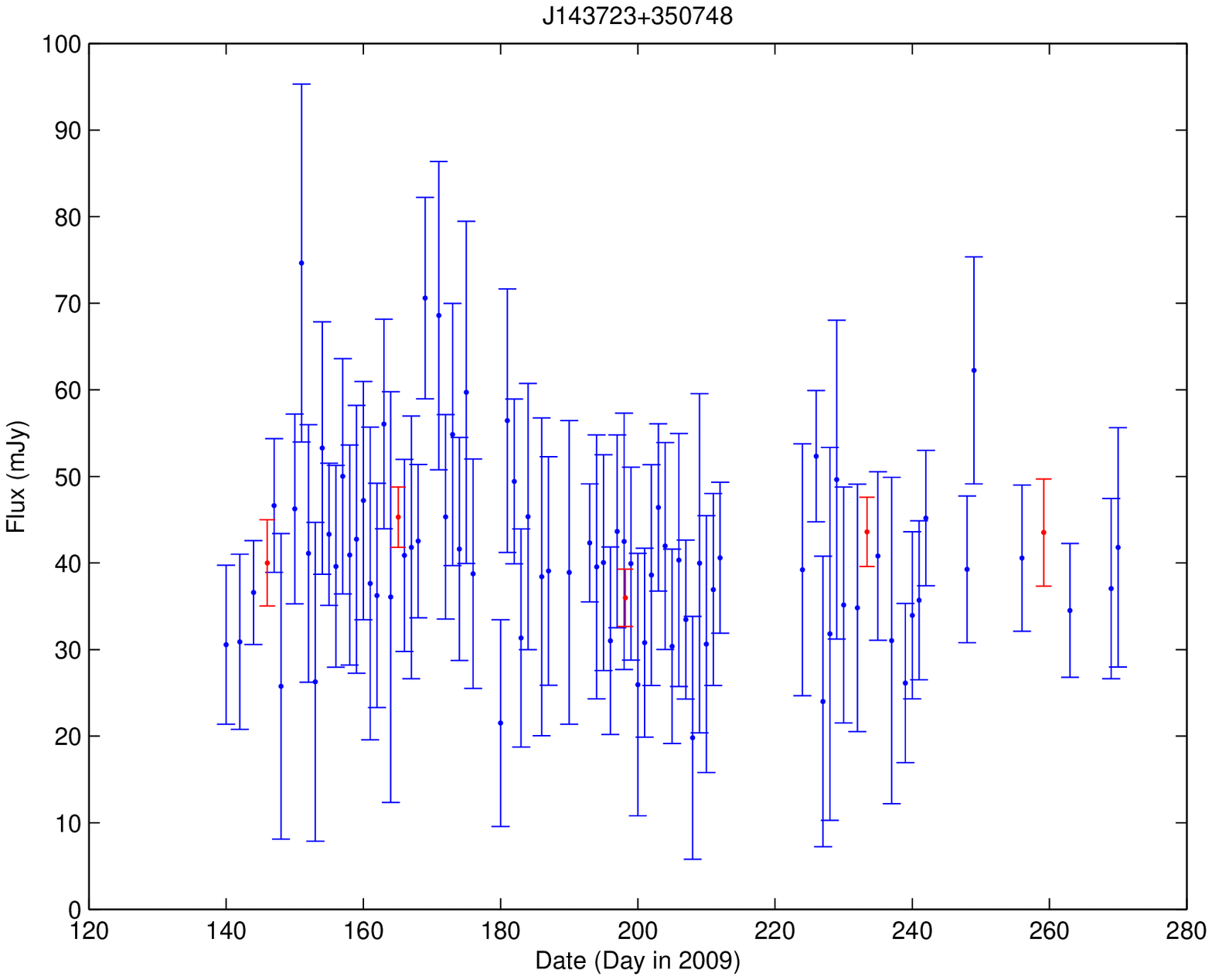,width=0.35\textwidth}}
\mbox{\psfig{figure=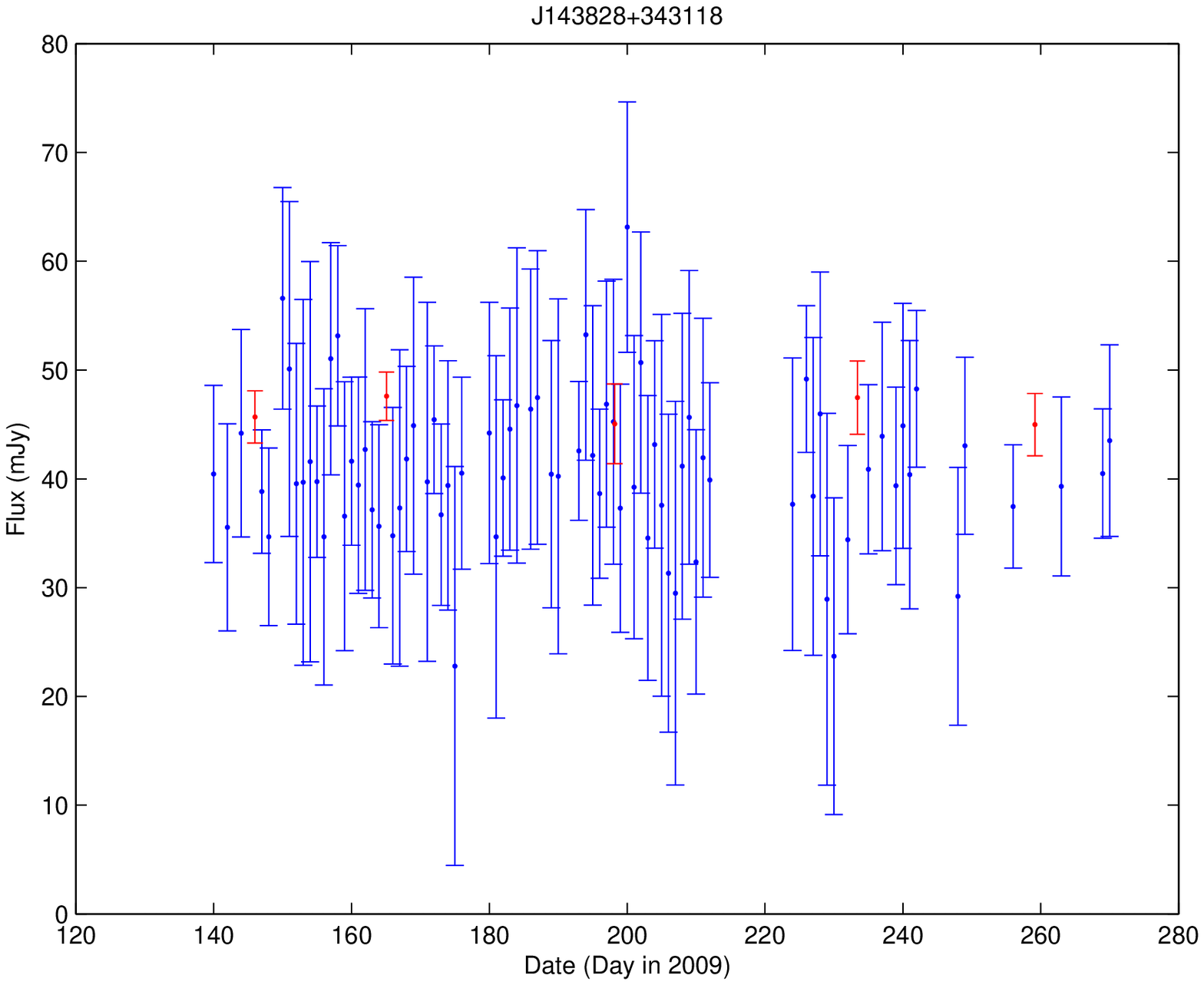,width=0.35\textwidth}\psfig{figure=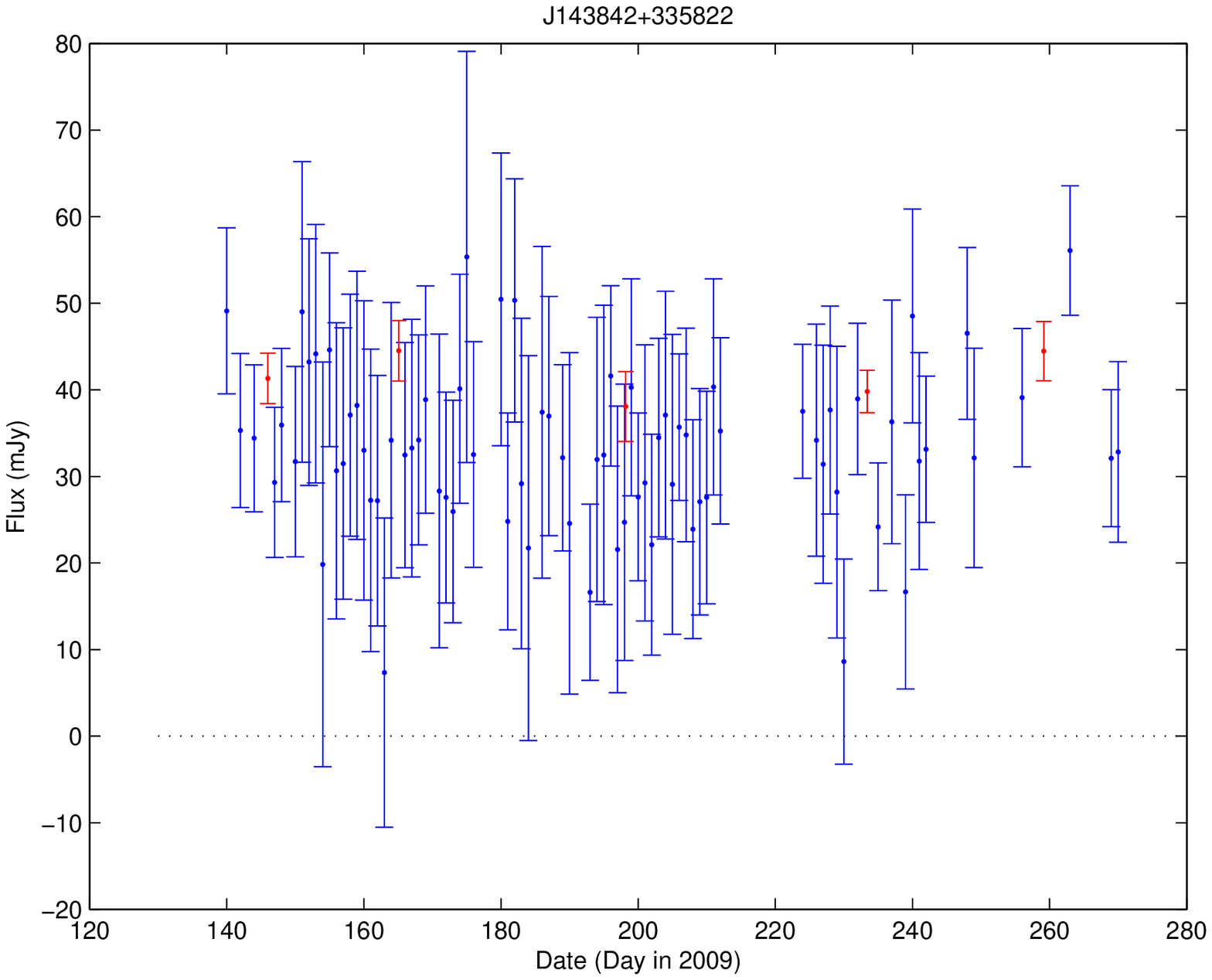,width=0.35\textwidth}}
\caption{Light curves for eight sources with the highest daily RMS modulation fractions and
$S > 30$ mJy and primary beam gain correction factor $<4.0$.  Blue
dots indicate daily flux densities; red dots indicate monthly flux densities.
\label{fig:var}}
\end{figure}

In Figure~\ref{fig:var} we show examples of sources with the strongest variability as measured
by $\sigma_{m,d}$.  We restrict this sample to sources with mean flux density $>30$
mJy and within the inner part of the region such that the primary beam correction is less than 4.
These restrictions provide a clean set of data in which pointing effects and low signal to noise
ratio will not introduce significant variations in the flux density.  The maximum $\sigma_{m,d}$
in these subset is 0.30 for J142632+350831.

\subsection{Fractional Modulation}

\begin{figure}[H]
\psfig{figure=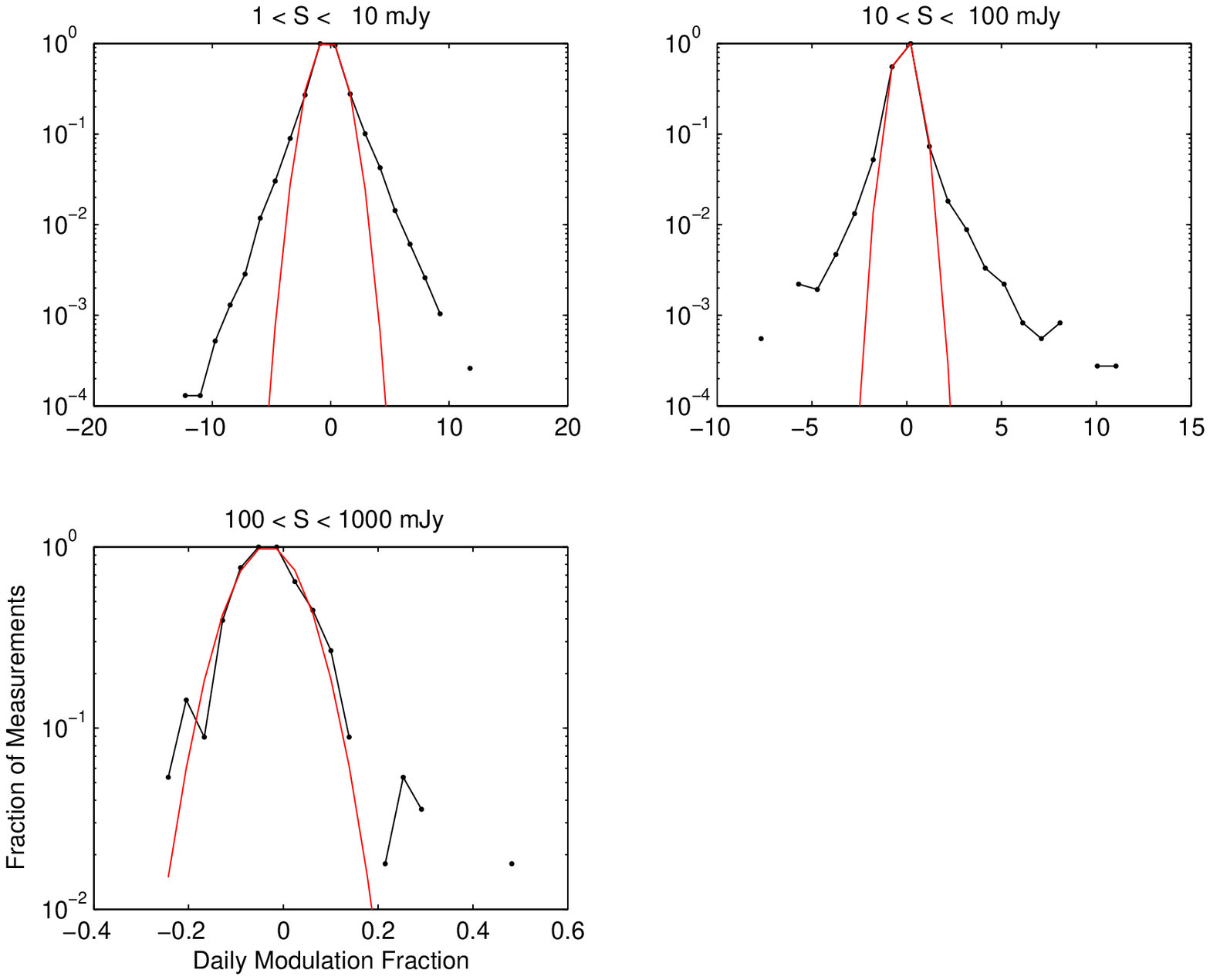,width=\textwidth}
\caption[]{
Distribution of daily modulation fraction $m_{i,d}$ for all sources and all epochs.  The three panels represent flux densities
of 1 to 10 mJy, 10 to 100 mJy, and 100 to 1000 mJy.  Red curves indicate Gaussian fits to the distribution.
\label{fig:fluxhistdaily}}
\end{figure}

\begin{figure}[H]
\psfig{figure=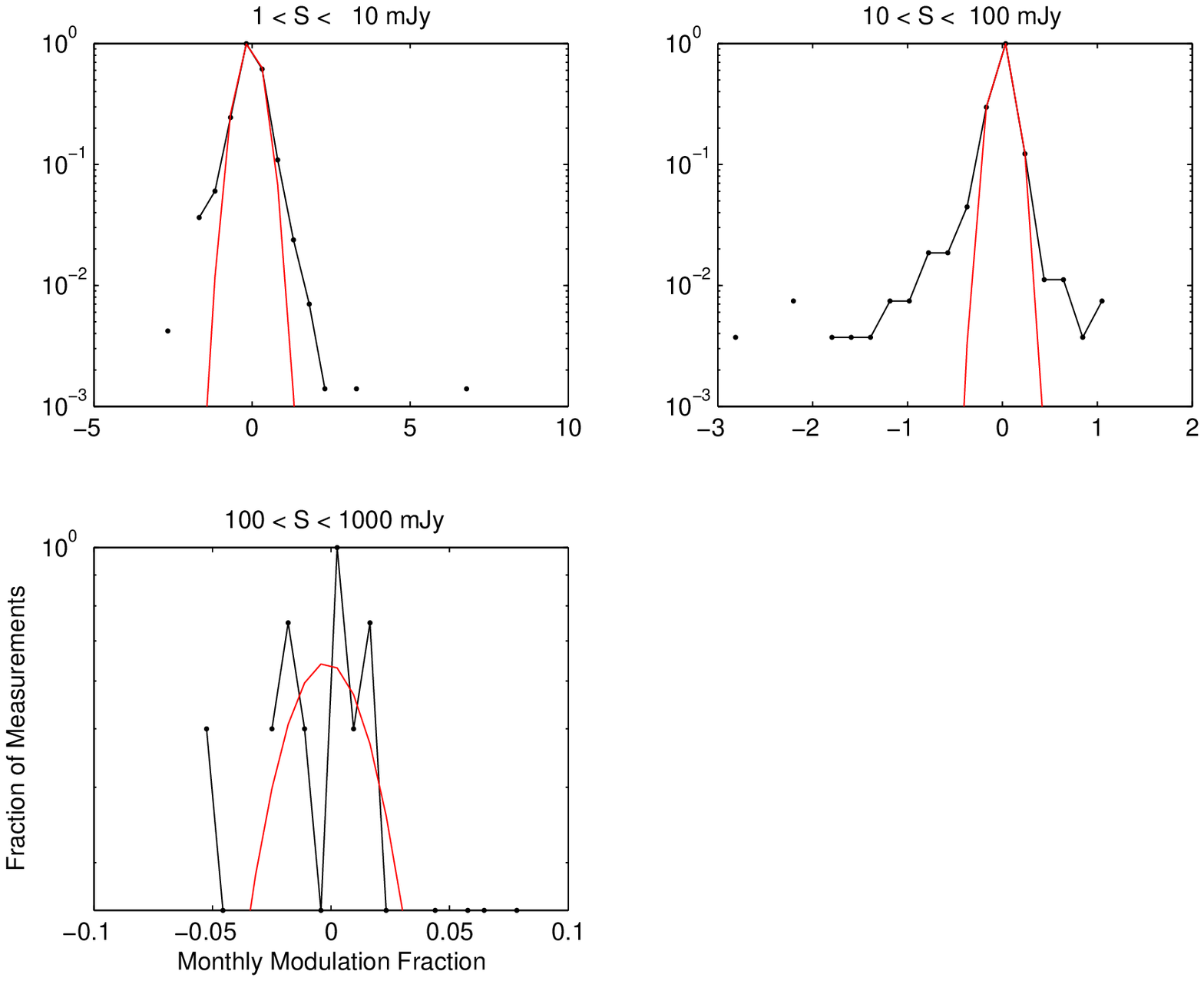,width=\textwidth}
\caption[]{
Distribution of monthly modulation fraction $m_{i,m}$ for all sources and all epochs.  The three panels represent flux densities
of 1 to 10 mJy, 10 to 100 mJy, and 100 to 1000 mJy.   Red curves indicate Gaussian fits to the distribution.
\label{fig:fluxhistmonthly}}
\end{figure}

The RMS fractional modulation is a good statistic for characterizing overall variability but it 
may not capture episodic variability.  For instance, a factor of 2 flare with a duration of a few days
would have a small impact on the RMS fractional modulation over the full data set.  
Such considerations are important when thinking about the identification of rare events.  
The fractional modulation is defined as the ratio of variability of each epoch to the deep flux density
for that source:
\begin{equation}
m_i = { (S_i - \bar{S}) \over \bar{S} }.
\end{equation}  
We create histograms for the daily ($m_{i,d}$; Fig.~\ref{fig:fluxhistdaily}) and
monthly fractional modulations ($m_{i,m}$; Fig.~\ref{fig:fluxhistmonthly}), summarizing
the variability for all sources and all epochs together.
These histograms are averaged
over decade ranges in flux density.  We also tabulate the probability of exceeding a
fractional modulation in Table~\ref{tab:fluxhist}.  These results incorporate $425 {\rm\ sources} \times 78 {\rm\ epochs} \approx 3 \times 10^4$
flux density measurements.  In the case of the monthly modulations for $S > 100$ mJy, there are a
limited number of measurements, leading to a rather poorly characterized distribution.

It is again clear from these data that the
monthly modulations are significantly lower than the daily modulations.  This is in part due
to the lower signal to noise ratio and possibly the poorer $(u,v)$-coverage and
greater systematic errors of the daily detections.  This effect is very clearly evident in
the daily modulations for sources with $1 < S < 10$ mJy.  The majority of these sources are
not detected in most, if not all, epochs.  The histogram therefore represents the noise statistics
more than the results.  We fit a Gaussian to the distribution as an estimate of the noise
contribution.  We see that this fit reasonably well represents the central portion of the
distributions.  There are long tails to all the fits that are either the results of 
individual measurements with large errors, systematic errors,
 or real variability in the source population.  Very large variations indicated in the histogram,
for instance, could be the result of individual epochs with a flux density measurement that is not 
statistically significant in its difference from the mean but shows up as a large fractional 
modulation relative to the deep flux.  Another way of stating this is that the Gaussian fit to
the distribution estimates the noise properties only under the assumption of uniform errors
for all sources in all epochs.  It is also important to note that $m_i  < -1$ corresponds
to a negative flux density, which occurs for sources near the flux density limit 
threshold.  Thus, the distribution of negative values in the histogram can be considered 
a bound on real activity in the sources.  Conservatively, we state that the
variability characterized here sets an upper limit to the variable source population.

The limits are nevertheless quite strong.  Among the brightest sources $S> 100$ mJy, the 
probability
of a daily variation $m_{i,d} > 1$ is $< 4\times 10^{-3}$.  For the faintest sources,
$1 < S < 10$ mJy, the probability of $m_{i,d} > 10$ is $<10^{-4}$.  A similar
limit holds for sources in the range $10 < S < 100$ mJy.  Very large variations ($>10\times$) 
in the monthly data do not occur and we set an upper limit of fewer than $7 \times 10^{-4}$
such events.  These limits can be considered an alternative characterization of the 
transient rate for flux-limited surveys without a reference deep image in which strongly 
variable sources can exceed the flux threshold and appear as transient sources.  Variability
of this kind could account for some fraction of the population reported by 
\citep{2009AJ....138..787M}.

\begin{figure}[H]
\mbox{\psfig{figure=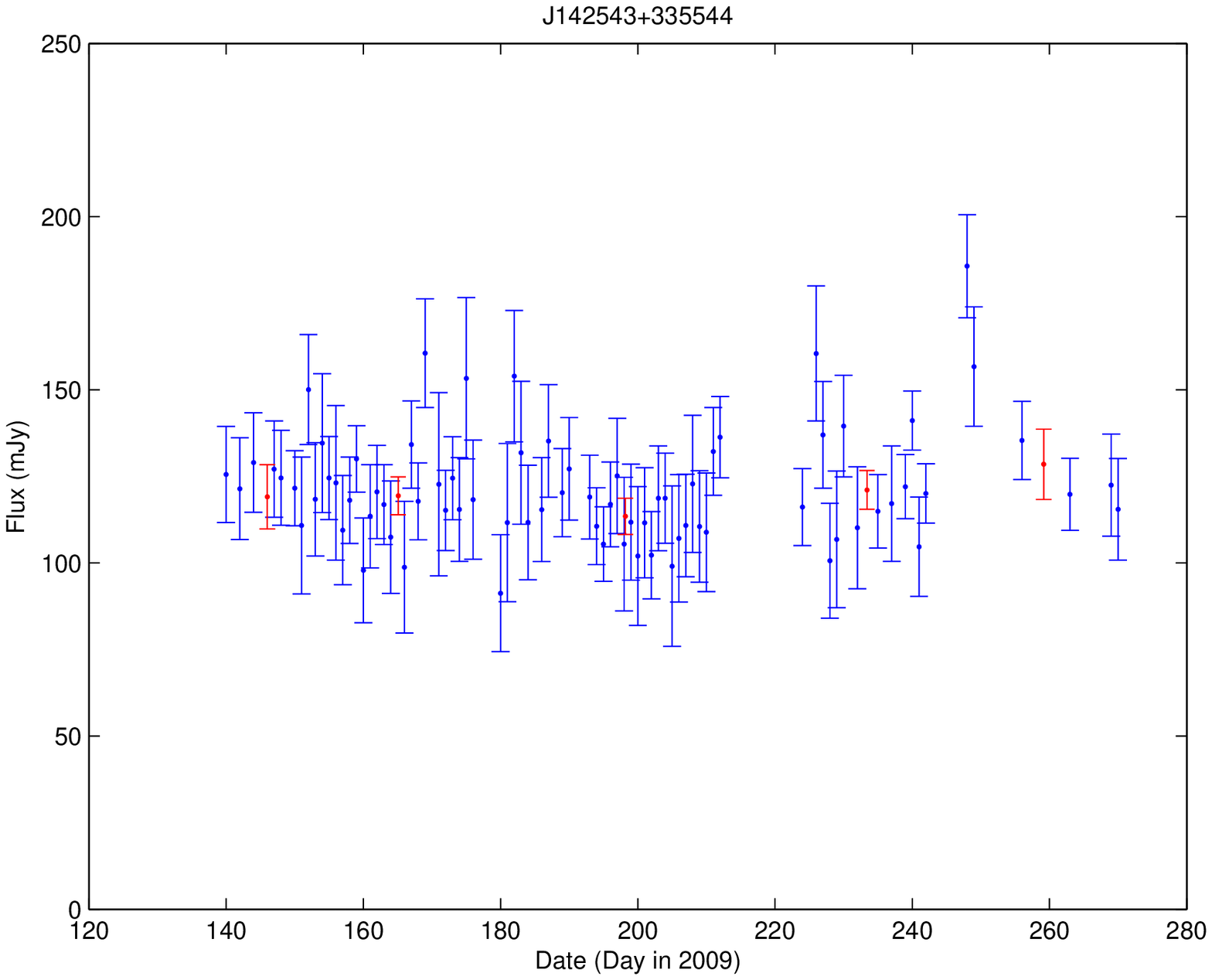,width=0.35\textwidth}\psfig{figure=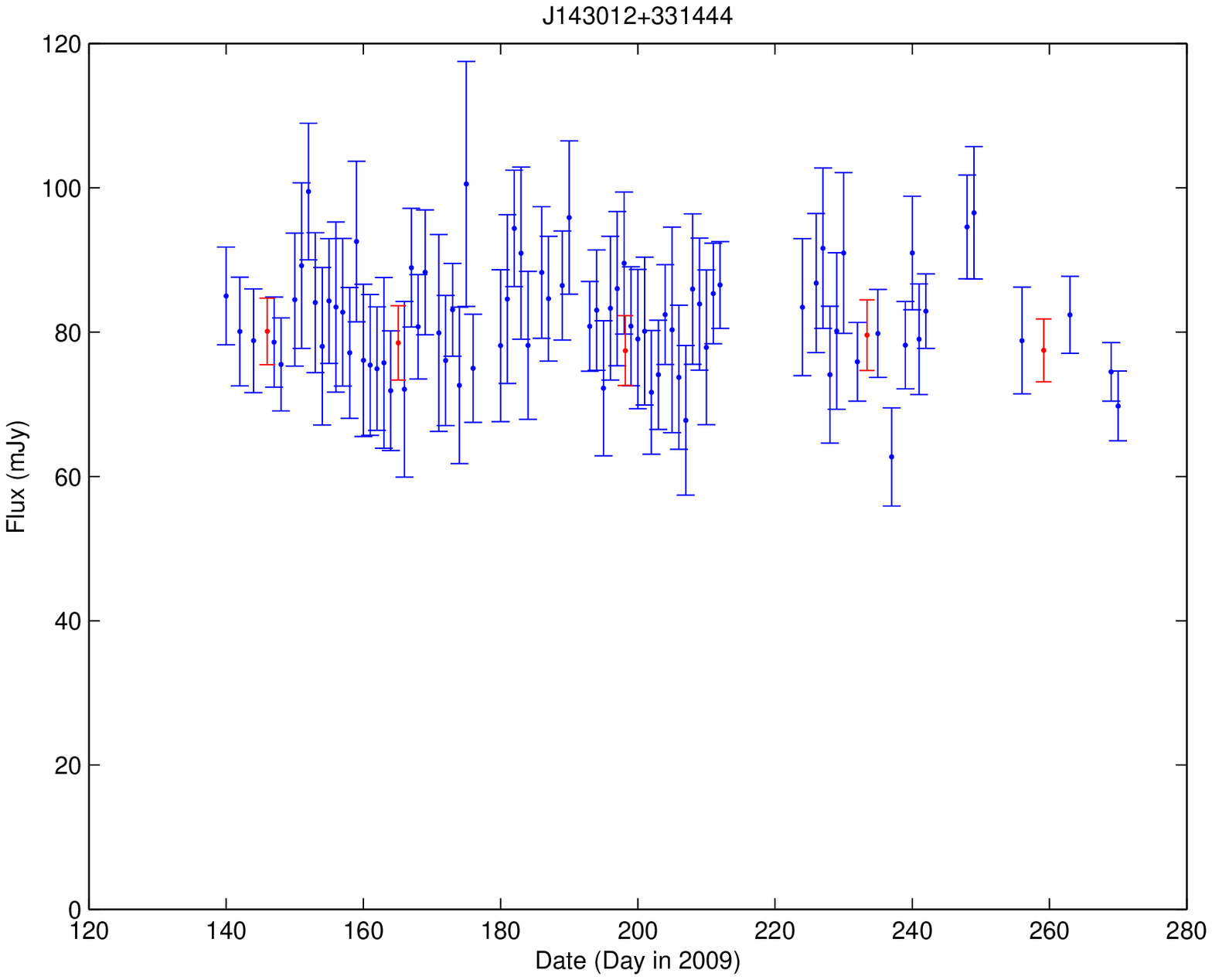,width=0.35\textwidth}}
\mbox{\psfig{figure=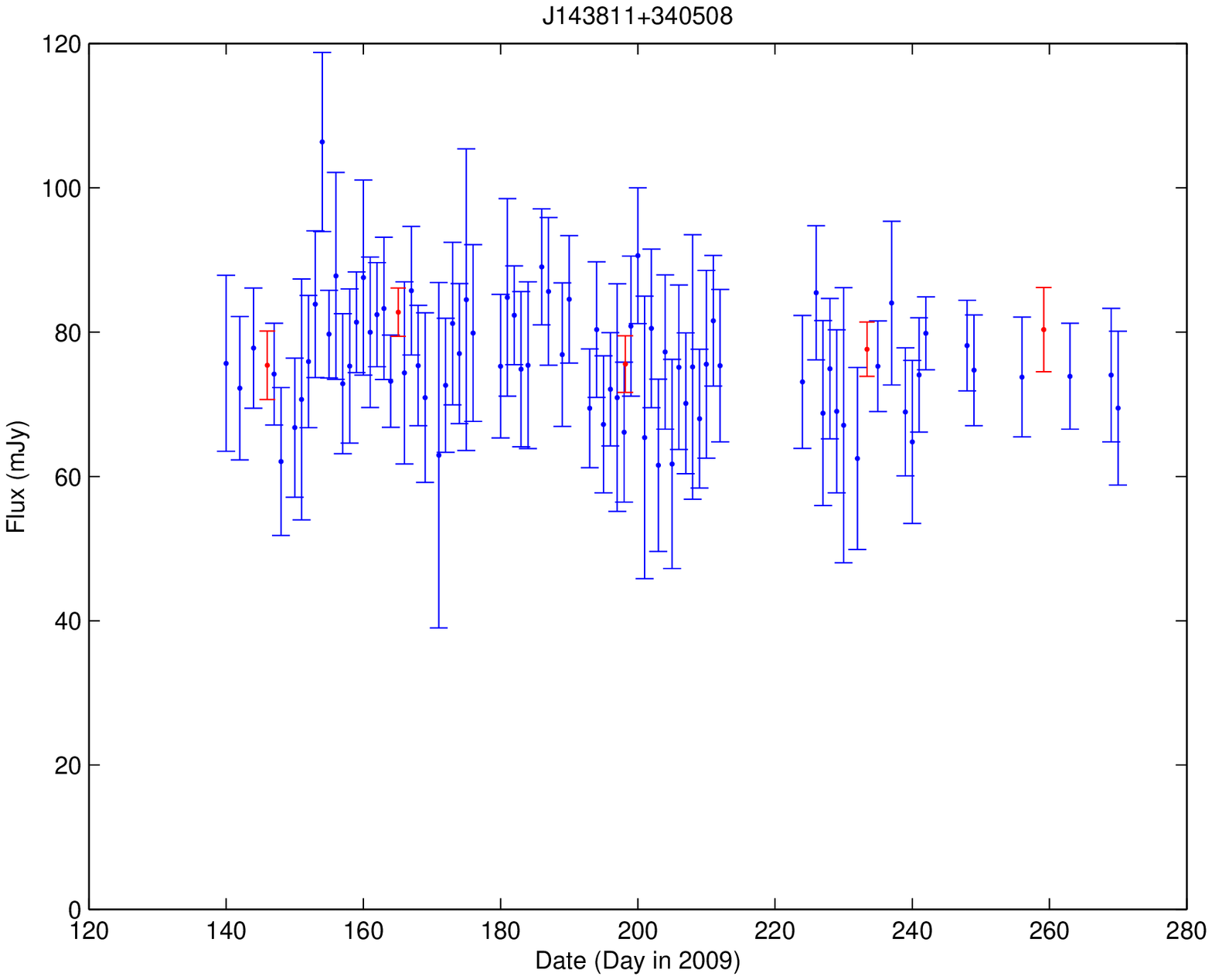,width=0.35\textwidth}}
\caption{Light curves for three sources with the highest daily modulation fractions with $S > 30$ mJy 
and primary beam gain correction factors $<4.0$
but not including
sources with high daily RMS modulation fractions (Fig.~\ref{fig:var}).  Blue
dots indicate daily flux densities; red dots indicate monthly flux densities.
\label{fig:var1}}
\end{figure}

In Figure~\ref{fig:var1}, we plot light curves for variable sources selected on the basis of
their maximum daily fractional modulation and appying the same restrictions to the data
as for the RMS fractional modulation ($S > 30$ mJy and gain factor less than 4).
We select nine sources with $m_{i,d} > 0.2$.  Of these, 6 are in common with those selected by
the RMS fractional modulation.  We plot the remaining three sources in Fig.~\ref{fig:var1}.
These sources have a maximum $m_{m,d}$ of 0.81 but $\sigma_{m,d} \sim 0.1$.  One can see clear evidence
of episodic variations with a duration of days in the light curves of these sources that 
stands out from otherwise stable light curves.

\begin{figure}[H]
\mbox{\psfig{figure=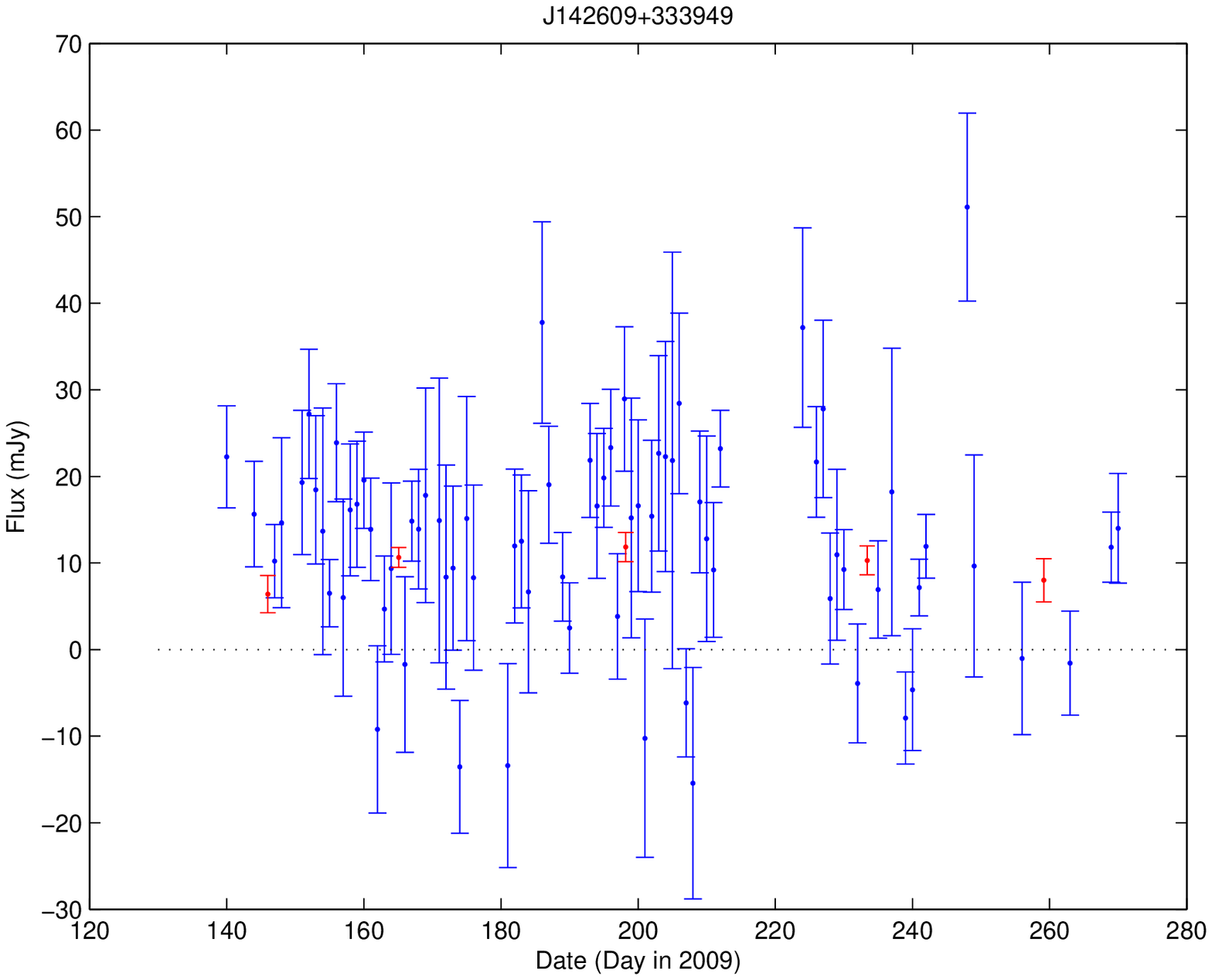,width=0.35\textwidth}\psfig{figure=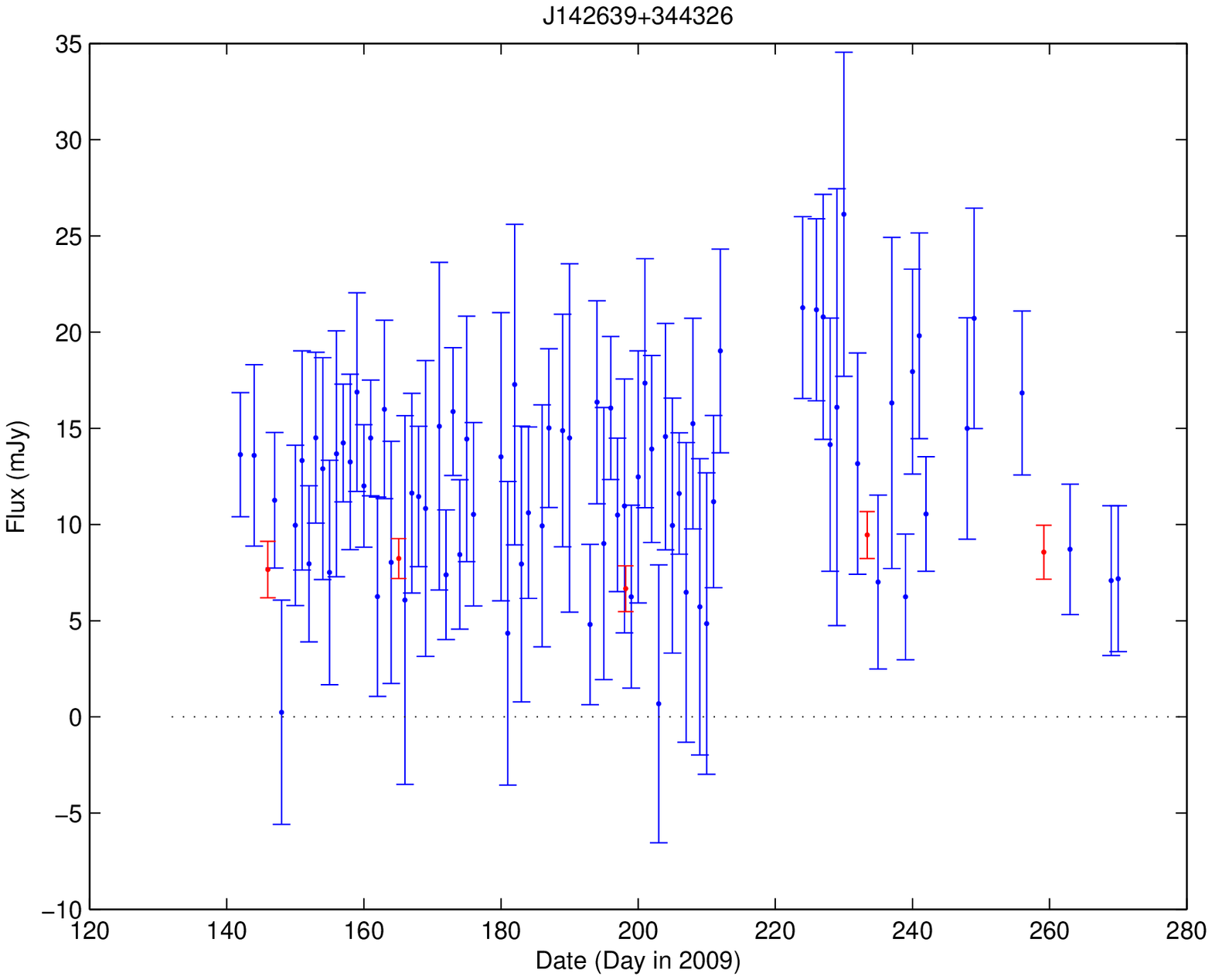,width=0.35\textwidth}}
\mbox{\psfig{figure=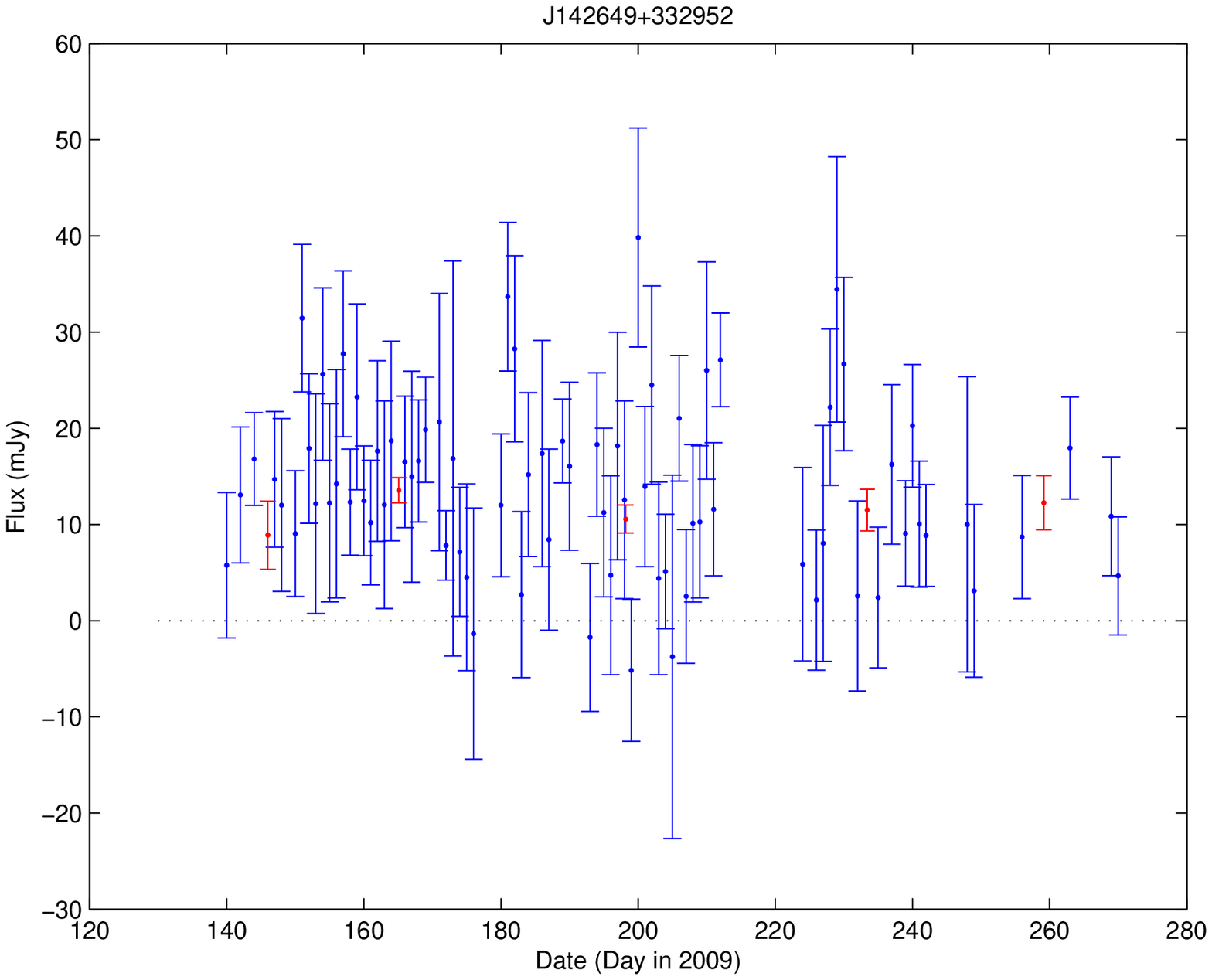,width=0.35\textwidth}\psfig{figure=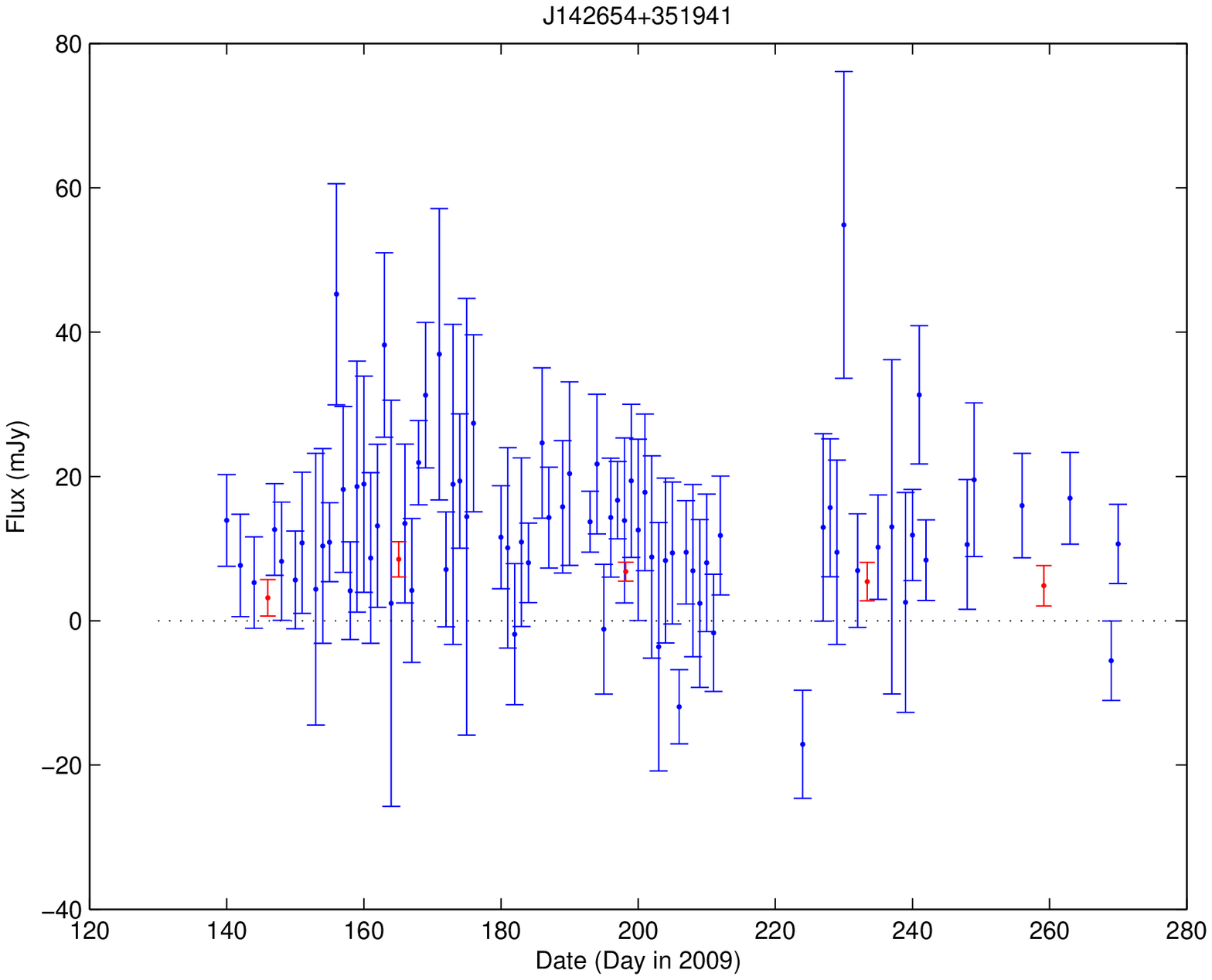,width=0.35\textwidth}}
\mbox{\psfig{figure=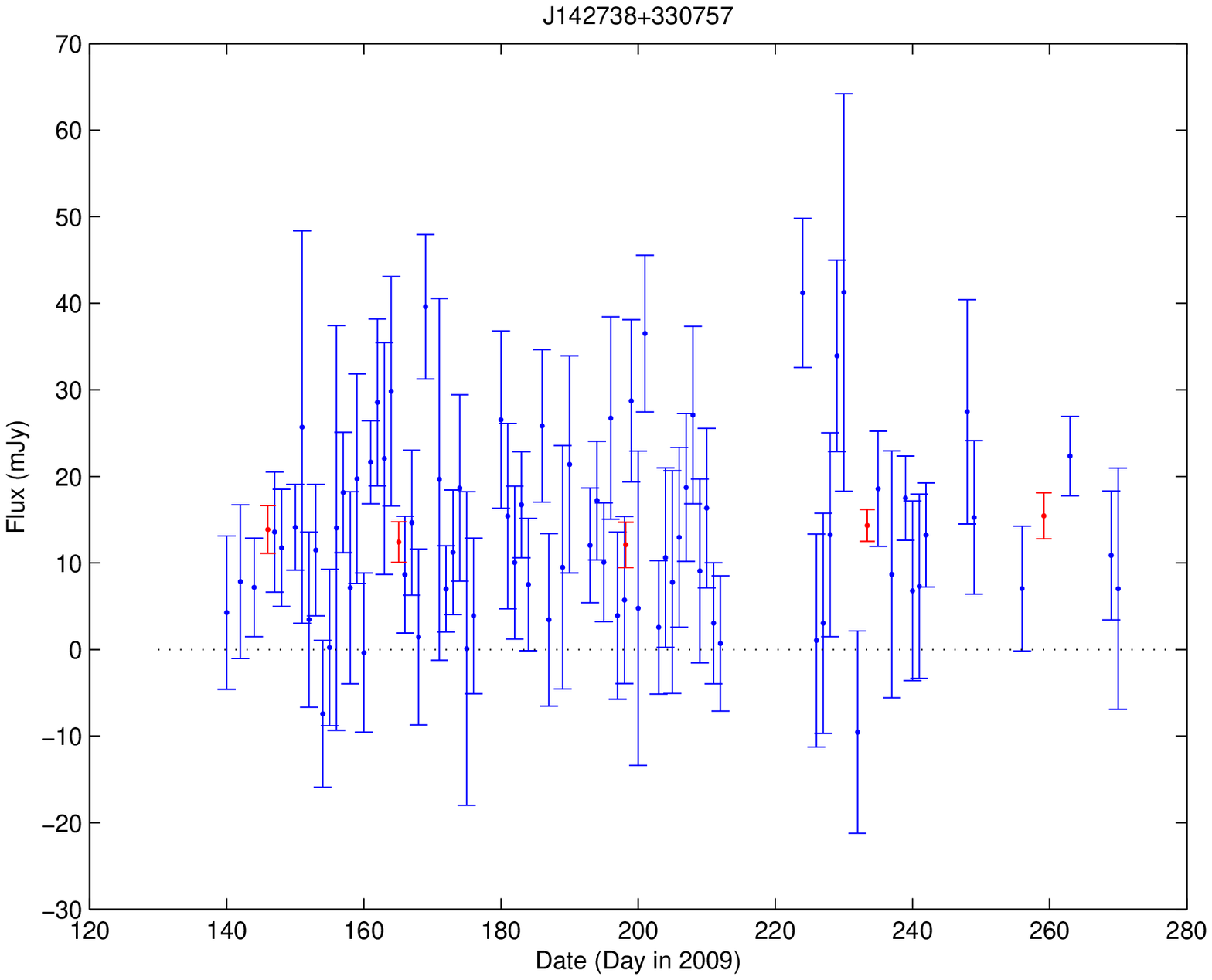,width=0.35\textwidth}\psfig{figure=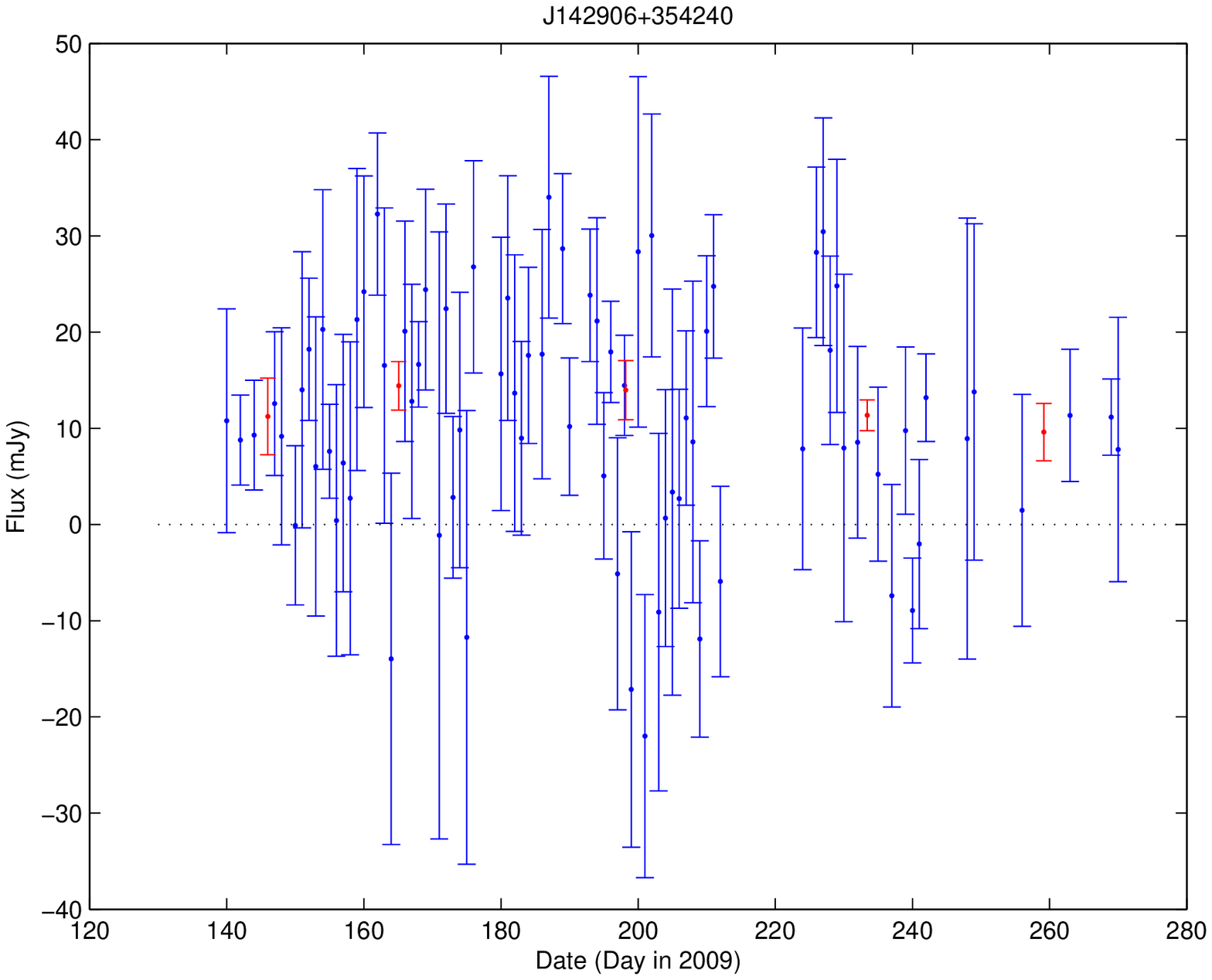,width=0.35\textwidth}}
\mbox{\psfig{figure=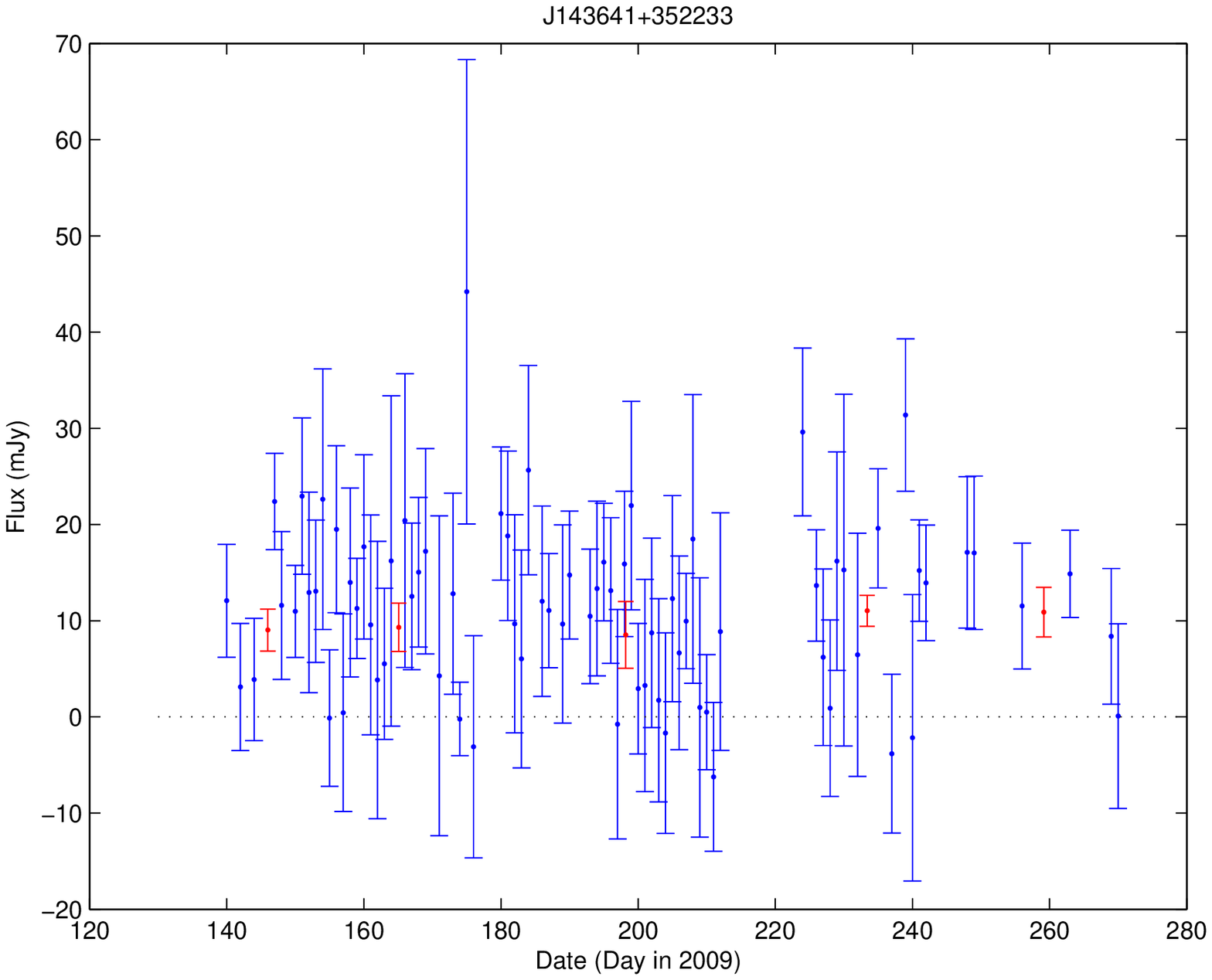,width=0.35\textwidth}}
\caption{Light curves for seven sources with the highest daily modulation fractions for $10 < S < 30$ mJy.  Blue
dots indicate daily flux densities; red dots indicate monthly flux densities.
\label{fig:var2}}
\end{figure}

This maximum value for $m_{m,d}$ is not the maximum seen in the histograms of $m_{m,d}$.  Thus, we 
conclude that the most variable sources fall outside of the
criteria we used for selecting these sources; that is, these sources are less 
than 30 mJy and/or have gain factors greater than 4.  Are these flux density variations real?
We plot light curves for sources with $m_{m,d} > 1$ and $10 < S < 30$ mJy in Fig.~\ref{fig:var2}.
These seven sources have a maximum $m_{m,d} = 3.9$, which is representative of the distribution
of the most variable sources.  We see strong modulation on a range of timescales in these sources.

\subsection{Structure Function Analysis}

\begin{figure}[H]
\mbox{\psfig{figure=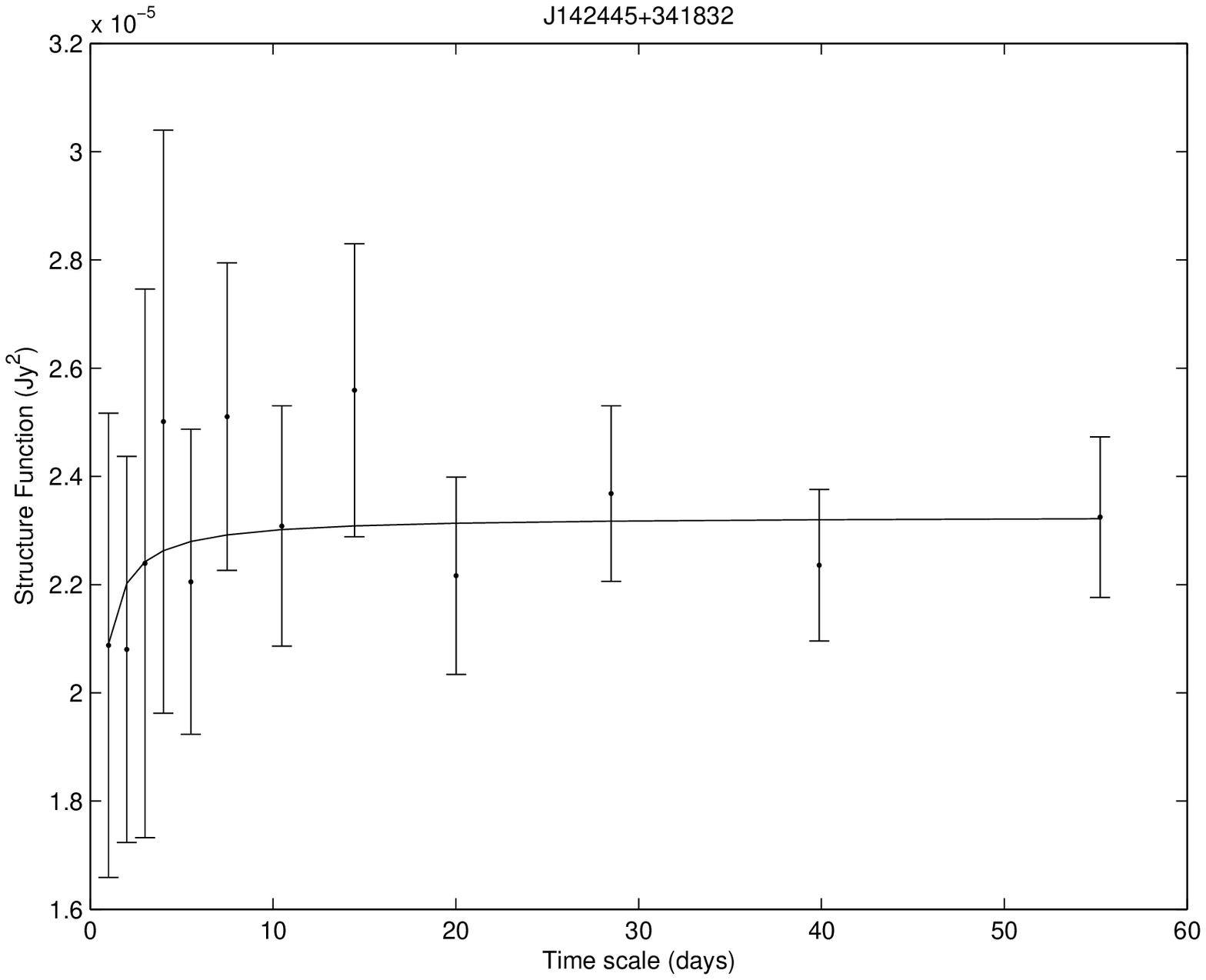,width=0.35\textwidth}\psfig{figure=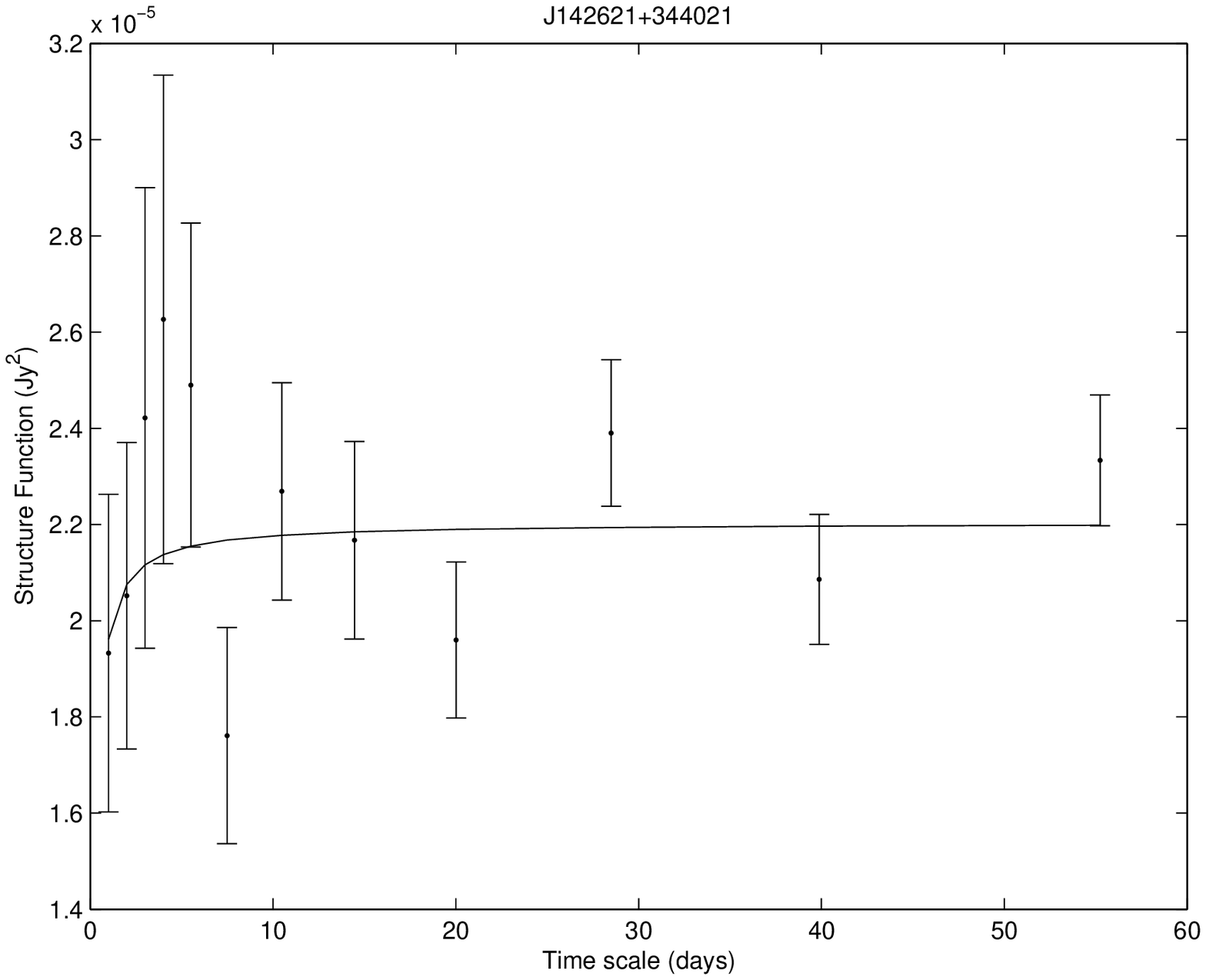,width=0.35\textwidth}}
\mbox{\psfig{figure=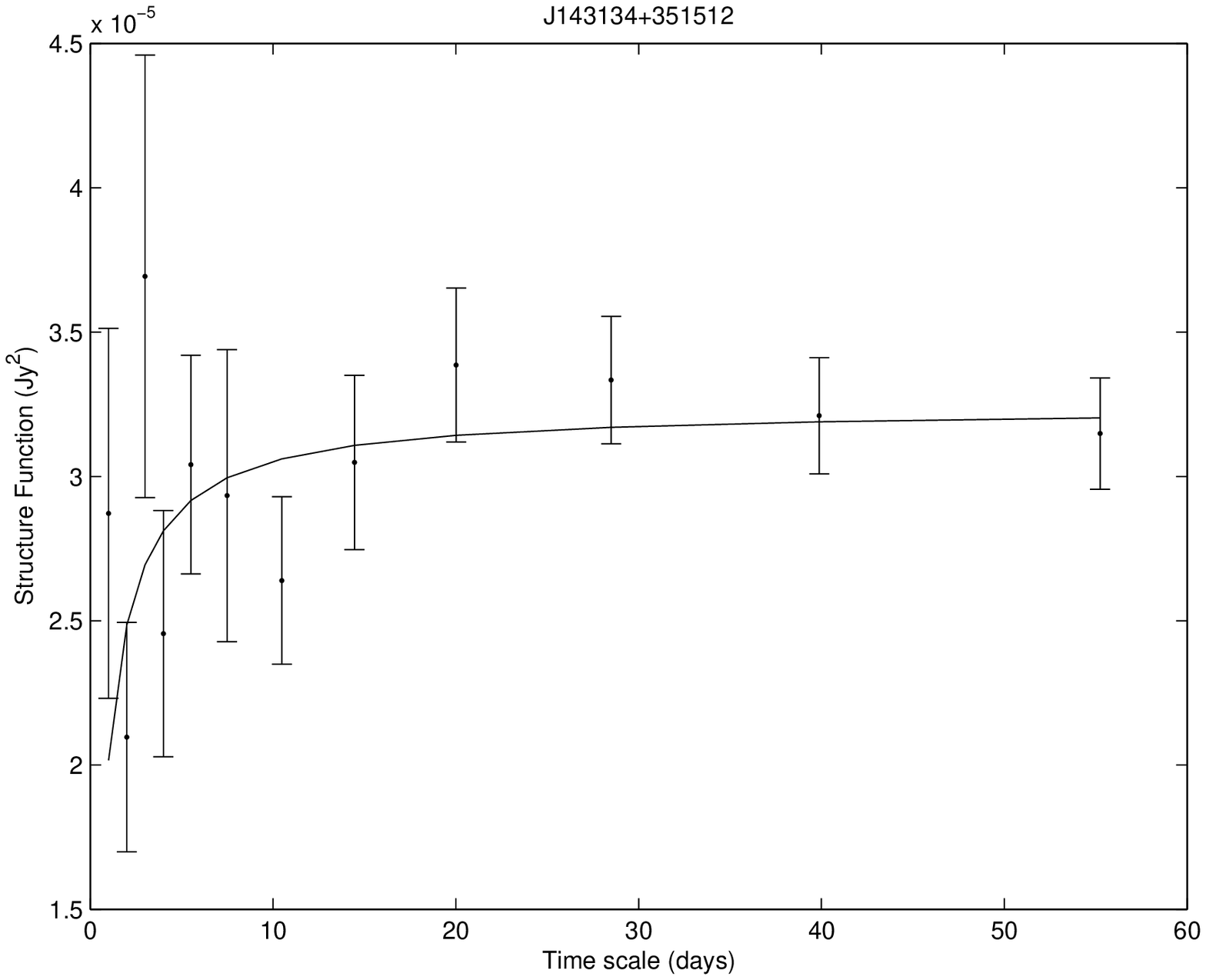,width=0.35\textwidth}\psfig{figure=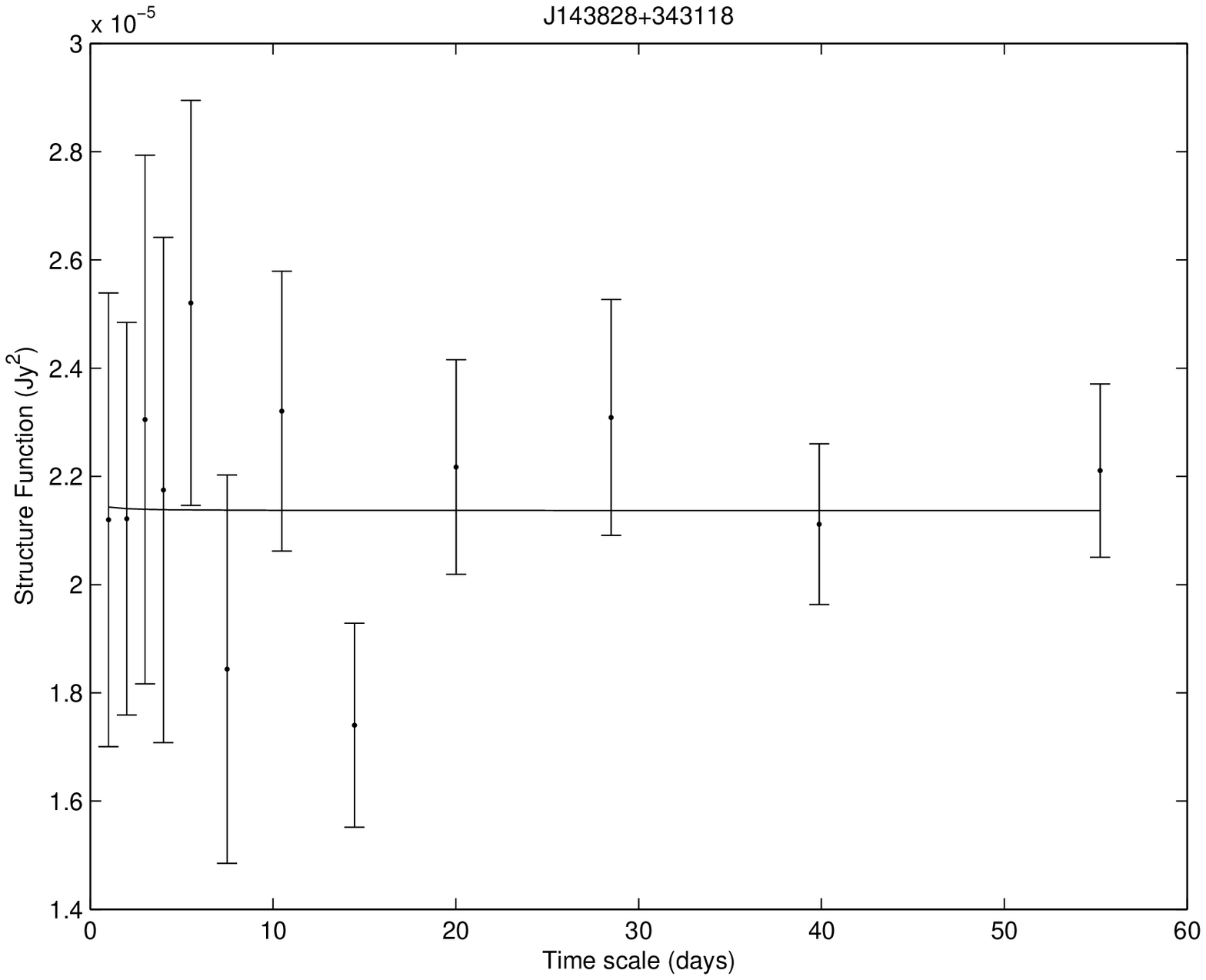,width=0.35\textwidth}}
\mbox{\psfig{figure=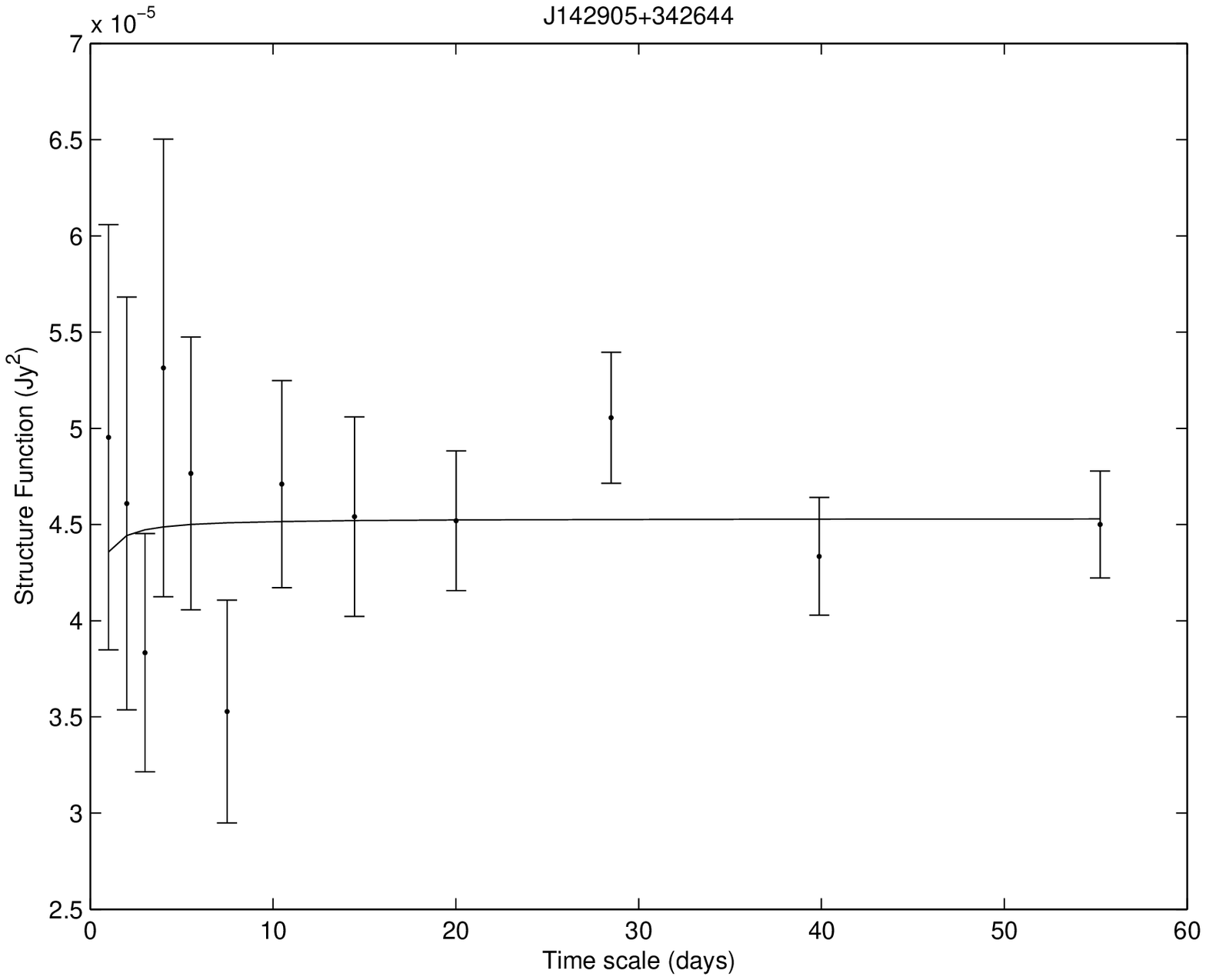,width=0.35\textwidth}\psfig{figure=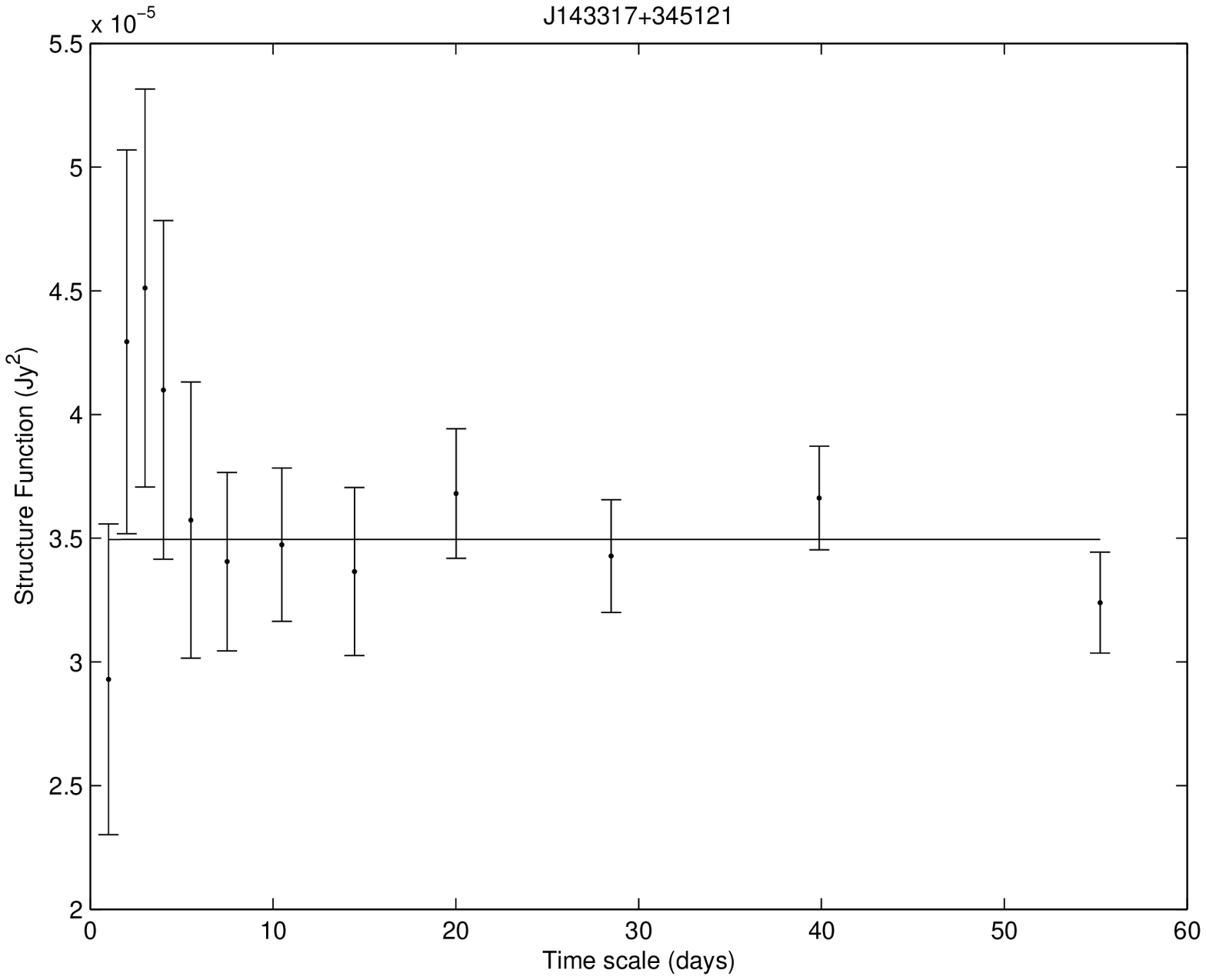,width=0.35\textwidth}}
\mbox{\psfig{figure=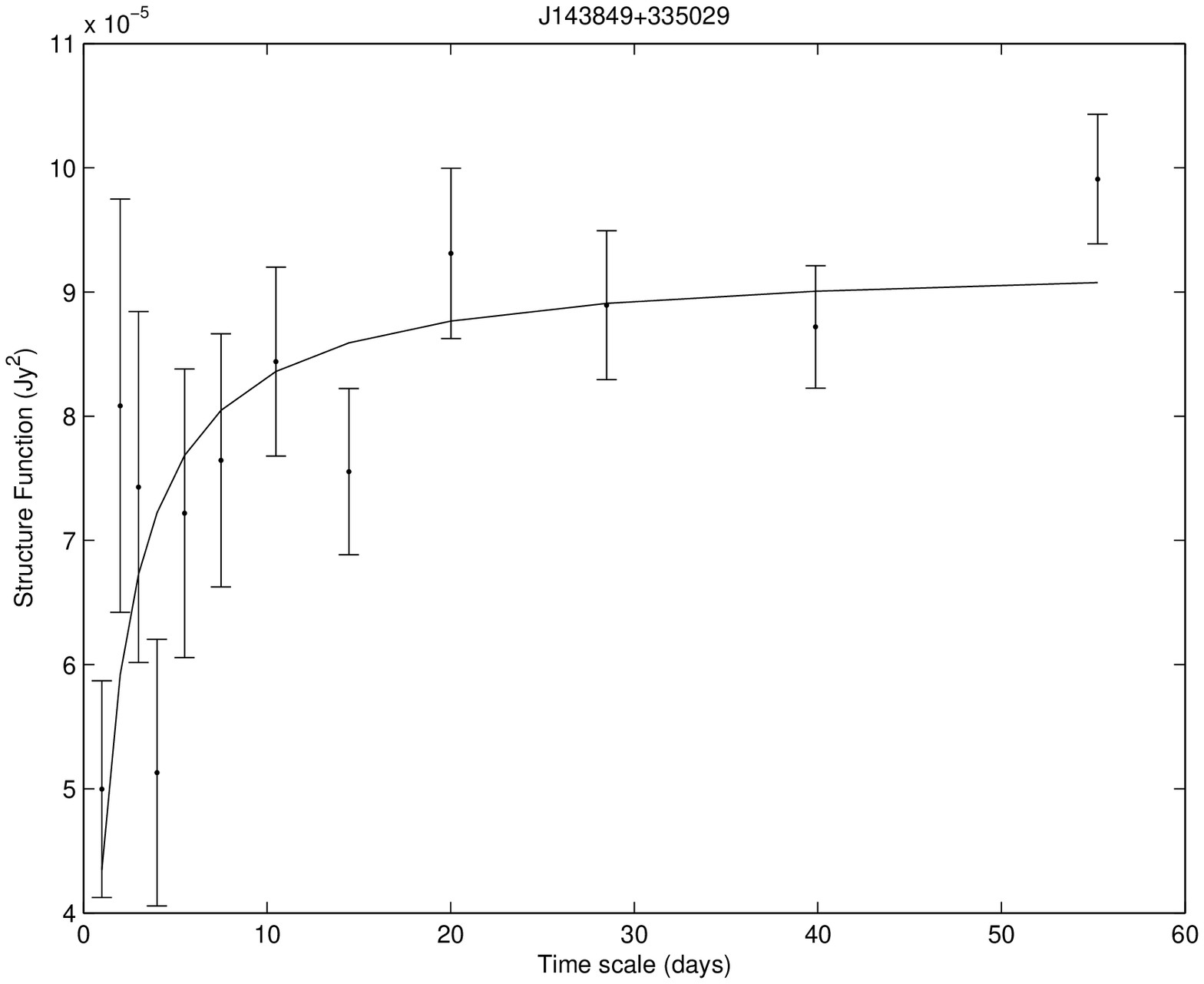,width=0.35\textwidth}\psfig{figure=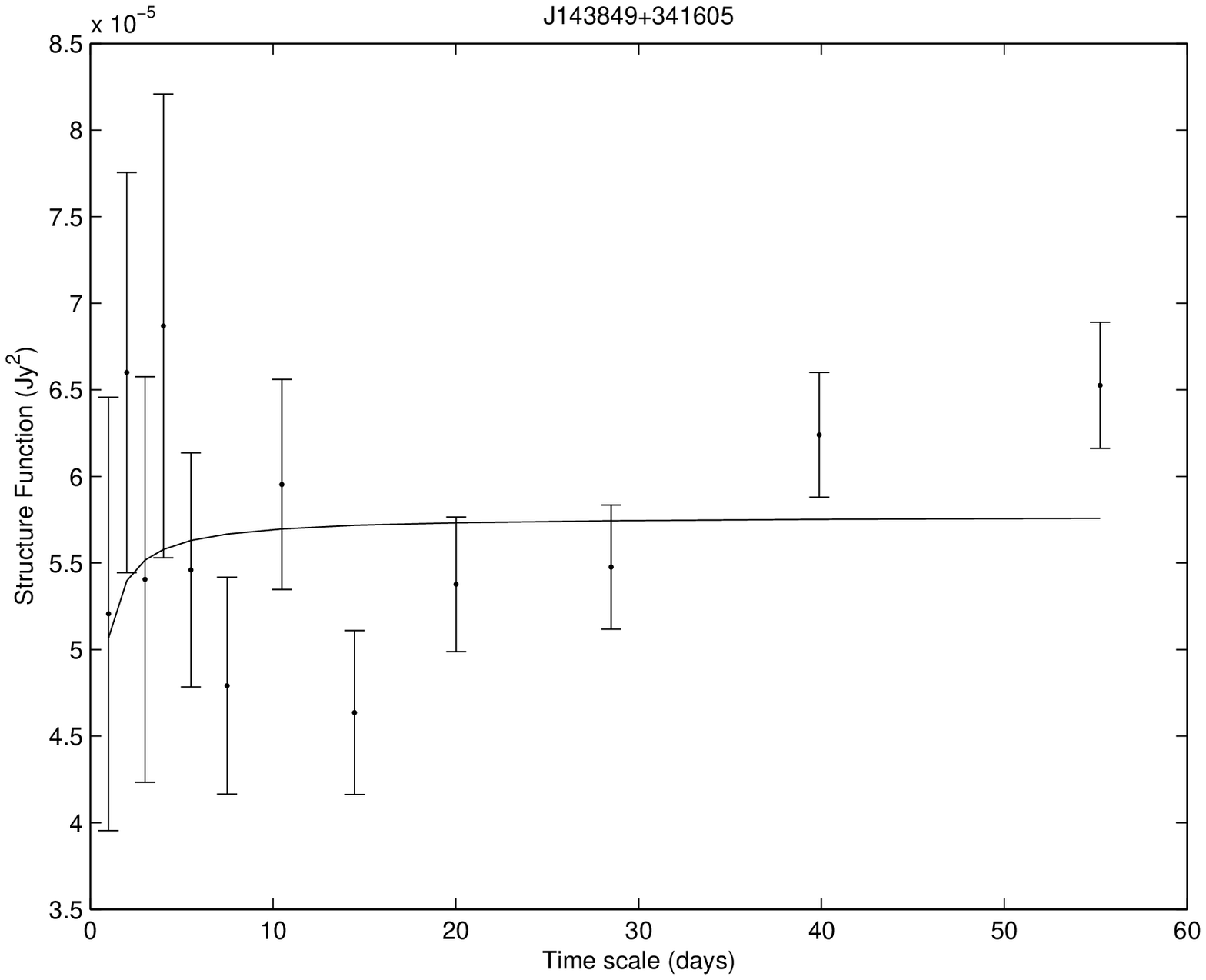,width=0.35\textwidth}}
\caption{Structure functions for selected sources.  The top four panels show sources selected 
with the highest daily RMS modulation fraction (Fig.~\ref{fig:var}).  The bottom four panels show sources selected as 
those with the lowest daily RMS modulation fraction (Fig.~\ref{fig:steady}).  The black curves are fits of
the ISS model described in the text.
\label{fig:sfunex}
}
\end{figure}

The structure function, $D(\tau)$, is the variance
over a range of characteristic time scale, $\tau$.
The RMS fractional modulation calculated on daily and monthly time scales can be 
construed as limiting cases of the structure function.  We use the same analytic form
for the structure function as \citet{2008ApJ...689..108L}.
We produced structure functions for each source over time scales of 1 to 55 days.  The maximum
time scale is half of the total duration of the observations.  In Fig.~\ref{fig:sfunex}, we show examples
of structures selected from sources with the highest and lowest daily RMS modulation fractions.
Since the RMS modulation fraction is an absolute quantity, we see evidence of variability 
timescales even for sources that have low RMS modulation fraction.

\begin{figure}[H]
\mbox{\psfig{figure=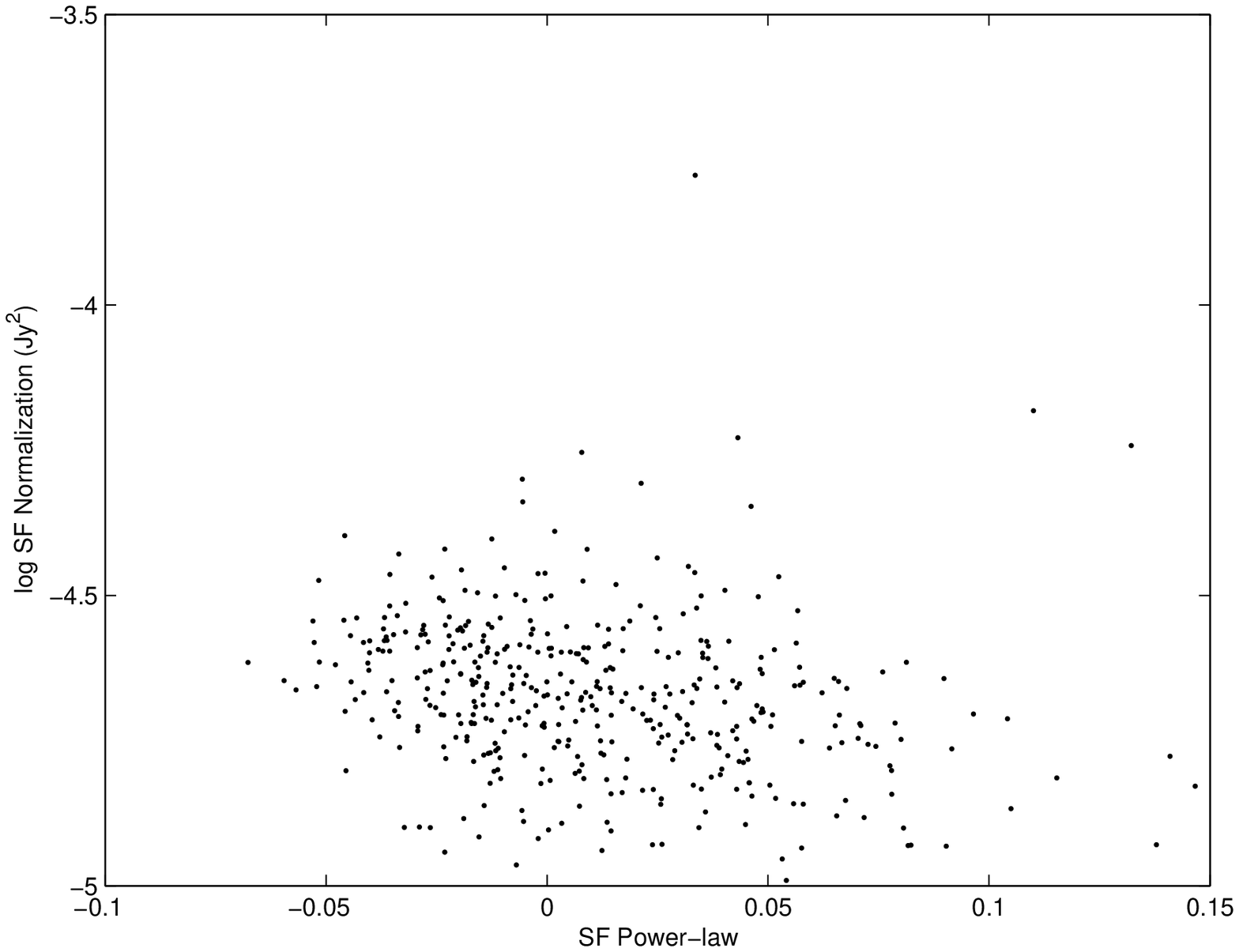,width=.45\textwidth}\psfig{figure=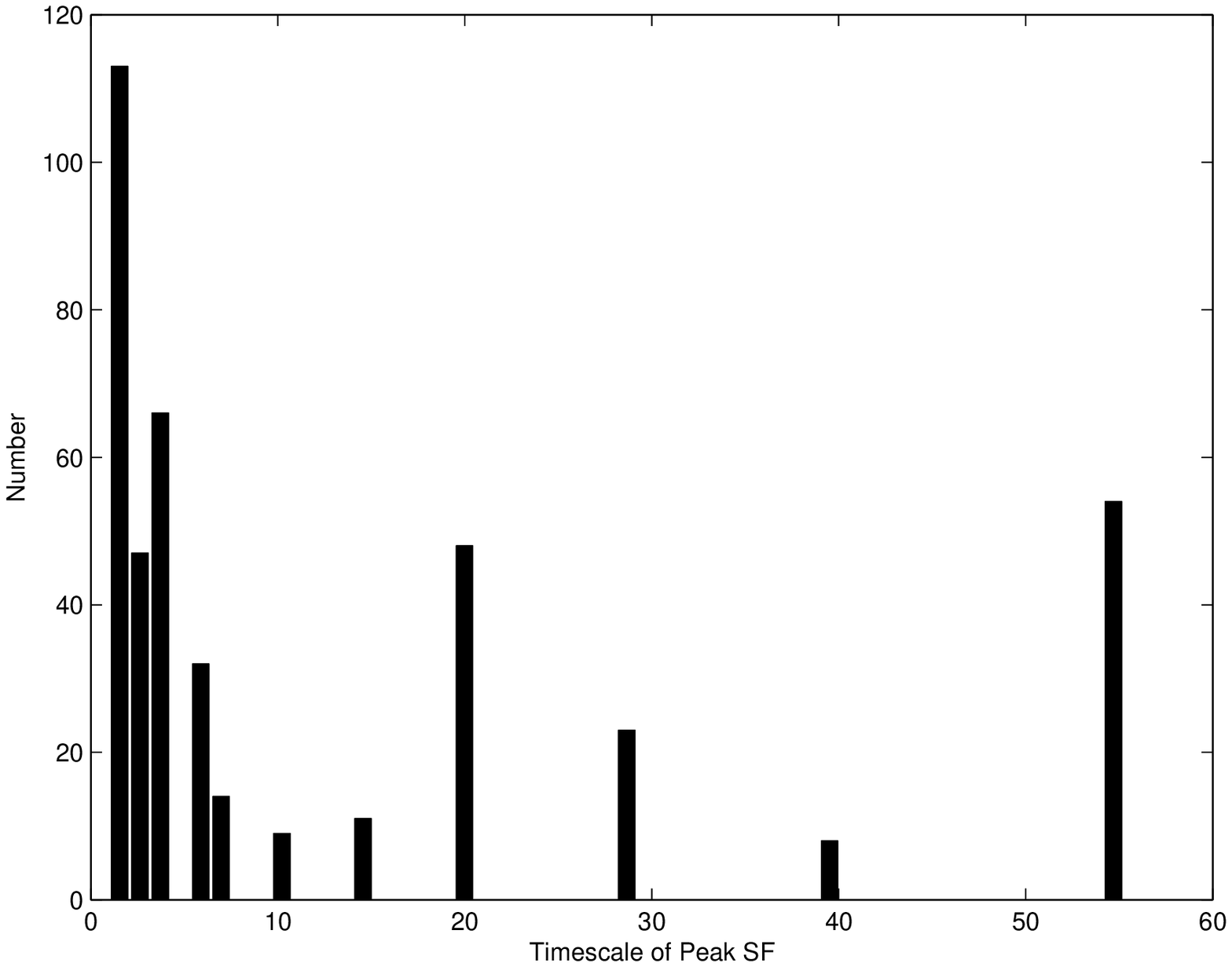,width=.45\textwidth}}
\caption[]{({\em left})  Results of power-law fits to structure functions for all sources.  The abscissa
represents the power-law index and the ordinate represents the normalization at 1 day.
({\em right}) Histogram of the time scale at the peak of the structure function for each source.
The histogram reveals the exponential spacing of time bins that we used in
construction of the structure function.
\label{fig:sf}}
\end{figure}

\begin{figure}[H]
\mbox{\psfig{figure=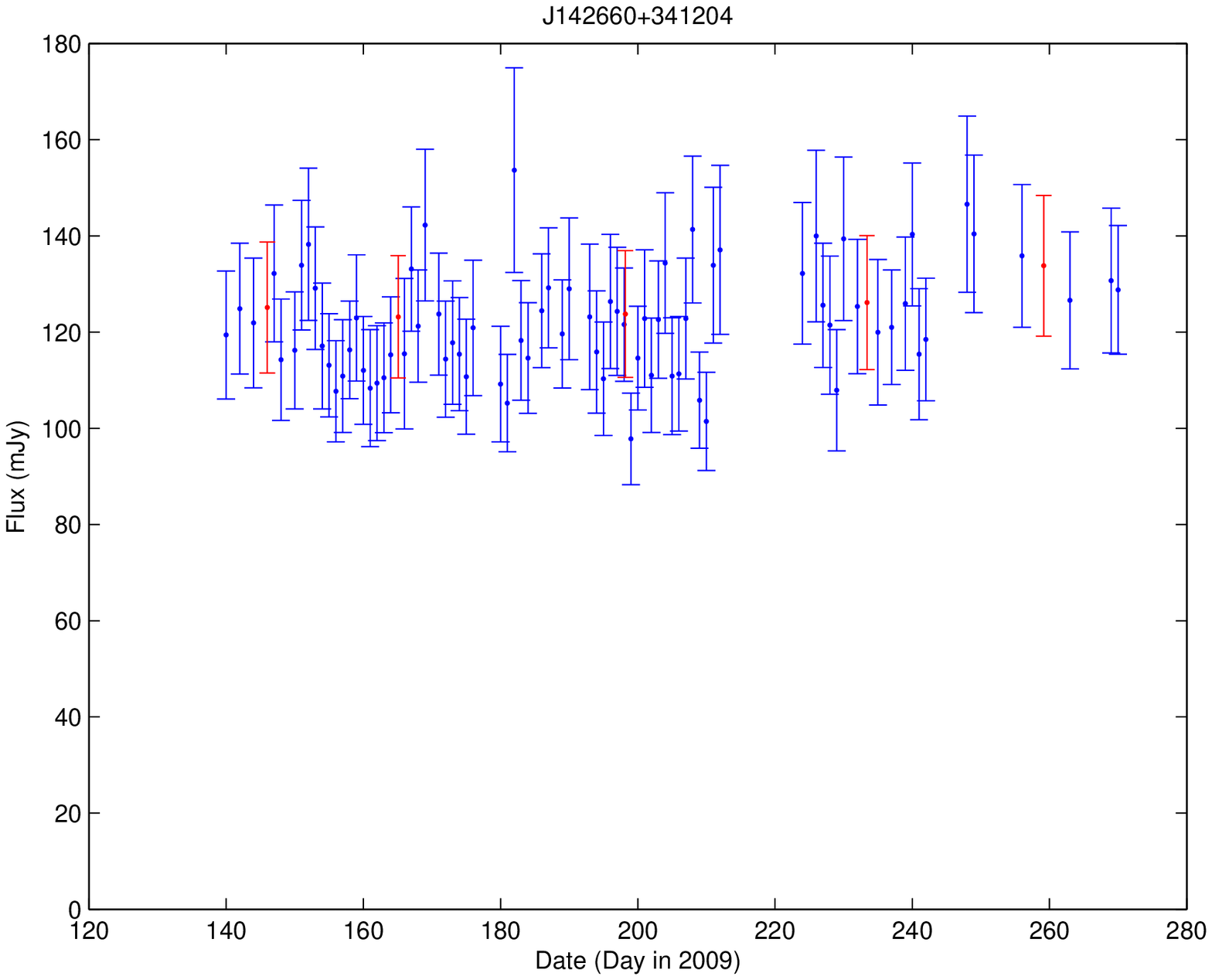,width=0.35\textwidth}\psfig{figure=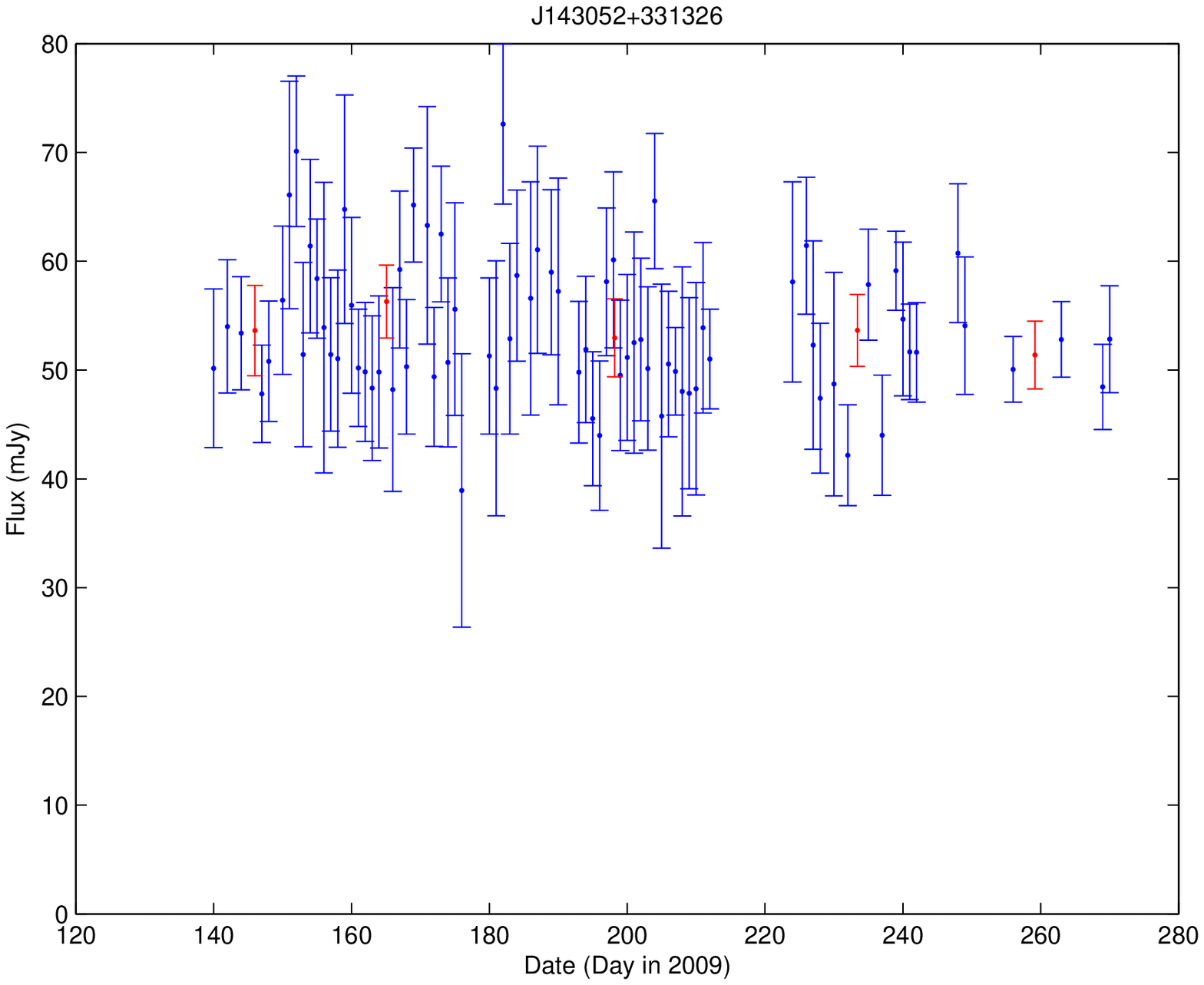,width=0.35\textwidth}}
\caption{Light curves for two sources with the highest structure function amplitude
but not including
sources with high daily modulation fractions (Fig.~\ref{fig:var}).  Blue
dots indicate daily flux densities; red dots indicate monthly flux densities.
\label{fig:var3}}
\end{figure}

The results are remarkably flat
and independent of time scale.  We calculated power-law fits to the structure function 
(Fig.~\ref{fig:sf}).  The range of power-law indices is -0.1 to 0.15, indicating 
that the light curves do not evolve significantly on time scales greater than a few days, typically.
Note that a linear increase in flux density over this period would have an index 2.  We also computed the
time scale of the peak of each structure function and find that those peaks are distributed over
the full range of time scales.  About half of the peaks are at timescales less than 5 days.
We identified sources with the largest peak structure function values.  Of the top eight,
only two were not identifed by any of our previous methods.  We plot these two in
Fig.~\ref{fig:var3}.  These have characteristic timescales at their peaks of a few days.
Others identified through this method have peak timescales as long as 55 days.   Given
the flat distribution of $D(\tau)$, many of these long timescale fluctuations may not
be statistically significant.

We also fit the function $D(\tau) = 2 m^2 \tau \ (\tau + \tau_{char})$, which is
characteristic of the fluctuations expected from interstellar scintillation 
\citep[ISS;][]{2008ApJ...689..108L}.  We find that $\tau_{char} < 3$ days for all sources, with
a median value $<1$ day.  Overall, the bias to short timescales is consistent with 
the findings of \citet{2008ApJ...689..108L} for the bright source population probed by the MASIV
survey and to the fainter source population studied by \citet{2011arXiv1103.3010O}.
The result is also consistent with analysis of pulsar flux variability due to ISS
\citep{2000ApJ...533..304R}.

\subsection{Multiwavelength Comparisons}

A region covering 8.1 sq.\ deg.\ in the Bootes field was imaged four times using the {\em Spitzer} Space Telescope by the Spitzer Deep Wide Field Survey (SDWFS) team
\citep{2009ApJ...701..428A}. Variability statistics for the 3.6 and 4.5\,\micron\ data for 474,179 sources were 
presented by \citep{2010ApJ...716..530K}. To explore whether sources that are highly variable in PiGSS are also highly variable in SDWFS, we wish to match the two catalogs. Since the SDWFS source density is much higher than that for PiGSS, however, we must first improve the localization of PiGSS sources to avoid the effects of source confusion.

We first match the 66 sources from the PiGSS variability catalog with mean flux density $\geq 10$\,mJy and in regions of the PiGSS image with primary beam gain $\leq 4.0$, to the Faint Images of the Radio Sky at Twenty-cm (FIRST) catalog \citep{1995ApJ...450..559B}. FIRST's astrometric precision ($\lesssim 1$\arcsec) enables us to then match our radio sources to the SDWFS catalog with little ambiguity. We searched for the closest FIRST source within 30\arcsec\ of the PiGSS sources. For the 64 PiGSS sources with a FIRST match, we searched the SDWFS catalog for the closest source within 10\arcsec\ of the FIRST position. This resulted in 50 PiGSS sources with a match, of which all had good photometry in all four Infra-Red Array Camera (IRAC) bands.

We can investigate the infrared variability of the matches using $\sigma_{12}$, a dimensionless
measure of the joint significance of variability at 3.6 and 4.5\,\micron\ from SDWFS \citep{2010ApJ...716..530K}. Of the sources with a SDWFS match, 5/50 have $\sigma_{12} \geq 2$ and Pearson correlation coefficient between the variability in the two {\em Spitzer} bands, $r > 0.8$, and hence qualify as variable according to the definition of \citet{2010ApJ...716..530K}. Our 10\%\ fraction of infrared-variable sources is comparable to the fraction of sources identified as AGNs by their infrared colors which are also variable in the infrared from \citeauthor{2010ApJ...716..530K}. This is consistent with the expectation that most PiGSS sources are AGNs. 

\begin{figure}[H]
\mbox{\psfig{figure=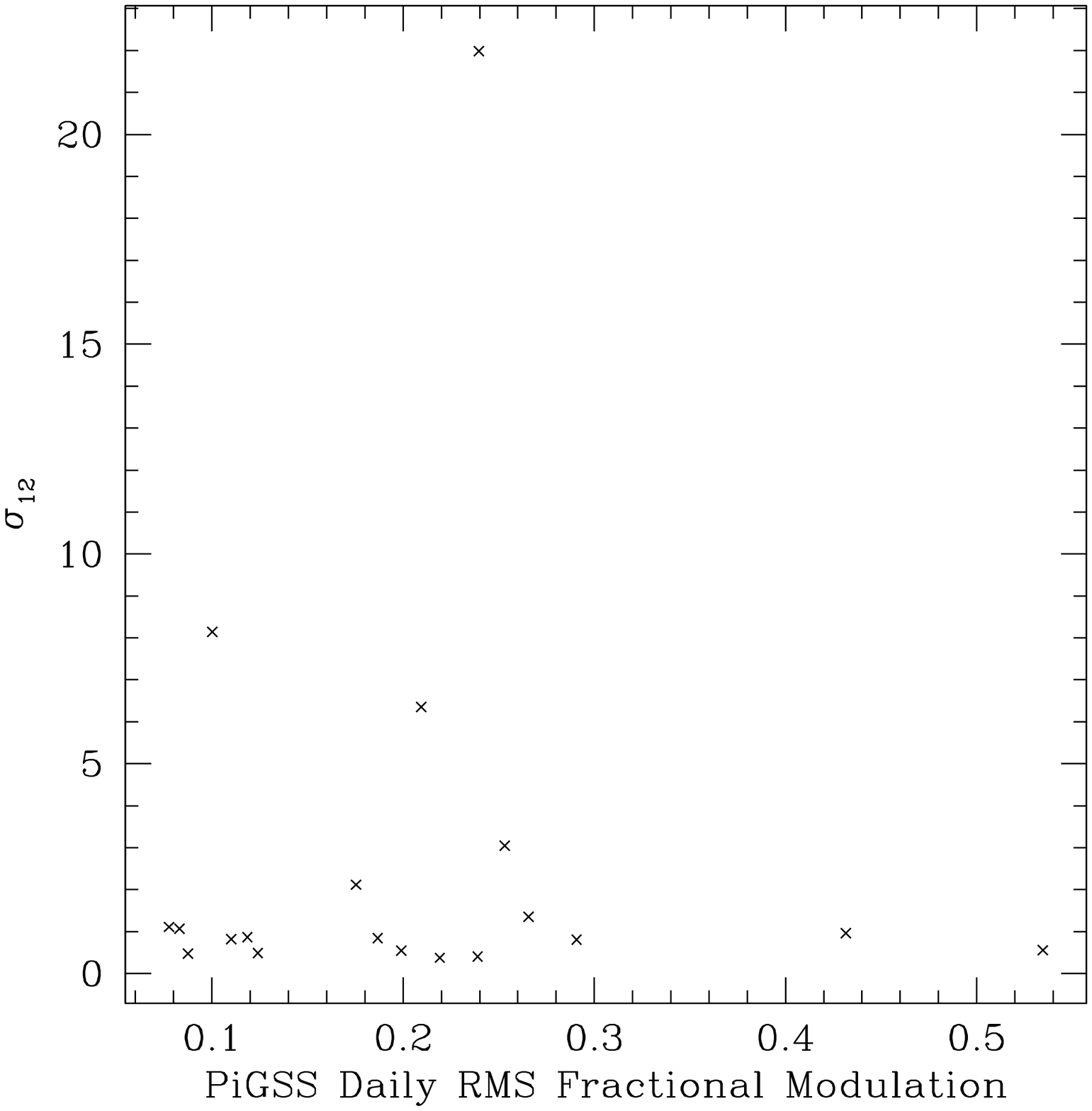,width=0.45\textwidth}\psfig{figure=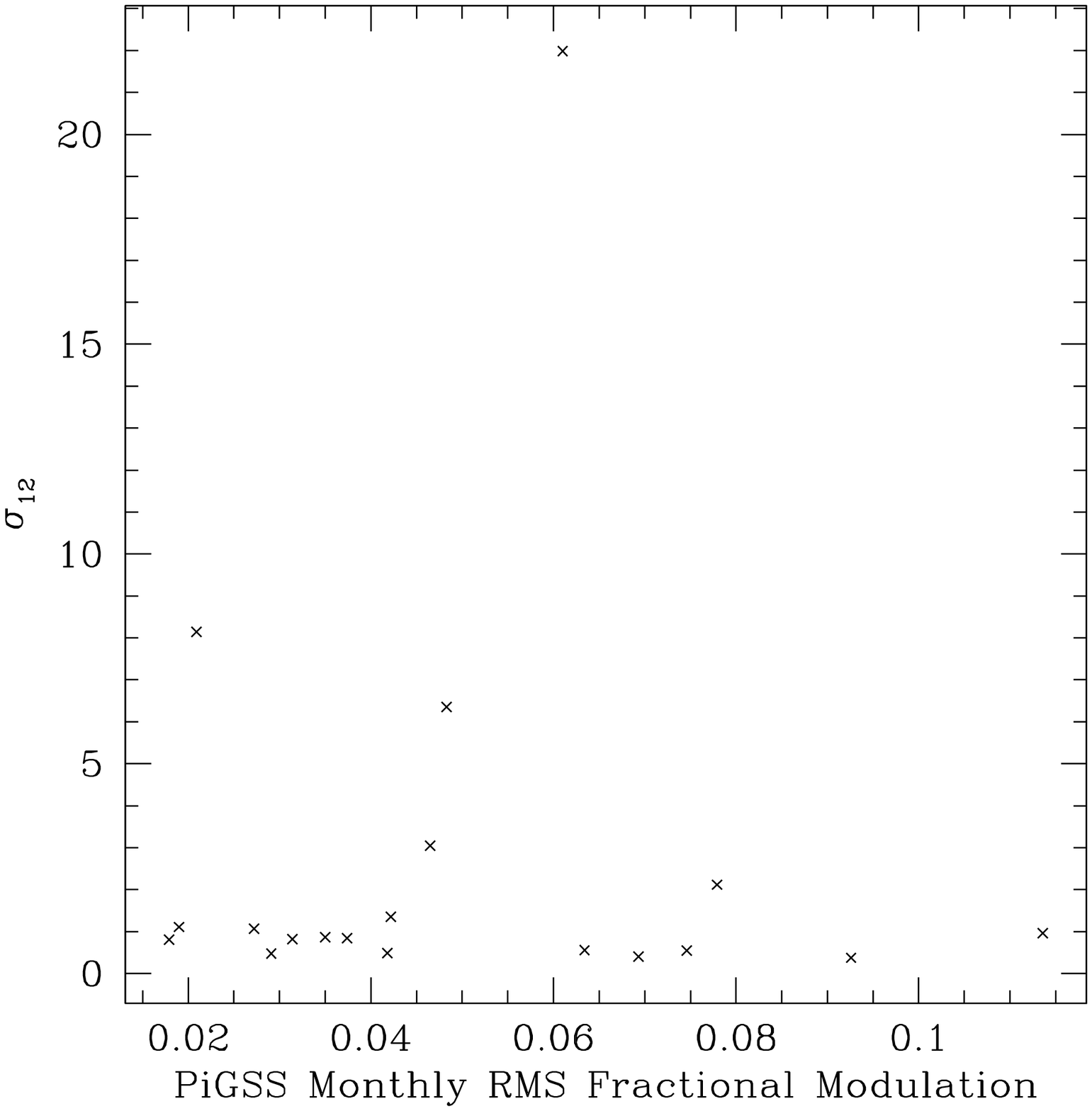,width=0.45\textwidth}}
\caption{
({\em left})
PiGSS RMS daily fractional variability versus joint variability at 3.6 and 4.5\,\micron, $\sigma_{12}$, from SDWFS. Plotted are the 19 sources with PiGSS mean flux density $\geq 10$\,mJy, PiGSS primary beam gain $\leq 4.0$, a match in FIRST within 30\arcsec\ of the PiGSS position, a match in SDWFS within 10\arcsec\ of the FIRST position, and Pearson correlation coefficient between the variability in the two {\em Spitzer} bands, $r > 0.5$. 
({\em right})PiGSS RMS monthly fractional variability versus $\sigma_{12}$.
There is little correlation between radio and infrared variability, aside from the four sources with large infrared variability, $\sigma_{12} > 3$, which all have monthly comparatively low radio variability, $\leq 0.061$. 
\label{fig:sdwfs}}
\end{figure}

In Fig.~\ref{fig:sdwfs}, we plot the PiGSS RMS monthly fractional variability versus $\sigma_{12}$ for the 19 sources with $r > 0.5$. The interpretation is limited by the small-number statistics, despite our more relaxed cut in $r$ here, but we note that sources with high variability in the infrared appear to show low variability at 3\,GHz. The four sources with $\sigma_{12} > 3$ all have monthly variability $\leq 0.061$. 

The outlier source in Fig.~\ref{fig:sdwfs}, which has $\sigma_{12} = 22.0$ and $r = 0.97$, is \,J142607+340433, which is the BL Lac SDSS J142607.71+340426.2.  This source does not show
significant variability in the radio.  Differences between radio and IR variability, of course,
may be due to episodes of variability during the different observing epochs.
The other four sources with $\sigma_{12} > 2$ and $r  > 0.5$ are not listed as blazars in the NASA Extragalactic Database.

We also compare the PiGSS catalog to the X-ray catalog generated for this field \citep{2005ApJS..161....9K}.
Chandra observations identify 3293 sources with high confidence over a 9 deg$^2$ region that largely overlaps the
PiGSS survey area.  We find 36 matches to PiGSS sources with $\geq 10$ mJy and primary beam gain $\leq 4.0$.
We compare the 0.5 to 7 keV flux ($F_x$) and the hardness ratio (HR) for these matches against the parent population of all
X-ray sources with a Kolmogorov-Smirnov test. Neither distribution differs from the parent population at 95\%
confidence.  We also search for a dependence of $F_x$ and HR on the daily and monthly RMS modulation fractions
and identify no clear dependence.  There is no information on variability of the X-ray sources in the field.

\section{Application to Source Classes}

We consider the application of current and future PiGSS results
to several source classes here.  These provide guidance towards understanding our
sensitivity on the range of timescales that different sources are active on.
It is not possible to address all possible source classes that are of interest.  Both 
\citet{2007ApJ...666..346B} and \citet{2010ApJ...711..517O} provide discussion of a range of source classes 
and their likely properties
as radio transient sources.  We steer the discussion here primarily to extragalactic 
origins for radio transients but include low mass stars and brown dwarfs due to their 
ubiquity.

\subsection{Tidal Disruption Events}

Stars that pass close within the tidal radius of a massive black hole can produce
a transient accretion disk and may produce a relativistic jet in some case
\citep[e.g.,][]{1988Natur.333..523R}.
X-ray, ultraviolet, and optical detections of flares from quiescent galactic nuclei
are consistent with the tidal disruption hypothesis 
\citep{2009ApJ...698.1367G,2010arXiv1008.4131S,2010arXiv1009.1627V}.  Until recently,
there have been no clear radio detections of tidal disruption events 
although there are two candidate sources from the MOST survey that meet the basic
criteria of variable emission originating in the nucleus of a nearby galaxy
\citep{2011arXiv1103.4328B,2010arXiv1009.1627V}.  
\citet{2011arXiv1102.1429G} proposed a model for the radio 
emission based on the interaction of the jet with the dense interstellar medium.
Radio sources are expected to appear approximately 1 year after the disruption event
and have a duration $\lsim 1$ year.
\citet{2011arXiv1103.4328B} used this model and existing radio surveys to constrain the rate
of relativistic jets created to be $\lsim 10^{-6}$ Mpc$^{-3}$ y$^{-1}$ at $2\sigma$
confidence.  Very recently, a radio source associated with a long-duration
gamma-ray burst with long-duration X-ray flaring in the nucleus of a nearby galaxy 
has been interpreted as beamed emission associated with a tidal disruption event
\citep{2011arXiv1104.3257B,2011arXiv1104.3356L}.  \citet{2011arXiv1104.4105V}
proposed a model in which the radio emission emerges from the jet on prompt timescales
based on standard accretion-disk and jet coupling.

Based on the reverse-shock model of \citet{2011arXiv1102.1429G}, we can use the 
comparison of the PiGSS-I deep image with archival data about the field
to set a constraint of $\lsim 10^{-5}$ Mpc$^{-3}$ y$^{-1}$ at $2\sigma$ confidence.
If the events have a duration of $\sim 1$ month, then the monthly results from 
this paper provide a similar sensitivity to the event rate.  If no events are
detected in $10^4$ square degrees at 10 mJy sensitivity from the full PiGSS campaign,
then the limit on event rates will be $\lsim 1 \times 10^{-7}$ Mpc$^{-3}$ y$^{-1}$ at $2\sigma$
confidence.  This limit is comparable to the rate inferred from UV and X-ray studies,
suggesting that we are in the domain of potential discovery.  We find similar limits
on and predictions for the rates using the ``always radio loud model'' (model a) 
from \citet{2011arXiv1104.4105V}.  Models b and c have peak luminosities 1 to 2 orders of magnitude less than model
a and, therefore, provide significantly lower constraints on the rate.

\subsection{Radio Supernovae}

Radio emission from supernovae originates from the interaction of the
stellar ejecta with circumstellar and interstellar material 
\citep{2002ARA&A..40..387W}.  
For a peak RSNe luminosity of $1 \times 10^{27}$
erg\,s$^{-1}$\,Hz$^{-1}$ and a characteristic duration of $\sim 1$ month, 
we are sensitive to RSNe out to a distance of 180 Mpc
in the monthly images.  Based on the observed optical Type IBc and II SNe rate of $5 \times 10^{-5}
{\rm y}^{-1} {\rm Mpc}^{-3}$, we estimate that these observations 
have an expectation of $\sim 0.2$ RSNe \citep{2011MNRAS.tmp..413L}.  
The full PiGSS $10^4$ deg$^2$ will, therefore, have
sensitivity to $\sim 10$ RSNe.  Deeper and more sensitive surveys can probe 
the apparent discrepancy between the rate of optical SNe and cosmic
star formation \citep{2011arXiv1102.1977H}.

\subsection{Orphan Gamma-ray Burst Afterglows}

Orphan GRB afterglows (OGRBAs) are GRB afterglows that occur when the 
relativistic beam of gamma-ray emission is directed away from the observer
\citep{1997ApJ...487L...1R}.  The resulting afterglow will radiate isotropically, or 
less anisotropically, than the relativistic beam, making it visible from many
directions.  A number of OGRBA models have been proposed
\citep{1998ApJ...509L..85P,2002ApJ...576..120T}.  The most recent
provides specific predictions for number and flux density of radio afterglows
\citep{2008MNRAS.390..675R}.  The limits from these models are a factor of 10 to
$10^3$ lower than in the previous models.  At a flux density threshold of 1 mJy, their model
predicts $\sim 10$ sky$^{-1}$ OGRBAs with durations of $\sim 100$ days.  
The long-duration of these events implies that we should make use of the 
deep image limit.  We use the PiGSS-I transient limit from the deep image 
of $\lsim 1$ deg$^{-2}$ at 1 mJy for transients with timescales greater than 
$\sim 100$ days.  Thus, the expectation for PiGSS-I is $\sim 10^{-3}$ OGRBA events.

Extrapolating to the full PiGSS survey with $10^4$ deg$^{2}$ at 10 mJy threshold,
we find an expectation of $\sim 0.1$ events.  Thus, there is an opportunity for a rare OGRBA 
to appear in the full PiGSS survey based on the model of 
\citet{2002ApJ...576..120T}.  The more optimistic models of \citet{2008MNRAS.390..675R}
and  \citet{1998ApJ...509L..85P} predicts $\sim 10$ to  100  detections; absence
of any detection will strongly reject these models.

\subsection{Radio Counterparts to Gravitational Wave Sources}

Neutron star mergers are potentially the hosts to short gamma-ray bursts
\citep[e.g.,][]{2006ApJ...638..354B} and also may produce gravitational wave (GW) events that could be detected
by Advanced LIGO and other ground-based GW detectors.
\citet{2011arXiv1102.1020N} have proposed a model for radio emission from the shock resulting
from energetic, sub-relativistic outflows.  The authors argue that one
long-duration event detected by \citet{2007ApJ...666..346B} fits the signature of this outflow.
Given the long-duration of these events ($\gsim 1$ month), the 
PiGSS-I deep data provide the best constraint yet.  For a flux density threshold
of 1 mJy, a beaming factor $f_b^{-1} = 30$ and a burst energy of $10^{51}$ erg, the authors estimate 
a surface density $\sim 1$ sky$^{-1}$.  A comparison with PiGSS-I deep rate limits, gives
an expectation of $\sim 10^{-4}$ events.  Extrapolation to the full PiGSS survey
at 10 mJy sensitivity gives an expectation of $\sim 0.01$ events.

\subsection{Extreme Scattering Events}

Extreme scattering events (ESEs) are sudden and sustained drops in the flux density of
compact extragalactic radio  sources as a consequence of refractive effects
\citep{1994ApJ...430..581F}.  ESEs typically have durations that range from 0.25
to 1.2 y and have an amplitude from 6\% to 100\% at 2.7 GHz.  From a sample of
flat spectrum objects, \citet{1994ApJ...430..581F} estimate from a Green Bank 
Interferometer (GBI) survey a rate of 0.008 event-years
per source-year.  2 out of 10 events in the GBI study have amplitudes of $\sim 100\%$,
implying a rate of $\sim 0.001$ event-years per source-year at this large amplitude.
PiGSS-II sensitivity is predominantly to events of duration less than 0.2 years,
shorter than the shortest duration known ESE.  The large number of sources monitored,
however, does permit us set to upper limits to the ESE event rate on these time scales.
If we consider the fractional modulation limits for monthly data, we can place an
upper limit on the number of ESE events 
at 100\% amplitude on 1 month time scales of 1.9\%  of all PiGSS events for sources
brighter than 10 mJy.  This translates directly to an upper limit of 0.019 events-year
per source-year.  Note that this is limit differs from the GBI limit, which is based on a
sample of flat spectrum sources.  In Paper I, we estimated that 20\% of sources in
this field are flat spectrum.  Thus, our limit on ESEs with 1 month duration is $\sim 0.1$
events-year per source-year.  Matched filters can provide an improved method for 
searching for ESEs on a range of time scales; \citet{2001ApJS..136..265L} applied a wavelet analysis to
Green Bank Interferometer data in search of ESEs.

\subsection{Intra-Day Variable Sources}

Diffractive scintillation in the ISM can lead to flickering on timescales of $\sim 1$ day
in compact, flat-spectrum radio sources \citep{2008ApJ...689..108L}.  The MASIV
survey found that 37\% of such sources exhibited fluctuations greater than 1.4\%
over 2 days.  The larger amplitude fluctuations that PiGSS is sensitive to are more rare.
Fractional variations $\gsim 0.1$ occur in approximately 1\% of the MASIV sources.
For the majority of PiGSS sources, variations on a 1-day time scale at this level
are difficult to detect in  our data.  
What is clear that very large 
amplitude fluctuations such as those seen for J1819+3845 and PKS 0405-385 are rarely
observed \citep{1997ApJ...490L...9K,2000ApJ...529L..65D}.  \citet{2008ApJ...689..108L} found
no objects with this level of variability out of over 400 flat spectrum sources
investigated.  

Could some of the faint
sources with strong variations be IDV?  
The large amplitdue, but low probability variations that are seen in Fig.~\ref{fig:fluxhistdaily}
imply rather large brightness temperatures.  A variation of 10 mJy with a timescale of 1
day implies a brightness temperature for a source at 1 Gpc of $\sim 10^{16}$ K.  If real,
brightness temperatures this large require ISS as an explanation.
As Figure~\ref{fig:var3} shows,
there are some sources that exhibit strong variability that is statistically
significant.  What remains to be seen from this list of highly variable objects is
whether that variability is intrinsic or due to systematic errors.  EVLA follow-up
of these sources near the PiGSS flux threshold can be an effective method
for determining the reality of these fluctuations.

\subsection{Low Mass Stars and Brown Dwarfs}

Low mass stars and brown dwarfs make up a large fraction of optical transients
\citep{Becker04,Kulkarni06} and are also known as impulsive radio transients
\citep{2008ApJ...674.1078O,2007ApJ...663L..25H}.  \citet{2007ApJ...666..346B}
found that these cool objects could provide a significant component of the
radio transient sky if they can produce rare magnetic outbursts with 
$L_{\nu} \gsim 10^{18} {\rm\ erg\ s^{-1}\ Hz^{-1}}$.  If we make use of the space density of
late type stars \citep{1999ApJ...521..613R}, the probability of radio bursts
with $L_{\nu} \gsim 10^{15} {\rm\ erg\ s^{-1}\ Hz^{-1}}$ \citep{Berger06},
and use the probability distribution of solar flares as a proxy for activity in
magnetic stars \citep{Nita04}, then one can estimate an expectation of the number
of transients per epoch as
\begin{equation}
N_{tot} \sim 10^{-2} \Omega \left( D_{lim} \over {\rm 1\ pc} \right)^{1.6} \left( S \over {\rm 1\ mJy} \right)^{-0.7},
\end{equation}  
where $\Omega$ is the survey field of view and $D_{lim}$ is the limiting distance at which transients can be seen.
These events are likely to be short-duration, so each daily epoch of PiGSS is an independent measurement.
For $D_{lim} \sim 300$ pc and the PiGSS-II sensitivity at 15 mJy, we find an expectation of $\sim 0.1$ 
events.  Extrapolating to the full PiGSS survey, we estimate that we may find 10s of events.
Most of these events will be dominated by very luminous, very distant flares.
Thus, the full PiGSS survey will provide a probe of extreme magnetic flares on the
surfaces of low mass stars and brown dwarfs.

\subsection{The Candidate RT J143621+334120}

We note that the candidate RT J143621+334120 found in the deep image with a duration $>100$ days
fits the expected radio properties of long duration transients such as
tidal disruption events, OGRBAs, and GW counterparts.  The absence of a host galaxy rules out
an RSNe origin.  The proximity to two SDSS stars suggests a late type star origin, however,
the duration of the event is much longer than expected.

\section{Conclusions \label{sec:conc}}

We report on daily observations of an 11-deg$^2$ region in the \boo constellation with
the ATA at 3.1 GHz as part of the PiGSS project.  The 78 epochs of observation
provide a systematic look at daily and monthly variations in the radio sky.  We find no 
radio transients.  There is a single event that was detected in the deep image
without a counterpart in pre-existing surveys such as NVSS 
that, if real, has a duration $\gsim 4$ months.  We place a new constraint on
the surface density of 1-day duration transients with $S > 15$ mJy of 0.025 deg$^{-2}$ and
monthly duration transients with $S > 5$ mJy of 0.18 deg$^{-2}$.  These limits are
consistent with previous efforts with the ATA and with other instruments.  

We also explored variability of the 425
radio sources in the deep field through a variety of statistics.
Largely, these sources show weak or no variability with RMS fractional modulations
less than 1 for sources brighter than 10 mJy.  20\% of the sources have statistically
significant variability.
The combination of 425 sources
and 78 epochs provides over $3 \times 10^4$ flux density samples from which we can
place limits on rare, large amplitude fluctuations.  The largest amplitude fluctuations 
($>10 \times$) occur fewer than 1 in $10^4$ source-days.  EVLA characterization of a
subset of these sources is important for determining the effect of systematic
errors in some of the larger variations reported here.  While these statistics are of
the most use for broadly characterizing the population, certain classes of objects
are better identified with matched filtering techniques.  For instance, extreme scattering
events and radio supernovae have reasonably well-defined and parametrizable light curves which 
could be used to explore these data sets in greater depth.  Future explorations of
data of this kind should include matched filter analysis.

The PiGSS results in the \boo field
represent approximately 20\% of the data from the PiGSS project and
were obtained in the early phases of routine telescope operations.
Subsequent analysis of higher quality 
daily observations in the Lockman Hole, ELAIS, and the Coma
cluster will probe significantly deeper into the radio transient and variable
population.  PiGSS observations will also characterize variability on time scales
of 1 year for $\sim 10^5$ radio sources in the North Galactic Cap.

The limits we have placed are useful constraints on the population of active high energy
objects in the Universe.  These limits explore the populations of tidal disruption events,
radio supernovae,
orphan gamma-ray burst afterglows, radio counterparts to GW sources,
extreme scattering events, intra-day variable sources,
and flare stars.

\acknowledgments

The authors would like to acknowledge the generous support of the Paul
 G. Allen Family
 Foundation, who have provided major support for design, construction,
 and operations of
 the ATA. Contributions from Nathan Myhrvold, Xilinx Corporation, Sun
 Microsystems,
 and other private donors have been instrumental in supporting the ATA.
 The ATA has been
 supported by contributions from the US Naval Observatory in addition
 to National Science
 Foundation grants AST-050690, AST-0838268, and AST-0909245.


\begin{deluxetable}{lrrrrrrrrrr}
\rotate
\tabletypesize{\scriptsize}
\tablecaption{Source Statistics \label{tab:stats}}
\tablehead{ 
\colhead{Source} & \colhead{$S$} & \colhead{Gain} & 
\colhead{$\chi^2_d$} & \colhead{$\chi^2_m$} & 
\colhead{$\sigma_{m,d}$} & \colhead{$\sigma_{m,m}$} & 
\colhead{$\min{m_{i,d}}$} & \colhead{$\max{m_{i,d}}$} & 
\colhead{$\min{m_{i,m}}$} & \colhead{$\max{m_{i,m}}$} }
\startdata
J142318+344210 & $   18.0 \pm    4.4$ & 15.14 &  1.69 &  4.62 &  9.04 &  1.31 & -6.97 &  6.81 & -2.91 &  1.15 \\ 
J142402+344518 & $   16.5 \pm    2.6$ & 8.82 &  1.43 &  1.03 &  4.03 &  0.57 & -5.56 &  4.38 & -0.93 &  0.57 \\ 
J142421+343857 & $    6.2 \pm    1.7$ & 5.70 &  1.49 &  8.99 &  1.60 &  0.95 & -8.17 & 11.54 & -1.51 &  1.83 \\ 
J142426+343602 & $   30.3 \pm    1.5$ & 5.01 &  0.94 &  1.68 &  0.54 &  0.16 & -1.16 &  0.86 & -0.23 &  0.20 \\ 
J142430+341914 & $    4.0 \pm    1.1$ & 3.85 &  1.83 &  0.82 &  5.76 &  0.63 & -12.92 & 12.39 & -1.28 &  0.78 \\ 
J142440+343757 & $   10.0 \pm    1.3$ & 4.47 &  1.25 &  1.14 &  1.75 &  0.47 & -3.81 &  2.60 & -0.73 &  0.77 \\ 
J142445+341832 & $   42.0 \pm    0.9$ & 3.21 &  0.72 &  0.76 &  0.29 &  0.06 & -0.84 &  0.49 & -0.07 &  0.10 \\ 
J142447+345317 & $    8.5 \pm    1.9$ & 6.40 &  1.31 &  3.33 & 36.67 &  0.26 & -7.50 &  7.29 & -0.31 &  0.52 \\ 
J142448+340957 & $    6.2 \pm    0.9$ & 3.19 &  1.21 &  1.27 &  1.93 &  0.48 & -4.13 &  4.71 & -0.75 &  0.55 \\ 
J142458+342527 & $    5.9 \pm    0.8$ & 2.92 &  1.80 &  1.02 & 11.58 &  0.61 & -4.93 &  4.12 & -0.46 &  1.06 \\ 
J142503+334405 & $    4.9 \pm    1.2$ & 4.68 &  1.37 &  3.29 & -16.59 &  1.32 & -7.45 &  8.84 & -0.33 &  7.03 \\ 
J142516+345310 & $   38.9 \pm    1.3$ & 4.56 &  0.75 &  2.45 &  0.44 &  0.11 & -1.65 &  0.51 & -0.21 &  0.11 \\ 
J142517+341606 & $    9.7 \pm    0.7$ & 2.32 &  0.99 &  2.39 &  0.94 &  0.27 & -2.34 &  2.23 & -0.48 &  0.20 \\ 
J142523+340944 & $    8.8 \pm    0.6$ & 2.23 &  1.56 &  0.46 &  1.57 &  0.23 & -4.25 &  1.50 & -0.18 &  0.48 \\ 
J142536+331215 & $   16.4 \pm    2.3$ & 10.13 &  1.45 &  0.87 &  2.89 &  0.51 & -6.13 &  4.93 & -1.18 &  0.76 \\ 
J142541+345848 & $  105.9 \pm    1.4$ & 4.19 &  1.11 &  3.35 &  0.20 &  0.16 & -0.56 &  0.21 & -0.23 &  0.19 \\ 
J142543+335544 & $  123.7 \pm    0.8$ & 2.33 &  1.08 &  0.84 &  0.12 &  0.03 & -0.26 &  0.50 & -0.04 &  0.08 \\ 
J142558+351311 & $    5.4 \pm    1.4$ & 5.33 &  1.32 &  1.68 &  2.23 &  1.01 & -6.15 &  6.75 & -1.21 &  1.50 \\ 
J142601+343129 & $    2.7 \pm    0.7$ & 1.79 &  1.26 &  0.97 &  2.31 &  0.35 & -5.80 &  7.47 & -0.59 &  0.52 \\ 
J142607+340433 & $   27.5 \pm    0.6$ & 1.66 &  0.82 &  5.05 &  0.24 &  0.06 & -0.60 &  0.50 & -0.10 &  0.06 \\ 
\enddata
\tablecomments{First 20 lines of the table; full table is included as data file.}
\end{deluxetable}

\begin{deluxetable}{rrrrrrrrrr}
\tabletypesize{\scriptsize}
\tablecaption{Probability of Fractional Modulation \label{tab:fluxhist}}
 \tablehead{ \colhead{$S_{min} $} & \colhead{$S_{max}$}  & \colhead{$ < -10.0 $}   & \colhead{$ < - 3.0 $}   & \colhead{$ < - 1.0 $}   & \colhead{$ < - 0.3 $}   & \colhead{$ >  0.3 $}   & \colhead{$ >  1.0 $}   & \colhead{$ >  3.0 $}   & \colhead{$ > 10.0 $}  \\
\colhead{(mJy)} & \colhead{(mJy)} }
\startdata
\multicolumn{9}{c}{Daily} \\\\
  1 &   10  &  0.0001 &  0.0428 &  0.2355 &  0.4851 &  0.2887 &  0.1587 &  0.0354 &  0.0001 \\ 
 10 &  100  & $ <  0.0002 $ &  0.0059 &  0.0669 &  0.3456 &  0.1227 &  0.0416 &  0.0084 &  0.0002 \\ 
100 & 1000  & $ <  0.0036 $ & $ <  0.0036 $ & $ <  0.0036 $ & $ <  0.0036 $ &  0.0036 & $ <  0.0036 $ & $ <  0.0036 $ & $ <  0.0036 $ \\ 
\multicolumn{9}{c}{Monthly} \\\\
  1 &   10  & $ <  0.0007 $ & $ <  0.0007 $ &  0.0439 &  0.2346 &  0.1661 &  0.0213 &  0.0013 & $ <  0.0007 $ \\ 
 10 &  100  & $ <  0.0024 $ & $ <  0.0024 $ &  0.0189 &  0.0615 &  0.0284 &  0.0047 & $ <  0.0024 $ & $ <  0.0024 $ \\ 
100 & 1000  & $ <  0.0400 $ & $ <  0.0400 $ & $ <  0.0400 $ & $ <  0.0400 $ & $ <  0.0400 $ & $ <  0.0400 $ & $ <  0.0400 $ & $ <  0.0400 $ \\ 
\enddata
\end{deluxetable}

\end{document}